\documentclass[twocolumn]{aastex631}

\usepackage{hyperref}
\usepackage{graphicx}
\usepackage{subcaption}
\usepackage{placeins}
\usepackage{caption}

\newcommand{\ha}{${\rm H\alpha}$}
\newcommand{\hb}{${\rm H\beta}$}

\newcommand{\simgt}{\,\rlap{\lower 3.5 pt \hbox{$\mathchar \sim$}} \raise1pt\hbox {${>}$}\,}
\newcommand{\simlt}{\,\rlap{\lower 3.5 pt \hbox{$\mathchar \sim$}} \raise1pt\hbox {$<$}\,}

\newcommand{\oiii}{[\textrm{O}~\textsc{iii}]}

\usepackage{threeparttable}

\begin{document}

\title{How I Wonder What You Are — JWST’s Little Red Dots do not TWINKLE}

\correspondingauthor{Zhaoran Liu}
\email{zrliu@mit.edu}
\author[0009-0002-8965-1303]{Zhaoran Liu}
\affiliation{MIT Kavli Institute for Astrophysics and Space Research, 70 Vassar Street, Cambridge, MA 02139, USA}

\author[0000-0003-3997-5705]{Rohan P. Naidu}
\altaffiliation{NASA Hubble Fellow, Pappalardo Fellow}
\affiliation{MIT Kavli Institute for Astrophysics and Space Research, 70 Vassar Street, Cambridge, MA 02139, USA}

\author[0000-0002-1174-2873]{Amy Secunda}
\affiliation{Center for Computational Astrophysics, Flatiron Institute, New York, NY 10010, USA
}

\author[0000-0002-5612-3427]{Jenny E. Greene}
\affiliation{Department of Astrophysical Sciences, Princeton University, Princeton, NJ 08544, USA}

\author[0000-0003-2871-127X]{Jorryt Matthee}
\affiliation{Institute of Science and Technology Austria (ISTA), Am Campus 1, 3400 Klosterneuburg, Austria}

\author[0000-0002-0302-2577]{John Chisholm}
\affiliation{Department of Astronomy, The University of Texas at Austin, Austin, TX, USA}
\affiliation{Cosmic Frontier Center, The University of Texas at Austin, Austin, TX 78712, USA}

\author[0000-0002-2380-9801]{Anna de Graaff}
\altaffiliation{Clay Fellow}
\affiliation{Center for Astrophysics, Harvard \& Smithsonian, 60 Garden St, Cambridge, MA 02138, USA}
\affiliation{Max-Planck-Institut f\"ur Astronomie, K\"onigstuhl 17, D-69117 Heidelberg, Germany}

\author[0000-0002-6265-2675]{Luke Robbins}
\affiliation{Department of Physics \& Astronomy, Tufts University, Medford, MA 02155, USA}

\author[0000-0002-0243-6575]{Jacqueline Antwi-Danso}
\altaffiliation{Dunlap Fellow}
\affiliation{David A. Dunlap Department of Astronomy \& Astrophysics, University of Toronto, 50 St George Street, Toronto, ON M5S 3H4, Canada}
\affiliation{Dunlap Institute for Astronomy \& Astrophysics, University of Toronto, 50 St George Street, Toronto, ON M5S 3H4, Canada}

\author[0000-0003-2680-005X]{Gabriel Brammer}
\affiliation{Cosmic Dawn Center (DAWN), Copenhagen, Denmark}
\affiliation{Niels Bohr Institute, University of Copenhagen, Jagtvej 128, K{\o}benhavn N, DK-2200, Denmark}

\author[0009-0007-3791-7890]{Wendy Q. Sun}
\affiliation{MIT Kavli Institute for Astrophysics and Space Research, 70 Vassar Street, Cambridge, MA 02139, USA}

\author[0000-0003-2895-6218]{Anna-Christina Eilers}
\affiliation{MIT Kavli Institute for Astrophysics and Space Research, 70 Vassar Street, Cambridge, MA 02139, USA}
\affiliation{Department of Physics, Massachusetts Institute of Technology, Cambridge, MA 02139, USA}

\author[0000-0001-7201-5066]{Seiji Fujimoto}
\affiliation{David A. Dunlap Department of Astronomy \& Astrophysics, University of Toronto, 50 St George Street, Toronto, ON M5S 3H4, Canada}
\affiliation{Dunlap Institute for Astronomy \& Astrophysics, University of Toronto, 50 St George Street, Toronto, ON M5S 3H4, Canada}

\author[0000-0001-6278-032X]{Lukas J. Furtak}
\affiliation{Department of Astronomy, The University of Texas at Austin, Austin, TX, USA}
\affiliation{Cosmic Frontier Center, The University of Texas at Austin, Austin, TX 78712, USA}

\author[0000-0003-0172-0854]{Erin Kara}
\affiliation{MIT Kavli Institute for Astrophysics and Space Research, 70 Vassar Street, Cambridge, MA 02139, USA}
\affiliation{Department of Physics, Massachusetts Institute of Technology, Cambridge, MA 02139, USA}

\author[0000-0002-5588-9156]{Vasily Kokorev}
\affiliation{Department of Astronomy, The University of Texas at Austin, Austin, TX, USA}
\affiliation{Cosmic Frontier Center, The University of Texas at Austin, Austin, TX 78712, USA}

\author[0000-0001-9002-3502]{Danilo Marchesini}
\affiliation{Department of Physics \& Astronomy, Tufts University, Medford, MA 02155, USA}

\author[0000-0001-5851-6649]{Pascal A.\ Oesch}
\affiliation{Department of Astronomy, University of Geneva, Chemin Pegasi 51, 1290 Versoix, Switzerland}
\affiliation{Cosmic Dawn Center (DAWN), Copenhagen, Denmark}
\affiliation{Niels Bohr Institute, University of Copenhagen, Jagtvej 128, K{\o}benhavn N, DK-2200, Denmark}

\author[0000-0002-2361-7201]{Justin D. R. Pierel}
\affiliation{Space Telescope Science Institute, 3700 San Martin Dr., Baltimore, MD 21218, USA}

\author[0000-0002-6196-823X]{Xuejian Shen}
\affiliation{MIT Kavli Institute for Astrophysics and Space Research, 70 Vassar Street, Cambridge, MA 02139, USA}

\author[0000-0003-3769-9559]{Robert A. Simcoe}
\affiliation{MIT Kavli Institute for Astrophysics and Space Research, 70 Vassar Street, Cambridge, MA 02139, USA}
\affiliation{Department of Physics, Massachusetts Institute of Technology, Cambridge, MA 02139, USA}

\author[0000-0001-5586-6950]{Alberto Torralba}
\affiliation{Institute of Science and Technology Austria (ISTA), Am Campus 1, 3400 Klosterneuburg, Austria}

\author[0000-0001-8593-7692]{Mark Vogelsberger}
\affiliation{Department of Physics and Kavli Institute for Astrophysics and Space Research, Massachusetts Institute of Technology, Cambridge, MA 02139, USA}
\affiliation{Fachbereich Physik, Philipps Universit\"at Marburg, D-35032 Marburg, Germany}

\begin{abstract}



Little Red Dots (LRDs) are a population of compact, red sources that have emerged as one of the most puzzling findings of JWST. Variability provides a direct probe of their central engines. Here we present the first joint spectroscopic and photometric time-domain study of LRDs undertaken with the JWST TWINKLE slitless spectroscopy program. Surveying the FRESCO GOODS-North legacy field, TWINKLE monitors a complete, \ha-flux-limited sample of 18 LRDs at $z=3.9 - 6.8$, achieving a rest-frame baseline of $\sim 140 - 220$ days. We detect no variability in photometry, \ha\ line flux, or line shape across the sample. If LRDs resembled AGN in reverberation mapping samples --- the foundation for black hole mass calibrations and luminosity scaling relations --- we would expect $>10$ sources to show measurable fluctuations. Observing none implies a $5.9\sigma$ deficit. The non-detections hold across all broad \ha\ emitters within TWINKLE's field of view -- the 18 V-shaped LRDs as well as 9 non-LRDs. Comparison with simulated light curves disfavors sub-Eddington accretion and is instead consistent with super-Eddington accretion, other mechanisms that suppress variability, or perhaps no AGN whatsoever. If LRDs do harbor black holes, calibrations derived from sub-Eddington systems may not apply, thereby explaining JWST's apparently ``overmassive” black holes. These observations provide unique constraints on the physics of one of the most enigmatic populations discovered by JWST.

\end{abstract}



\section{Introduction} \label{sec:intro}

JWST has revealed an unexpectedly abundant population of compact, red sources largely at $z \approx 4$--$9$, dubbed ``Little Red Dots'' (LRDs) \citep[e.g.,][]{matthee23, Kokorev24, Kocevski25, Labbe25, Zhang26, Hviding25}. LRDs display a characteristic V-shaped spectral energy distribution with a blue UV continuum transitioning to a steep red rest-frame optical slope. They also exhibit broad Balmer emission and point-like morphology, seemingly indicative of AGN activity. They were therefore initially interpreted as dust-reddened AGN, yet they are unlike any well-studied AGN population. They are remarkably X-ray faint \citep{kocevski23, Yue24, Sacchi25} and appear largely deficient in hot or cold dust emission \citep[e.g.,][]{Williams24, Xiao25, Setton25, Casey25}, though weak dust emission has been reported in some cases \citep[e.g.,][]{Delvecchio25,Barro26, Brazzini26}. While the high Balmer decrements (\ha/\hb\ $\sim 9$) at face value are consistent with significant reddening, the IR weakness argues against a classical dust origin. Steep Balmer breaks detected in some LRDs \citep[e.g.,][]{Kokorev24_balmerbreak, Labbe24a, Wang25, Hviding25, Setton25b, Leung25, Perez-Gonzalez26} have been interpreted as evidence for massive, evolved stellar populations \citep[e.g.,][]{Bruzual83, Wang24a, Furtak24, Labbe23, deGraaff25a, Morishita25, Witten25}, but clustering measurements point to host galaxy stellar masses of only $\sim10^{7}$--$10^{8}\,M_\odot$ \citep[e.g.,][]{Matthee25, Pizzati25, Lin26}, difficult to reconcile with this picture. Taken together, these observations suggest LRDs are neither classical dust-reddened AGN nor massive evolved stellar populations, and their physical nature remains an open question.


The discovery of MoM-BH*-1 and the Cliff provided new clues toward understanding this puzzle \citep{Naidu25, deGraaff25}. Displaying Balmer breaks far exceeding the theoretical maximum for any stellar population, these works proposed that LRDs are powered by ``Black Hole Stars'' (BH*) --- black holes embedded within dense gas cocoons whose emergent spectra bear classical signatures of black holes (e.g., broad Balmer lines) as well as stars (e.g., Balmer breaks). BH*s embedded in host galaxies explain a variety of LRD observables \citep[e.g.,][]{deGraaff25b, Barro26,Sun26}, including e.g., the high Balmer decrements via radiative transfer effects such as collisional excitation and resonant scattering without invoking large amounts of dust \citep[e.g.,][]{Torralba25, Yan25, Chang26}. Theoretical models and radiative transfer calculations have provided a physical basis for this framework \citep[e.g.,][]{Inayoshi25, Cantiello25, Liu25, Kido25, Begelman26, Santarelli26, Sneppen26, RomanGarza26}, and a number of observational studies have examined its predictions \citep[e.g.,][]{Ji25, Torralba25, Kokorev25, DEugenio25, deGraaff25b, Barro26, Rusakov26, Sun26, Torralba2026}. 

Alternate hypotheses for the central engines of LRDs include actual stars, in particular, supermassive ($\approx10^{5} M_{\rm{\odot}}$) stars (SMSs) \citep[e.g.,][]{Furtak23, Zwick25, Nandal26, Chisholm26} and traditional AGN accretion disks with modifications such as e.g., embedded stars, extremely dense gas along the line of sight, broad-line region stratification, and higher covering fractions \citep[e.g.,][]{Chen26, Scholtz26,Madau26}.



Regardless of the specific physical framework, a key tension persists: virial mass estimates based on local calibrations \citep[e.g.,][]{GreeneHo05} imply ``overmassive'' black holes whose mass is comparable to their host galaxies \citep[e.g.,][]{Chen25hostifany, Jones25, Maiolino25qso1}, pushing the implied black hole mass density at $z\sim5$ to uncomfortable levels \citep[e.g.,][]{Inayoshi24}. It is possible that the broad lines arise from non-virial processes such as radiative transfer in dense gas \citep[e.g.,][]{Naidu25,Torralba25, Rusakov26,Sun26,Chang26}, and that these systems host lower-mass black holes accreting at or above the Eddington limit \citep[e.g.,][]{King24, Lupi24, Lambrides24, Umeda25, Greene26, Secunda26, Liu26}. Distinguishing among these scenarios demands a probe that is sensitive to the accretion state yet independent of the assumed spectral model.



Variability offers precisely such a probe. Standard AGN accretion disks are expected to vary on hour-to-year timescales. AGN variability is traditionally interpreted using the lamppost model, where time-varying X-rays are reprocessed by the accretion disk, driving correlated UV-optical variability \citep[e.g.,][]{Clavel92, Ulrich97, McHardy06, Edelson15}, while the X-ray variability itself traces the innermost accretion flow (see \citealt{Kara25} for a recent review). There is also growing evidence for intrinsic UV-optical variability on month-to-year timescales, driven by thermal or magnetic fluctuations in the disk itself \citep[e.g.,][]{Lyubarskii97, Beard25, Hagen24, Secunda24, Secunda25}. The amplitude and timescales of variability encode fundamental properties --- black hole mass, accretion rate, and disk geometry --- making it a direct diagnostic of the detailed physics of the system. 

Crucially, the expected variability signatures depend on the nature of the system. If LRDs are powered by a standard AGN disk, we would expect stochastic UV-optical variations. If instead the central engine, whether an AGN or an SMS, is embedded within a dense gas envelope, short-timescale variations would be significantly damped as the envelope reprocesses and smooths the signal. The presence, absence, or character of variability therefore provides a powerful, model-independent constraint on the nature of LRDs.


Studies have sought LRD variability and found little to no signs \citep[e.g.,][]{Kokubo25, Tee25, Burke25, Zhang25a, Stone25} with a handful of exceptions \citep[][]{Furtak25, Ji25, zhang25century, Naidu25, DEugenio26}. However, these studies largely exploited serendipitous multi-epoch coverage with inhomogeneous depths and limited samples, had to overcome cross-instrument calibration (e.g., HST vs. JWST or NIRCam vs. NIRSpec), and have mostly relied on photometry alone rather than simultaneous line+continuum monitoring. Furthermore, repeated observations with NIRSpec are challenging -- small changes in slit position sampling different regions of the target can produce drastic changes in spectra. And unless the source is observed in the exact same slitlets at the exact same position, systematics such as variations in the path-loss and slit-loss corrections \citep[e.g.,][]{degraaff25rubies} complicate the interpretation of apparent variability.

Here we present the first results from TWINKLE (JWST Cycle~4, PID~7404, PIs: R. P. Naidu, J. Matthee and J. Chisholm), a NIRCam slitless spectroscopy program specifically designed to probe the LRD variability systematically. In contrast to NIRSpec multi-object spectroscopy, the slitless approach observes every source in the field simultaneously, yielding an \ha-flux-limited sample with a clean, unbiased selection function. Furthermore, this approach sidesteps systematics associated with the exact positioning of an object in a slit. By combining TWINKLE with the Cycle~1 FRESCO survey and two pure-parallel slitless programs in the same fields, we achieve the longest rest-frame baseline currently available with JWST in blank fields ($\sim$200 days) for a well-defined broad-line sample. Here we explore continuum variability for 27 sources and broad \ha\ line variability for the 10 sources with multi-epoch grism coverage. We compare these measurements against two benchmarks: luminosity-matched AGN from the SDSS Reverberation Mapping sample \citep{Shen24} and simulated light curves spanning sub- to super-Eddington accretion regimes \citep{Secunda26}.

We introduce our targets, observations, data reduction and sensitivity estimation in Section \ref{sec:data}. Our analyses of photometric and spectroscopic variability, as well as the comparison between empirical and theoretical expectations, are presented in Section~\ref{sec:results}. Throughout the paper, we adopt the AB magnitude system \citep{oke83, Fukugita96}, cosmological parameters of $\Omega_{\rm{m}}=0.3$, $\Omega_\Lambda=0.7$, $H_0=70$\,km\,s$^{-1}\,{\rm Mpc}^{-1}$, and the \citet{Chabrier03} initial mass function (IMF).

\section{Targets, Observations and Data Reduction} \label{sec:data}
\begin{figure*}[!htb]
\centering
\includegraphics[width=0.7\textwidth]{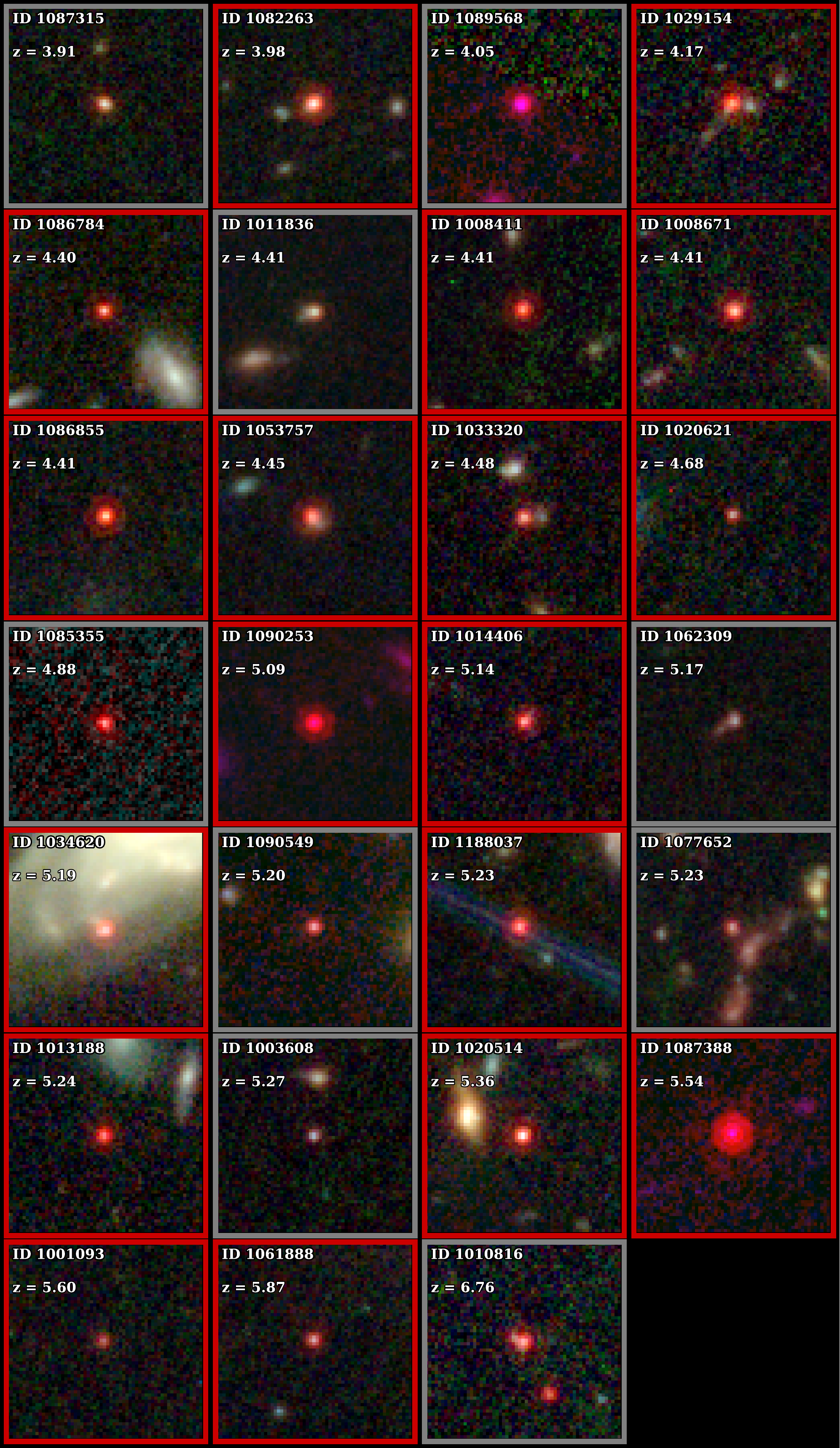}
\caption{RGB cutouts of the 27 broad \ha\ emitters, ordered by increasing spectroscopic redshift ($z = 3.91$--$6.76$). Each $3.\!''0\, \times\,3.\!''0$ stamp is constructed from JWST/NIRCam mosaics (F115W, F200W or F210M, F444W), where F210M is used when F200W is unavailable. Labels indicate the spectroscopic redshift and JADES DR5 ID \citep{Robertson26}. Sources satisfying the LRD selection criteria (Section \ref{sec:lrd_selection}; 18 out of 27 sources) are outlined in red.}
\label{fig:stamps}
\end{figure*}


We compile a parent sample of known broad H$\alpha$ emitters at $3.9 < z < 6.8$ in the GOODS-N field from previous spectroscopic surveys. A total of 19 broad \ha\ emitters at $3.9 < z < 5.5$ were identified from NIRCam grism surveys \citep{matthee23, CoveloPaz25, Zhang26}, where ``broad'' is defined as FWHM$_{{\rm H}\alpha} > 1000\ \rm km\ s^{-1}$, and 10 additional broad \ha\ emitters confirmed by JWST/NIRSpec spectroscopy at $4.4 < z < 6.8$ \citep{Maiolino24}, identified via two-component (narrow + broad) Gaussian fits to the Balmer lines, where (i) the broad component is at least twice as wide as the narrow component, (ii) the broad component is detected at ${>}5\sigma$ significance, and (iii) no corresponding broad emission is present in forbidden lines. 

We monitor this sample with NIRCam F444W imaging and WFSS data from FRESCO (Cycle~1, GO-1895, \citealt{Oesch23}; 2023 Feb) as the photometric and spectroscopic baseline, and two new epochs from TWINKLE (Cycle~4, GO-7404; 2026 Jan and 2026 Mar). TWINKLE is explicitly designed to match the spectroscopic and imaging depth of FRESCO at $\approx3.9-4.5\mu$m, the wavelengths where the broad H$\alpha$ line of the \citet{Matthee24} LRD sample falls. Additional intermediate epochs are provided by two archival pure-parallel WFSS programs, Slitless Areal Pure-Parallel HIgh-Redshift Emission Survey (SAPPHIRES; GO-6434; PI: E. Egami; \citealt{Sun25}) and the Public Observation Pure Parallel Infrared Emission-Line Survey (POPPIES; GO-5398; PI: J. Kartaltepe). The third and final TWINKLE epoch is scheduled for end-April 2026 and will be incorporated in an update to this paper.

Among the grism-selected sources, 18 out of 19 have photometric coverage in our first two epochs, with one source falling outside the footprint due to a slight position angle (PA) offset. Of the NIRSpec-confirmed sources, 9 out of 10 are covered by available imaging data, yielding a total photometric monitoring sample of 27 sources. \ha\ line monitoring is available for 10 sources based on overlapping multi-epoch grism observations, including eight grism-selected sources and two originally identified with NIRSpec. The remaining sources lack grism line monitoring because their \ha\ emission falls outside the wavelength coverage of the NIRCam grism at their respective redshifts. The false-color images of these 27 sources are shown in Figure~\ref{fig:stamps}.

\subsection{LRD Classification}\label{sec:lrd_selection}

We classify our sample into LRDs and non-LRDs based on three complementary criteria, and designate a source as an LRD if it meets at least one of them.        

The first criterion follows \citet{Kocevski25}, who identify LRDs by a V-shaped UV-to-optical SED: $\beta_{\rm UV} < -0.37$ and $\beta_{\rm opt} >0$, where the slopes are measured by weighted linear regression over redshift-dependent filter sets with $\mathrm{S/N}\geq 3$ per band. Uncertainties on $\beta$ are estimated by 1000 Monte Carlo realizations in flux space. The second criterion follows \citet{Kokorev24}, who apply broadband color cuts in two redshift regimes ($z<6$ and $z>6$) to select sources with blue UV and red optical colors. Both criteria additionally require morphological compactness $F(0.\!''2) / F(0.\!''1)<1.7$ \citep{Greene24, Labbe25}. For sources lacking F200W coverage, we adopt F210M as a substitute. Sources lacking F277W coverage are assessed by the \citet{Kocevski25} criterion alone. This can cause genuine LRDs to be misclassified as non-LRD, for instance, ID~1085355 lacks F277W and has F356W contaminated by \ha, leaving the optical slope unconstrained. The union of the photometric selections yields 80$\%$ completeness as estimated from tests on a NIRSpec-selected sample \citep{Hviding25}. 

The third criterion applies the V-shape definition to archival PRISM spectra following \citet{Hviding25}. A total of 18 sources have available PRISM spectroscopy from multiple JWST programs (JADES, GTO-1181, \citealt{Eisenstein26}; GO-5664, PI: J. Matthee; DIVER, GO-8018, PI: X. Lin). For sources whose rest-optical continuum S/N is insufficient to measure $\beta_{\rm opt}$, the spectroscopic criterion is inapplicable and their photometric classification is retained. Among the remaining sources with measurable continua, we identify 7 satisfying the V-shape criterion, of which one is not recovered by either of the two photometric selections.

\begin{figure*}[!htb]
\centering
\includegraphics[width=0.93\textwidth]{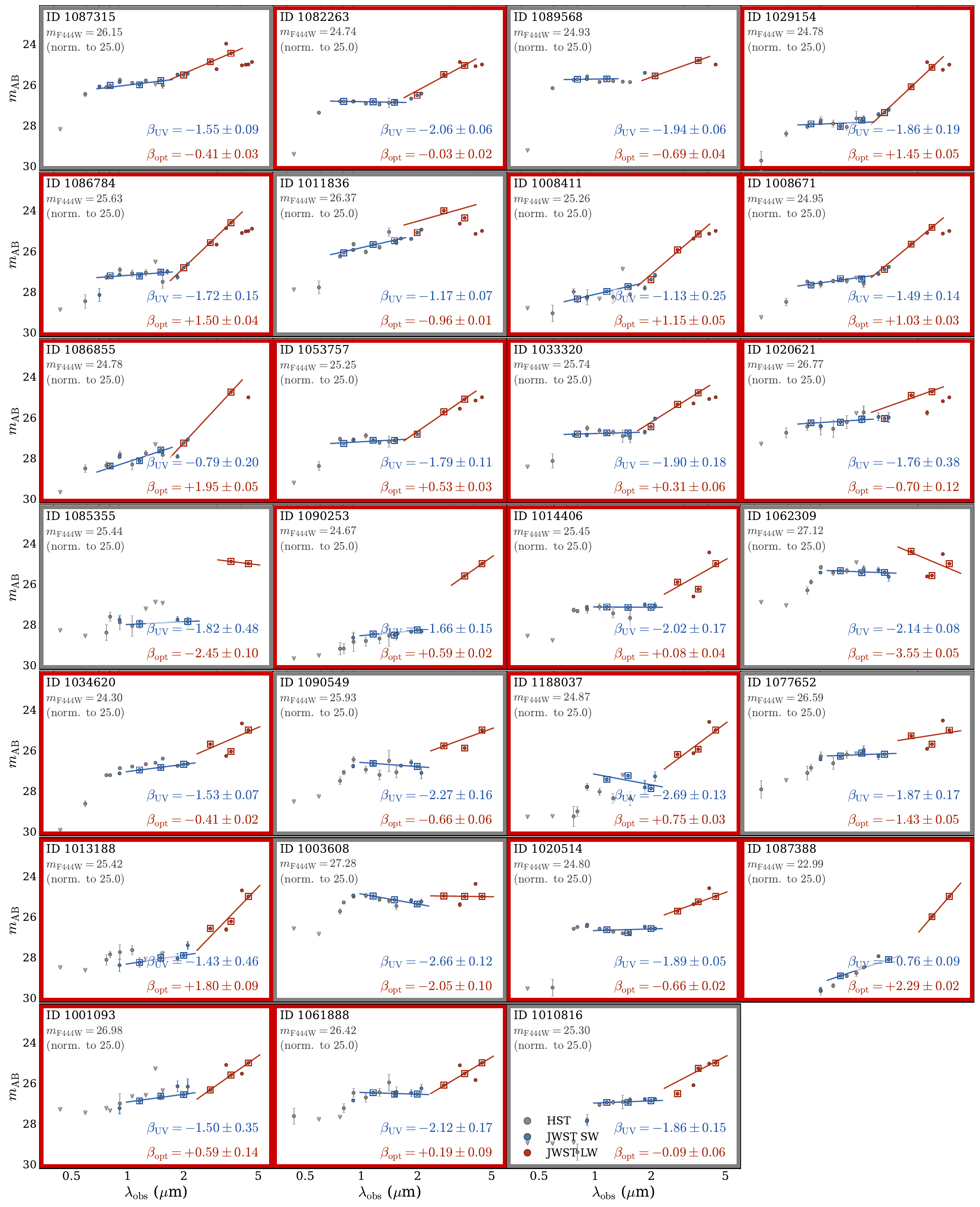}
\caption{
Observed-frame spectral energy distributions (SEDs) of the 27 broad-line \ha\ emitters in our sample. Each SED is normalized so that the F444W flux corresponds to $m_{\rm AB} = 25$. Filled circles show photometry from HST (black), JWST/NIRCam short-wavelength filters (blue), and long-wavelength filters (F277W and redder; dark red), based on aperture-corrected fluxes measured within a circular aperture of radius $r = 0.\!''15$ from the JADES photometric catalog \citep{Robertson26}. Open squares mark the photometric points used in the $\beta$-slope fitting. Blue and red lines indicate the best-fit UV and optical power-law slopes, $\beta_{\rm UV}$ and $\beta_{\rm opt}$, respectively. Sources satisfying any of the LRD criteria (Section~\ref{sec:lrd_selection}) are outlined in red.}
\label{fig:sed}
\end{figure*}

Of our 27 broad-line H$\alpha$ emitters, 18 are classified as LRDs by at least one criterion. The remaining 9 sources fail all criteria: 8 fail all three selections, and 1 has a red SED but does not satisfy the morphological compactness requirement. The spectral energy distributions of all 27 sources are shown in Figure~\ref{fig:sed}. We will return to the (lack of) distinction in terms of their variability between LRDs and non-LRDs in \S\ref{discuss3}.

\begin{figure*}
\centering

\begin{subfigure}{0.44\textwidth}
    \includegraphics[width=\linewidth]{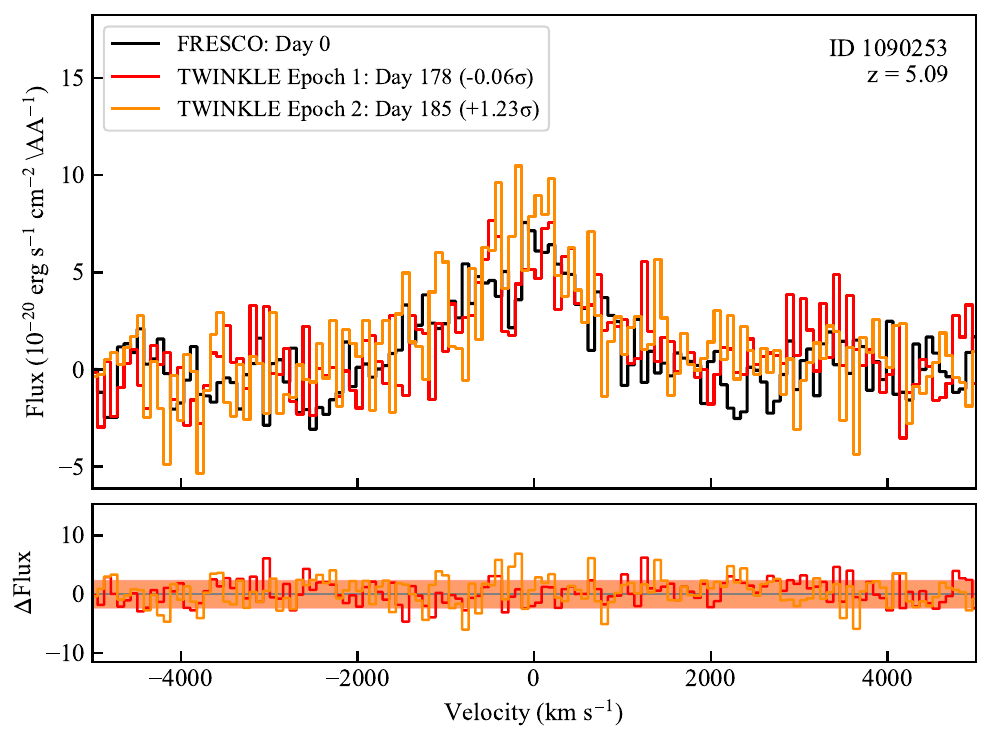}
\end{subfigure}
\begin{subfigure}{0.44\textwidth}
    \includegraphics[width=\linewidth]{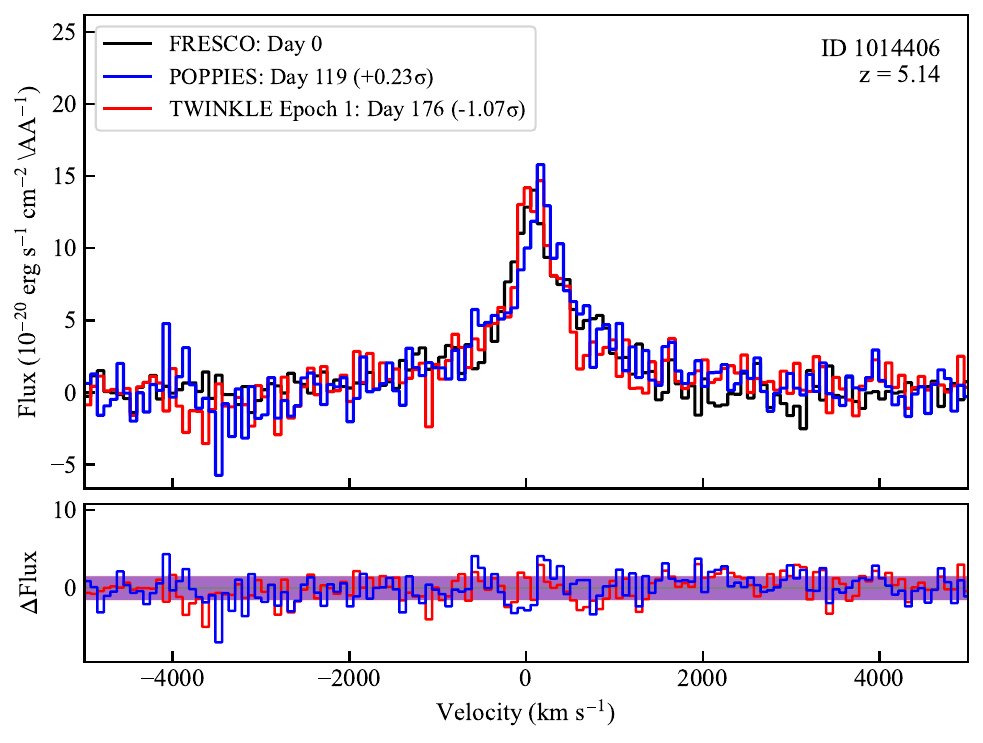}
\end{subfigure}

\vspace{-2.1em}

\begin{subfigure}{0.44\textwidth}
    \includegraphics[width=\linewidth]{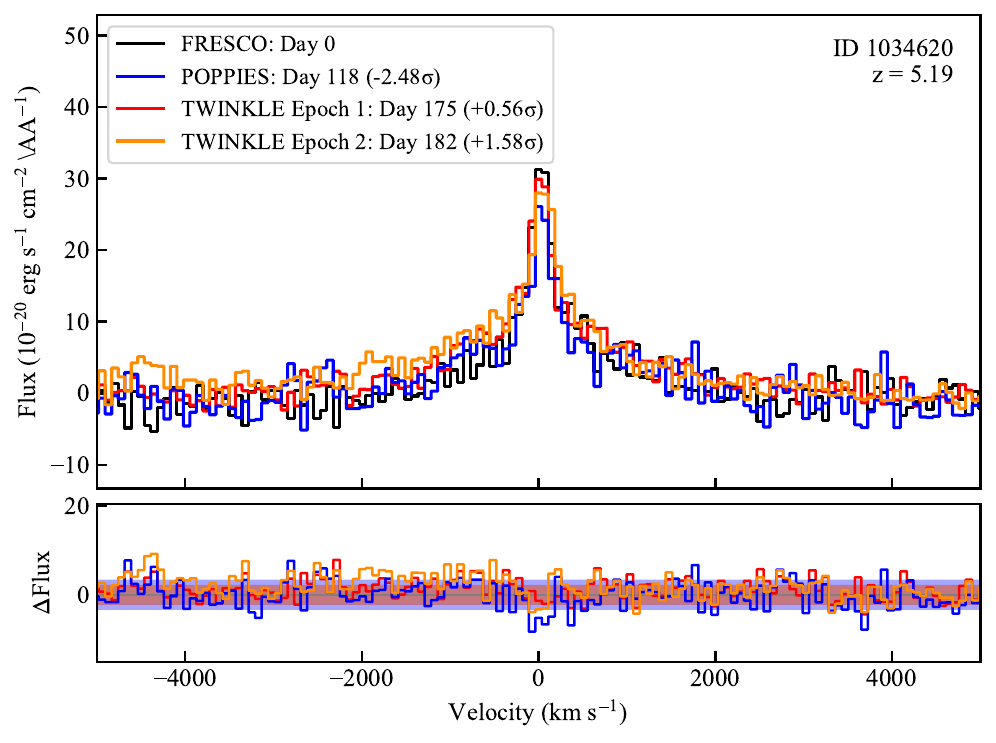}
\end{subfigure}
\begin{subfigure}{0.44\textwidth}
    \includegraphics[width=\linewidth]{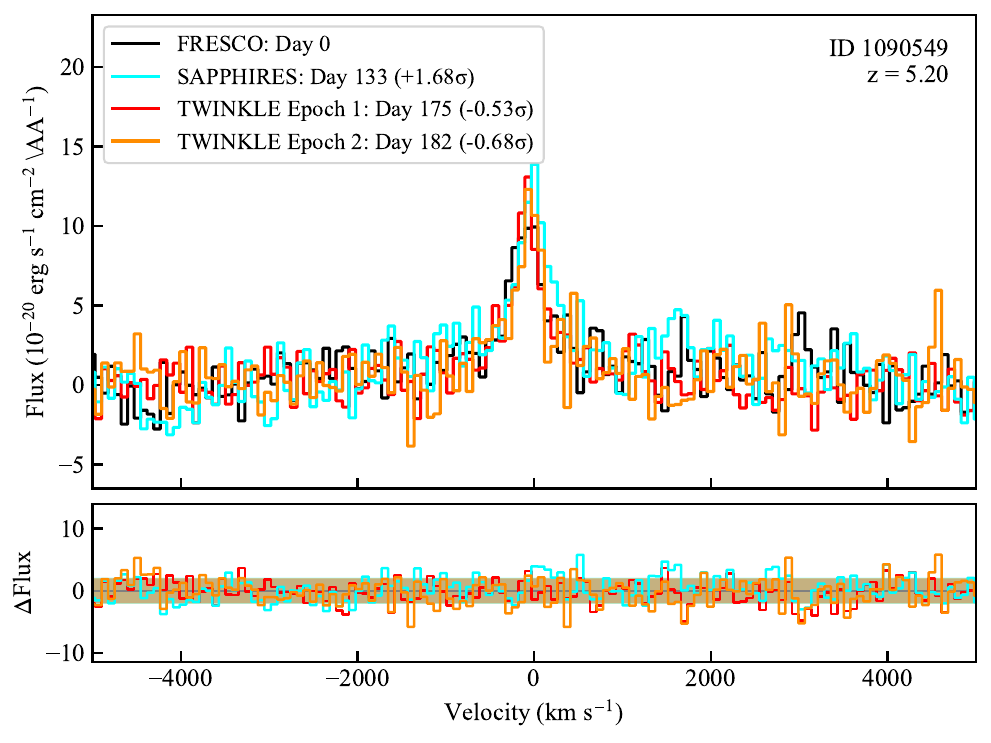}
\end{subfigure}

\vspace{-2.05em}

\begin{subfigure}{0.44\textwidth}
    \includegraphics[width=\linewidth]{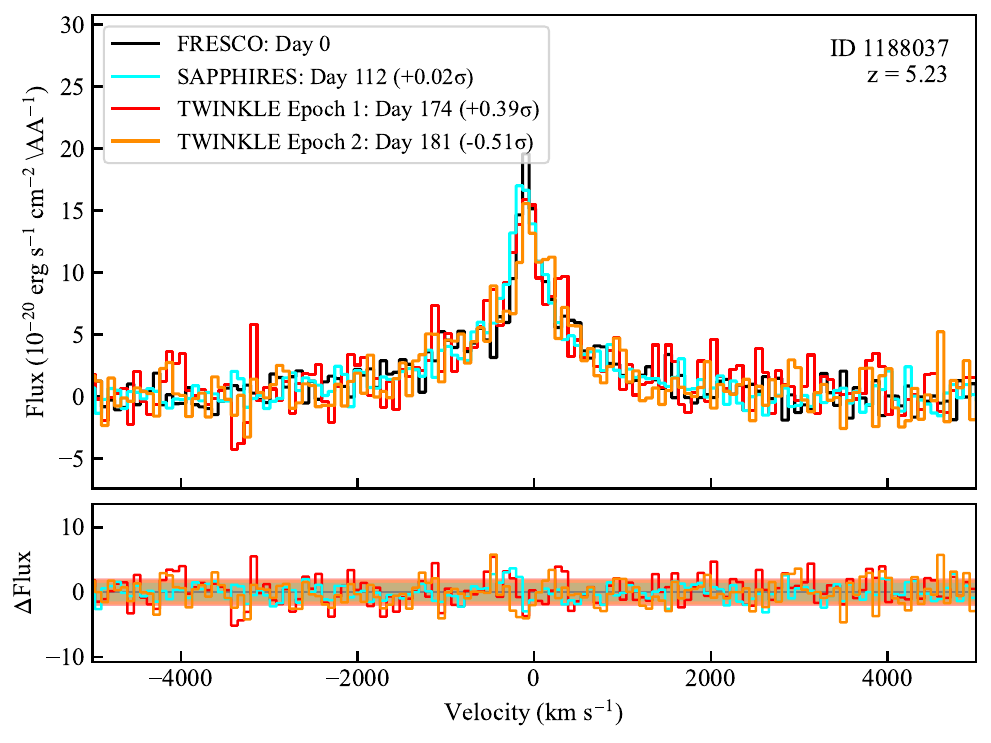}
\end{subfigure}
\begin{subfigure}{0.44\textwidth}
    \includegraphics[width=\linewidth]{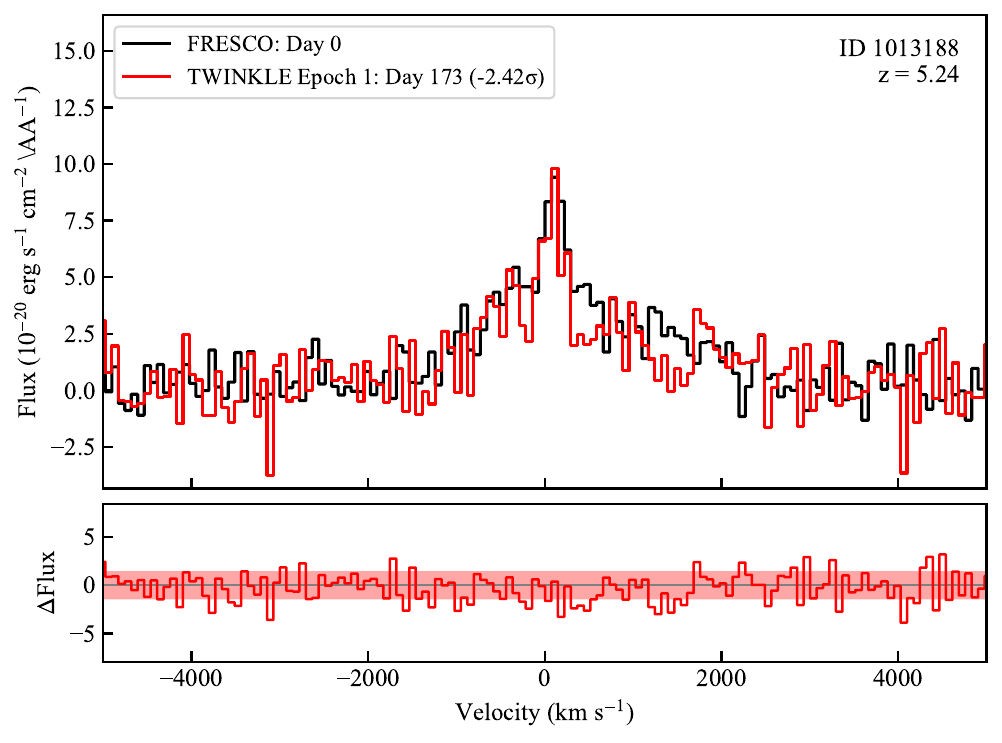}
\end{subfigure}

\vspace{-2.05em}

\begin{subfigure}{0.44\textwidth}
    \includegraphics[width=\linewidth]{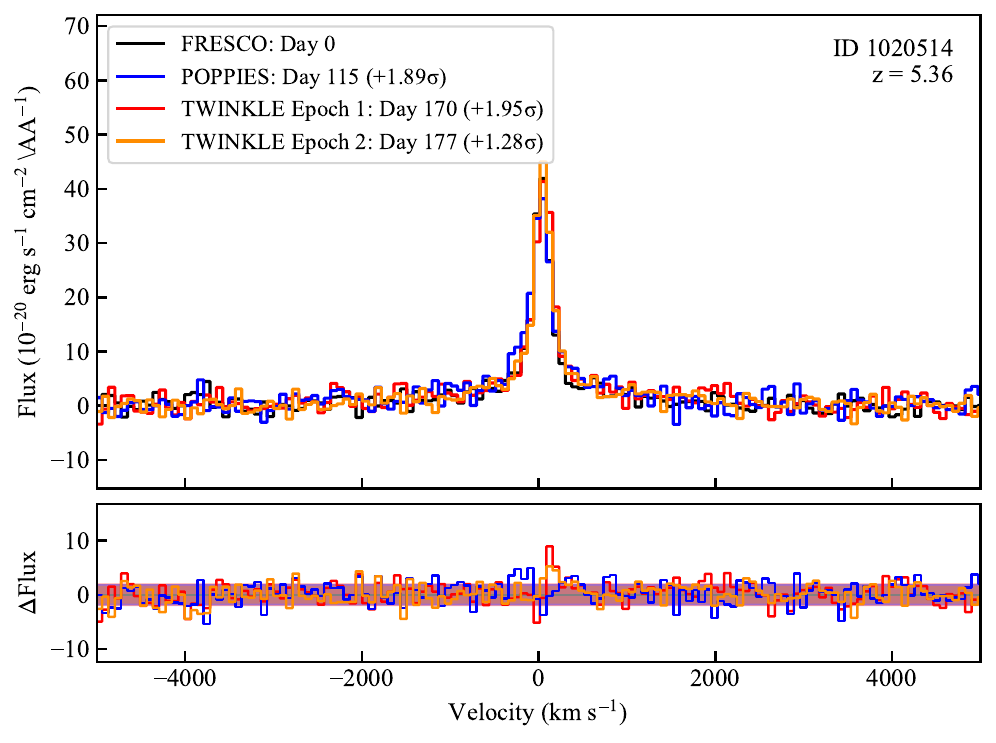}
\end{subfigure}
\begin{subfigure}{0.44\textwidth}
    \includegraphics[width=\linewidth]{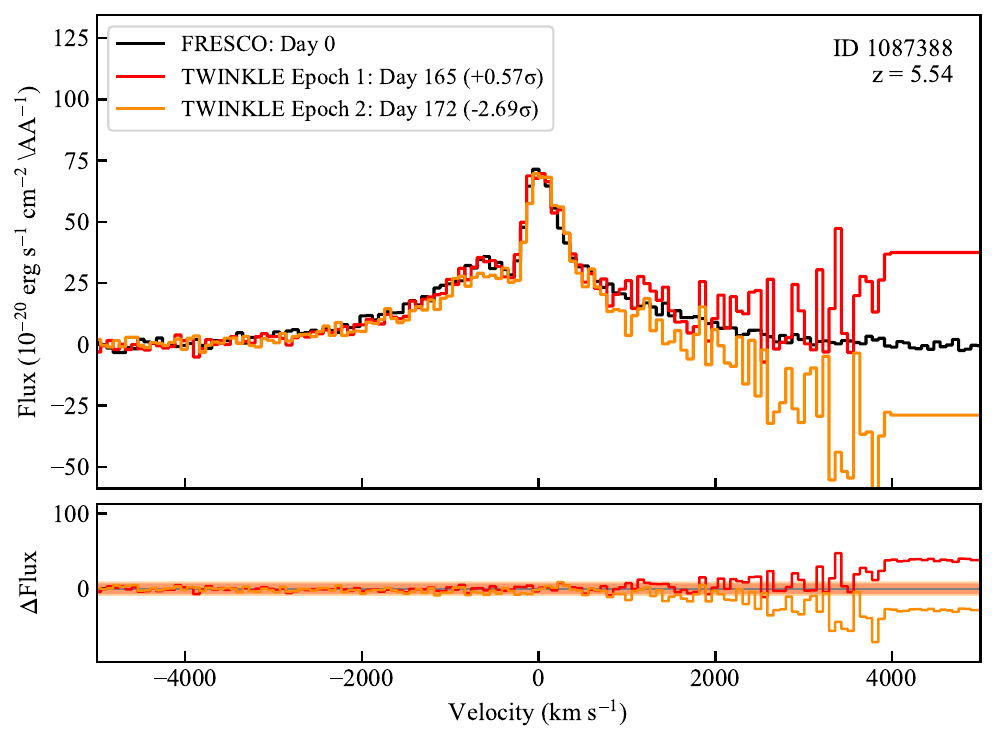}
\end{subfigure}

\caption{Multi-epoch H$\alpha$ line profiles of eight grism-selected sources. For each object, the top panel shows the optimally extracted 1D \ha\ spectrum observed by FRESCO (black) and TWINKLE (red for Epoch 1, orange for Epoch 2), with additional epochs from SAPPHIRES (cyan) and POPPIES (blue) where available. Rest-frame time is measured
relative to the FRESCO observations (February 2023). The bottom panel shows the flux difference spectra
($\Delta F = F_{\rm epoch} - F_{\rm FRESCO}$).
Shaded regions indicate the 1$\sigma$ empirical noise level estimated from line-free spectral sidebands. The integrated variability significance $S$ is labeled for each source as defined in Section~\ref{sec:wfss}.}
\label{fig:line_profile}
\end{figure*}

\begin{figure*}[!t]
\addtocounter{figure}{-1}  

\centering
\begin{subfigure}{0.44\textwidth}
    \includegraphics[width=\linewidth]{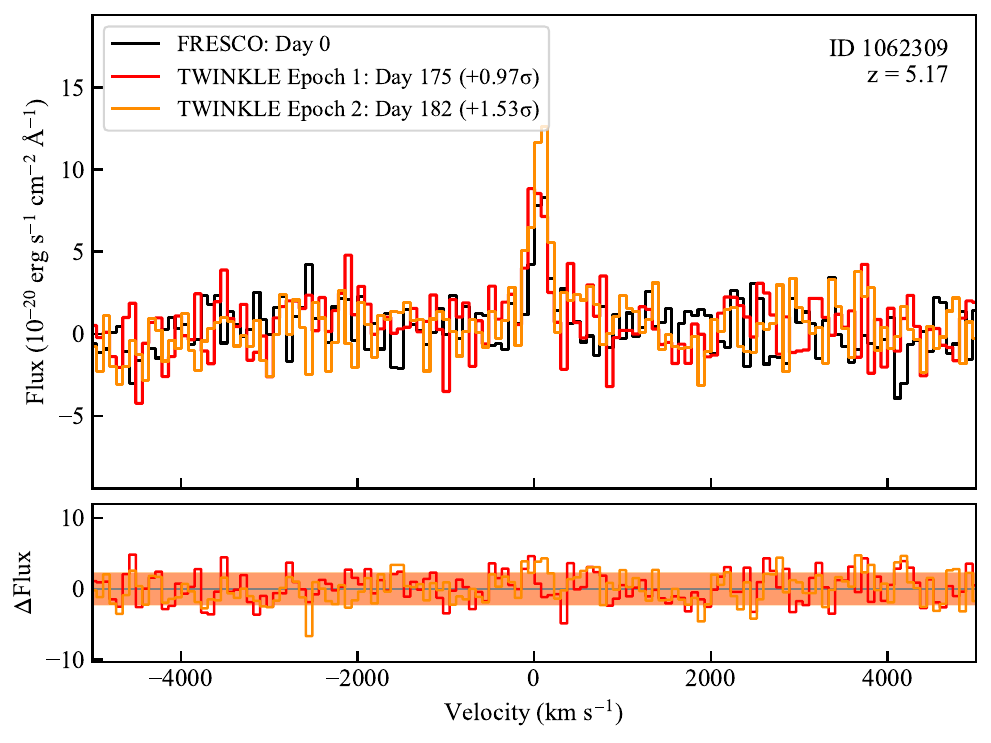}
\end{subfigure}
\begin{subfigure}{0.44\textwidth}
    \includegraphics[width=\linewidth]{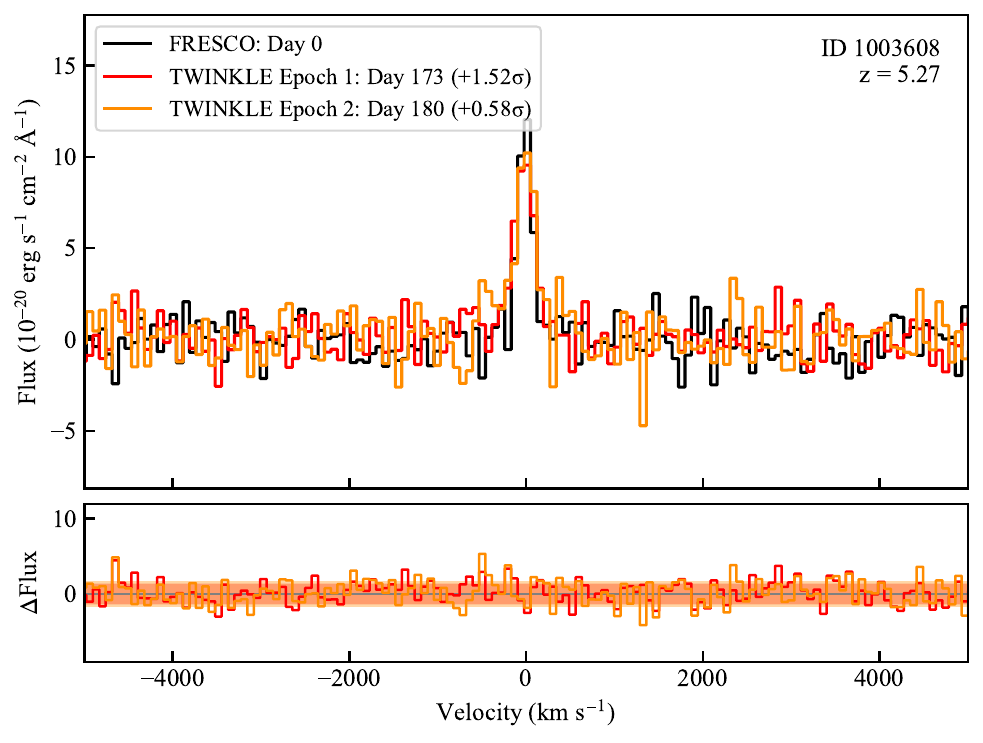}
\end{subfigure}

\caption{Continuation of Figure~\ref{fig:line_profile}, but for sources originally selected by NIRSpec.}
\end{figure*}




\begin{figure*}[!htb]
\centering
\includegraphics[width=0.91\textwidth]{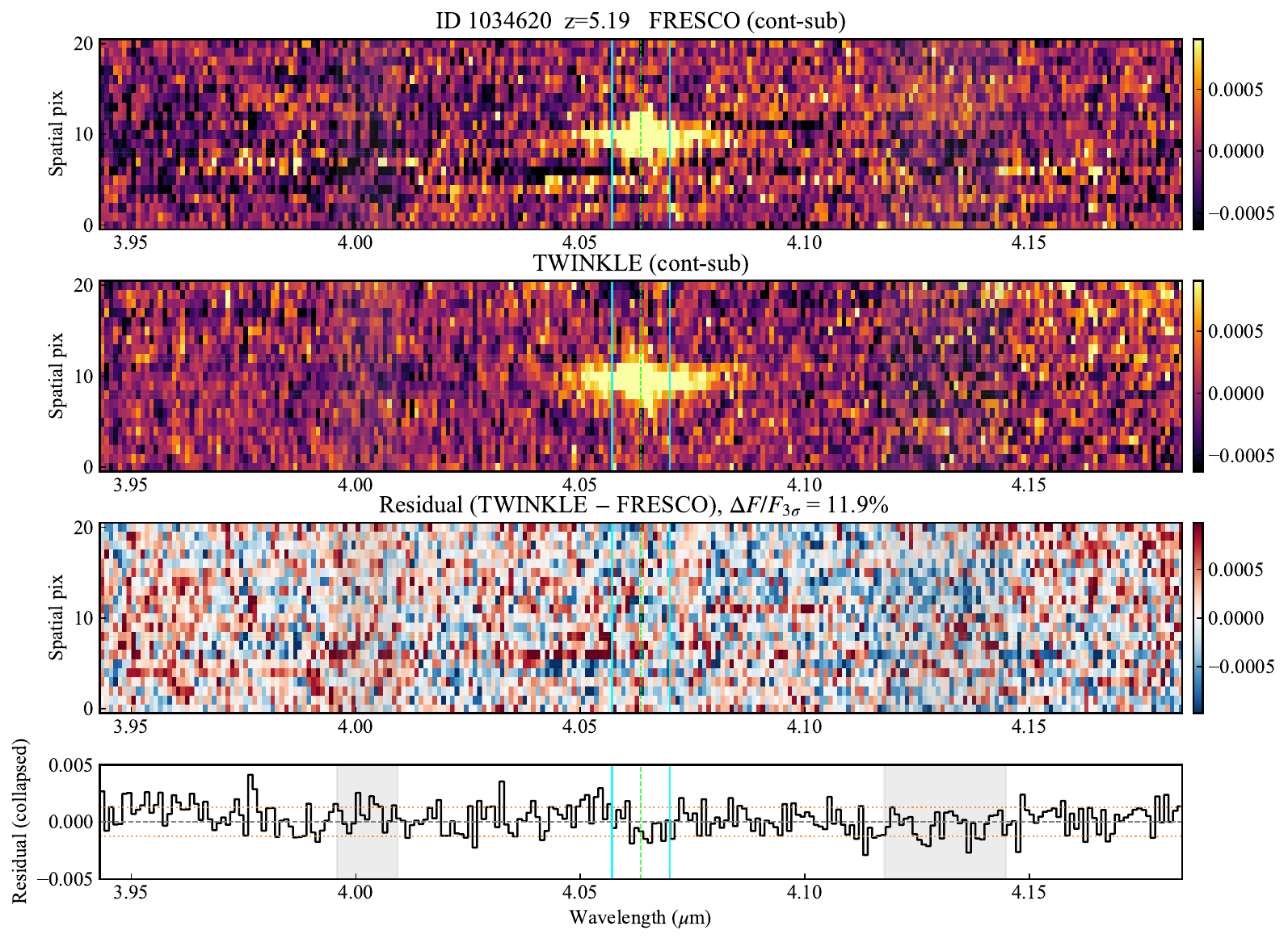}
\caption{Example 2D spectral residual analysis for ID1034620. Top to bottom: FRESCO continuum-subtracted 2D spectrum; TWINKLE continuum-subtracted 2D spectrum; residual map (TWINKLE $-$ FRESCO); optimally extracted 1D residual (black), with the orange dashed lines indicating $\pm1\sigma$ noise level estimated from the line-free regions on both sides of the \ha\ line (grey shading). Cyan vertical lines mark the \ha\ integration window ($|v| < \mathrm{FWHM}_{\rm broad}$, capped at 2000\,km\,s$^{-1}$); the dashed lime line marks the line center. The $3\sigma$ sensitivity limit on fractional \ha\ variability (Equation~\ref{equ:sensitivity}) is indicated in the header of the residual panel; for this source $|\Delta F/F|_{3\sigma} = 11.9\%$, indicating any \ha\ flux change exceeding this threshold between the FRESCO and TWINKLE epochs would have been detected at $\geq3\sigma$.}
\label{fig:sensitivity}
\end{figure*}
\subsection{JWST NIRCam WFSS}\label{sec:wfss}

\begin{deluxetable}{llrc}
\tablecaption{NIRCam F410M and F444W observations \label{tab:obs}}
\tablewidth{0pt}
\tablehead{
  \colhead{Program} & \colhead{Mode} &
  \colhead{Exposure (s)} & \colhead{Depth}
}
\startdata
FRESCO        & F444W (imaging) & 934 & 28.4 \\
TWINKLE Ep.\,1 & F444W (imaging) & 934 & 28.3 \\
TWINKLE Ep.\,2 & F444W (imaging) & 934 & 28.3 \\
FRESCO         & F444W (grism)   & 7043 & 2.2  \\
TWINKLE Ep.\,1 & F410M (grism)   & 2920 & 2.7  \\
TWINKLE Ep.\,2 & F410M (grism)   & 2920 & 2.7 \\
\enddata
\tablecomments{The $5\sigma$ sensitivity limit is reported for imaging in mag and grism in 10$^{-18}$ erg s$^{-1}$ cm$^{-2}$.}
\end{deluxetable}


Slitless spectroscopy is particularly well-suited for this experiment: it avoids slit losses and placement uncertainties, provides high multiplex over the full field, and simultaneously delivers imaging for robust continuum measurements. The F410M medium-band filter was chosen because the \ha\ emission of all broad-line sources in \citet{matthee23} falls within this bandpass at their respective redshifts, while its narrower width relative to F444W reduces sky background and improves observing efficiency. 


We adopt a consistent WFSS reduction approach for all four programs to ensure a fair comparison of the line profiles across all epochs. We reduce the WFSS data with the official JWST calibration pipeline \citep{Bushouse23} supplemented by several customized steps described below. We start with the stage 1 products ({\tt rate.fits}) and assign the world coordinate system (WCS) to each exposure. Flat fielding is performed using the JWST calibration pipeline, with direct imaging flat reference files from the JWST Calibration Reference Data System (CRDS), as dedicated WFSS flat reference files are currently not available. The reduction uses CRDS version \texttt{12.1.10} and the context {\texttt{jwst\_1322.pmap}}. Background subtraction is done in two steps, combining a median background subtraction and residual removal with {\tt SExtractor} \citep{bertin1996}. Finally, we employ the median-filter technique to subtract the continuum and isolate emission lines \citep{Kashino23_grism, Sun23_grism, Liu24}. To avoid over-subtraction from median-filter continuum removal, we mask the \ha\ line and extend the mask along the dispersion direction to capture its broad wings prior to continuum modeling. The spectral trace and dispersion functions are adopted from \citet{Sun25}. From the extracted 2D spectra, we obtain 1D spectra using optimal extraction \citep{Horne86}. The spatial kernel for optimal extraction is derived from the FRESCO observations of ID 1087388 (GN-9771 in \citealt{matthee23, Torralba25}), the brightest LRD in our sample. We collapse the continuum-subtracted spectrum over a narrow velocity window $v = \pm 200\,\mathrm{km\,s^{-1}}$ centered on \ha\ to obtain a spatial flux profile. Pixels with negative values arising from noise in the continuum-subtracted wings are set to zero before normalizing the profile to unity. This kernel is then applied uniformly to all sources for optimal extraction.   

Nine LRDs were designed to be covered by TWINKLE, with their \ha\ emission lines falling within the F410M bandpass \citep{matthee23, Zhang26}. We obtain spectra for eight of them from the Epoch 1 and Epoch 2 observations; full coverage will be achieved at different PA across all three epochs. We also extract spectra from POPPIES and SAPPHIREs whenever the sources fall in these surveys' field of view.

With the extracted 1D spectra, we treat the FRESCO observations (2023 February 11) as the Day 0 baseline and compare the spectra from each subsequent epoch against it. The spectroscopic variability is estimated based on the difference spectrum, $\Delta f_\lambda = f_{{\rm epoch},\lambda} - f_{{\rm FRESCO},\lambda}$. The integrated flux difference is measured within a velocity window $|v| < \mathrm{FWHM}_{\rm broad}$ centered on \ha, where $\mathrm{FWHM}_{\rm broad}$ is the FWHM of the broad \ha\ wing obtained from the spectral line fitting using a narrow Gaussian component and a broad exponential component following \citet{Torralba25}. This choice reflects the non-Gaussian structure of the broad wings \citep[e.g.,][]{Labbe24a, deGraaff25, Rusakov26, Scholtz26}. To avoid the flux measurement being dominated by noisy outer pixels, the integration window is capped at 2000 km\,s$^{-1}$ for the two sources with very broad wings.

To estimate the noise in $\Delta f_\lambda$, we use the line-free sideband $4000 < |v| < 6000\,\mathrm{km\,s^{-1}}$ and compute the robust per-pixel standard deviation of the difference spectrum within this window, estimated via the median absolute deviation ($\sigma_{\rm side} = 1.4826\times\mathrm{MAD}$). Because $\sigma_{\rm side}$ is measured directly from the residual difference spectrum, it naturally captures both the formal statistical noise encoded in the pipeline weight maps and any additional systematic variance introduced by grism contamination and continuum-subtraction residuals. We adopt $\sigma_{{\rm eff},\,\lambda} = \max\!\left(\sigma_\lambda,\,\sigma_{\rm side}\right)$ as the effective per-pixel uncertainty. Here $\sigma_\lambda = \sqrt{\sigma_{{\rm epoch},\lambda}^2 + \sigma_{{\rm FRESCO},\lambda}^2}$ is the uncertainty propagated from the weight maps of the two epochs. The formal pipeline uncertainty thus sets a noise floor in spectral regions where the sideband estimate is poorly constrained, while the empirical $\sigma_{\rm side}$ dominates wherever systematic effects --- such as grism contamination from neighbouring sources --- exceed the formal prediction. The detection significance is then
\begin{equation}
S = \frac{\sum_{|v|<\mathrm{FWHM}_{\rm broad}} \Delta f_\lambda\,\Delta\lambda}
         {\sqrt{\sum_{|v|<\mathrm{FWHM}_{\rm broad}}
         \bigl(\sigma_{\rm eff,\,\lambda}\,\Delta\lambda\bigr)^{2}}},
\end{equation}
where $\Delta\lambda$ is the width of each spectral element. The final extracted spectra and their detection significances across different epochs are shown in Figure \ref{fig:line_profile}.

In the event that we do not detect significant variability ($>3\sigma$), we quantify our detection sensitivity by deriving per-source $3\sigma$ upper limits on the \ha\ line flux variability from the 2D spectral pair $\mathrm{Epoch}_i - \mathrm{Epoch}_0$, where $\mathrm{Epoch}_i$ denotes any subsequent observation epoch (TWINKLE, SAPPHIRE, or POPPIES) and $\mathrm{Epoch}_0$ denotes the FRESCO baseline.
\begin{itemize}

\item \textbf{Construct a wavelength-aligned 2D residual map:} We construct the 2D residual map $R_{\rm 2D} = S_{\rm TWINKLE} - S_{\rm FRESCO}$ from the wavelength-aligned continuum-subtracted 2D spectra.

  \item \textbf{Noise estimation:} We then collapse the residual spatially using optimal extraction with the ID 1087388 spatial profile as the kernel. The
  per-pixel noise $\sigma$ is estimated from the optimally extracted 1D residual
  in the sideband $|v| = 4000$--$6000\,\mathrm{km\,s^{-1}}$.

\item \textbf{Matched-filter upper limit:} We integrate within the same \ha\ velocity window ($|v| < \mathrm{FWHM}_{\rm broad}$, capped at $2000\,\mathrm{km\,s^{-1}}$) as defined above, and model the residual as
\begin{equation}
R_i = \alpha\,P_i + \epsilon_i ,
\end{equation}
where $P_i$ is the optimally extracted FRESCO line profile normalised to the total \ha\ line flux $F$ ($\sum_i P_i,\Delta\lambda = F$), $\epsilon_i$ denotes noise, and $\alpha \equiv\Delta F/F$ directly represents the fractional flux variability. Assuming approximately constant per-pixel variance $\sigma^2$, the minimum-variance estimator for $\alpha$ is
\begin{equation}
\hat{\alpha}
= \frac{\sum_i P_i R_i}{\sum_i P_i^2},
\qquad
\sigma_\alpha
= \frac{\sigma}{\sqrt{\sum_i P_i^2}}.
\end{equation}
The per-source $3\sigma$ upper limit on the fractional variability is then:
\begin{equation}\label{equ:sensitivity}
\left|\frac{\Delta F}{F}\right|_{3\sigma}
= 3\,\sigma_\alpha .
\end{equation}

\end{itemize}
The resulting $|\Delta F/F|_{3\sigma}$ values for all sources with available spectra are listed in Table~\ref{tab:sources}; each entry represents the minimum fractional \ha\ flux variability detectable at $3\sigma$ for that source given the depth of the TWINKLE/SAPPHIRE/POPPIES--FRESCO pair.

For ID 1087388, \ha\ falls near the red edge of the F410M bandpass; we therefore retain only the blue half of the spectrum ($-\mathrm{FWHM}_{\rm wing} < v < 0$) for the noise and flux estimation. 



\begin{figure*}[!htb]
\centering
\includegraphics[width=0.95\textwidth]{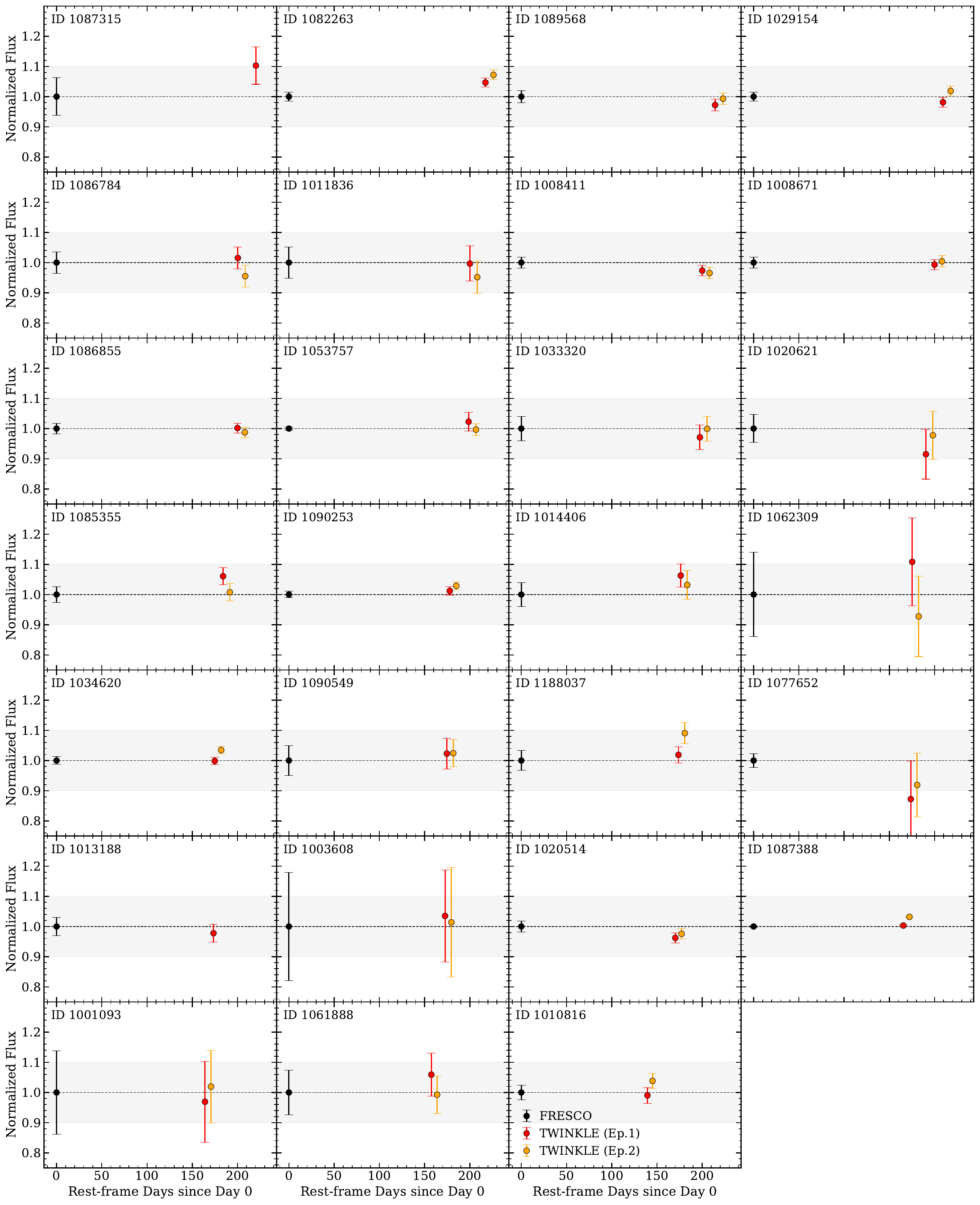}
\caption{F444W light curves of all 27 targets across multiple epochs, with fluxes normalised to the first epoch. For ID~1053757 and ID~1077652, which lack FRESCO coverage, we use the $r = 0.\!''15$ aperture photometry from the JADES catalog \citep{Robertson26} and divide out the JADES aperture correction to recover the raw aperture flux, ensuring consistency with the direct aperture photometry measured for the remaining sources. The gray shaded region indicates $|\Delta F/F| = 10\%$. All sources are consistent with variability below this level within their measurement uncertainties, however, for the faintest sources the per-source $3\sigma$ imaging sensitivity exceeds $10\%$ (Table~\ref{tab:sources}), meaning that variability at the $10\%$ level cannot be ruled out for those objects.}
\label{fig:lc}
\end{figure*}

\subsection{NIRCam Photometry}\label{sec:photometry}

NIRCam imaging in F182M, F210M, and F444W from FRESCO and TWINKLE was obtained simultaneously with the grism exposures. While additional NIRCam imaging from other programs exists over parts of the field, we do not incorporate these data into our analysis due to their non-uniform spatial and filter coverage. All data are reduced using the \texttt{grizli} pipeline \citep{Brammer19, Brammer23} with the \texttt{jwst\_1293.pmap} calibration context, following the procedures described in \citet{Valentino23} and \citet{Naidu26}. We process all three epochs --- FRESCO, TWINKLE Epoch 1, and TWINKLE Epoch 2 --- using an identical reduction and drizzling procedure, constructing a separate mosaic for each epoch for variability analysis. The final drizzled images have pixel scales of 30 mas for the SW channel and 60 mas for the LW channel. The per-epoch F444W mosaics are then used for the subsequent variability analysis. 

\begin{figure*}[!htb]
\centering
\includegraphics[width=0.99\textwidth]{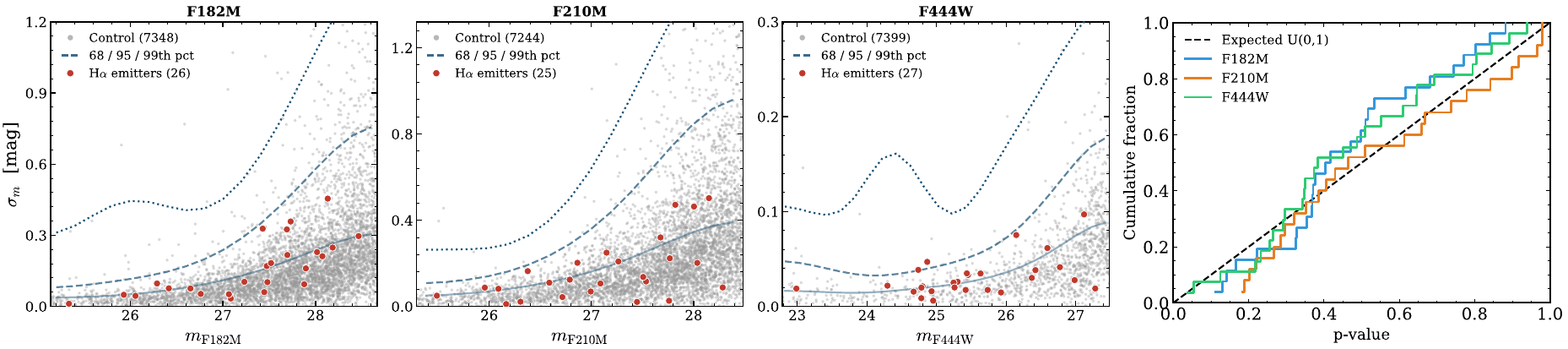}
\caption{\textit{Left three panels:} RMS magnitude scatter ($\sigma_{\rm m}$) versue JADES aperture ($r = 0.\!''15$) magnitude across three epochs (FRESCO + two TWINKLE visits) for our 27 broad \ha\ emitters (red point) and ${\sim}7{,}400$ compact control sources ($F(0.\!''2)/F(0.\!''1) < 2.0$) per band (grey). $\sigma_{\rm m}$ represents the sample standard deviation of their magnitude. Dashed curves mark the 68th, 95th, and 99th percentile envelope of the control distribution in a sliding 1-magnitude window. All \ha\ emitters lie within the 95th percentile envelope, consistent with no detected variability. \textit{Right panel:} Empirical cumulative distribution of per-source p-values defined as the fraction of magnitude-matched controls with $\sigma_m \geq \sigma_m^{\rm LRD}$. The dashed line shows the expected uniform distribution assuming \ha\ are drawn from the same population as controls. All three bands are consistent with the null hypothesis, showing no evidence that \ha\ emitters are more variable than field control sample.}
\label{fig:photometry_rms}
\end{figure*}

We perform aperture photometry on the F444W mosaics to probe the rest-frame optical emission of our sample. We use a circular aperture with a radius of $0.\!''15$, with local background subtraction from a source-masked annulus ($0.\!''6$–$2.\!''0$). To quantify the photometric stability between the two epochs, we construct a control sample of 102 isolated point sources. Applying the same photometric procedure, we measure a noise floor of $\sigma = 4.4\%$ between epochs, with a negligible zero-point offset of $-0.24\%$. We then estimate the per-source sensitivity of NIRCam photometry as 

\begin{equation}
\sigma = \sqrt{\left(\frac{\sigma_1}{f_1}\right)^2 + \left(\frac{\sigma_2}{f_2}\right)^2}
\end{equation}
where $f_i$ and $\sigma_i$ are the aperture flux and its uncertainty at each epoch. This yields 3-$\sigma$ sensitivities ranging from $0.7\%$ for the brightest source (ID 1087388) to $54.5\%$ for the faintest (ID 1003608, $m_{\mathrm{AB}} = 27.3$ mag) due to its intrinsic faintness. The median $3\sigma$ sensitivity is $9.0\%$. Given the wide magnitude range of our sample ($m_{\mathrm{AB}} = 23.0$--$27.3$), the sensitivity varies substantially from source to source. We therefore report per-source sensitivities (Table~\ref{tab:sources}) and incorporate these values in the following analysis. In addition to the measurement sensitivity derived from the photometric error, we construct a control sample of ${\sim}$7400 compact sources ($F(0.\!''2)/F(0.\!''1) < 2.0$) drawn from the same imaging data and compute the RMS magnitude scatter $\sigma_{\rm m}$ across epochs for each source to empirically characterize the noise floor. Figure~\ref{fig:photometry_rms} shows $\sigma_{\rm m}$ as a function of magnitude. The 68th, 95th, and 99th percentile envelopes of the control distribution define the expected scatter from measurement noise alone, and every \ha\ emitter falls within this envelope.

\section{Results}\label{sec:results}

\subsection{No Line, Line Profile, or Continuum Variability}\label{sec:results1}

We detect no significant variability across the full sample in any of the three diagnostics we measure: continuum flux, \ha\ line flux, and \ha\ line profile.                                 
\begin{itemize}
\item \textbf{Continuum.} For all 27 sources, we measure the F444W continuum flux in each epoch and compute the fractional change $\Delta F/F$ relative to the FRESCO baseline. The observed scatter spans $0.2$--$12.8$\%, with a median of $2.6$\%. Three sources show changes exceeding $10$\%: ID~1087315 ($10.3$\%), ID~1062309 ($10.8$\%), and ID~1077652 ($12.8$\%); however, all three are intrinsically faint and the apparent flux differences are significant at only $< 1.5\,\sigma$. Two of these sources have additional observations in a second epoch (ID~1062309, ID~1077652), where the measured photometric scatter remains $<10\%$. Consistently, all 27 sources fall within the 95th percentile of the empirical noise floor defined by the control sample. We therefore do not classify any of these sources as variable (see Figures~\ref{fig:lc}, \ref{fig:photometry_rms}).
\item \textbf{\ha\ Line Flux.} For the 10 sources with multi-epoch NIRCam WFSS coverage, we find no $>3\sigma$ detections in the 2D spectral residuals for any source. One source, ID~1013188, shows a $2.4\sigma$ flux decrement in the first epoch and was not observed in the second epoch due to a different position angle. Given the low signal-to-noise of the underlying spectrum and the absence of a corresponding change in the photometric flux, we interpret this as a noise fluctuation and do not claim a detection. We note that excluding ID~1013188, whose line flux sensitivity is limited to $\Delta F/F = 20.9\%$, does not affect our conclusions (see Figures \ref{fig:line_profile} and \ref{fig:sensitivity}).

\item \textbf{\ha\ Line Profile.} For the same 10 sources, we compare the broad \ha\ line profile across all available epochs using a narrow Gaussian plus broad exponential component fit (Section \ref{sec:wfss}). No source shows a significant change in FWHM or line centroid. Again, ID~1013188 shows a marginal profile change coincident with its $2.4\sigma$ flux decrement; for the same reasons outlined above, we regard this as tentative and do not claim a detection.
\end{itemize}

After accounting for measurement uncertainties, zero sources in our sample satisfy our detection criterion in any epoch --- namely a continuum variation or an \ha\ flux change exceeding $3\sigma$.

\subsection{Expected AGN Variability from SDSS-RM}\label{sec:sdss}

To test whether the observed photometric and spectroscopic variability is consistent with the assumption that LRDs host standard AGN engines, we perform a Monte Carlo simulation to estimate the expected number of variability detections if our LRDs were drawn from the SDSS-RM AGN population \citep{Shen24}. This dataset provides multi-epoch light curves for 849 AGN over a $\sim$7-year baseline, including 53 sources with broad \ha\ light curves; we use these for our variability simulation.


\begin{figure}[!htb]
\centering
\includegraphics[width=0.45\textwidth]{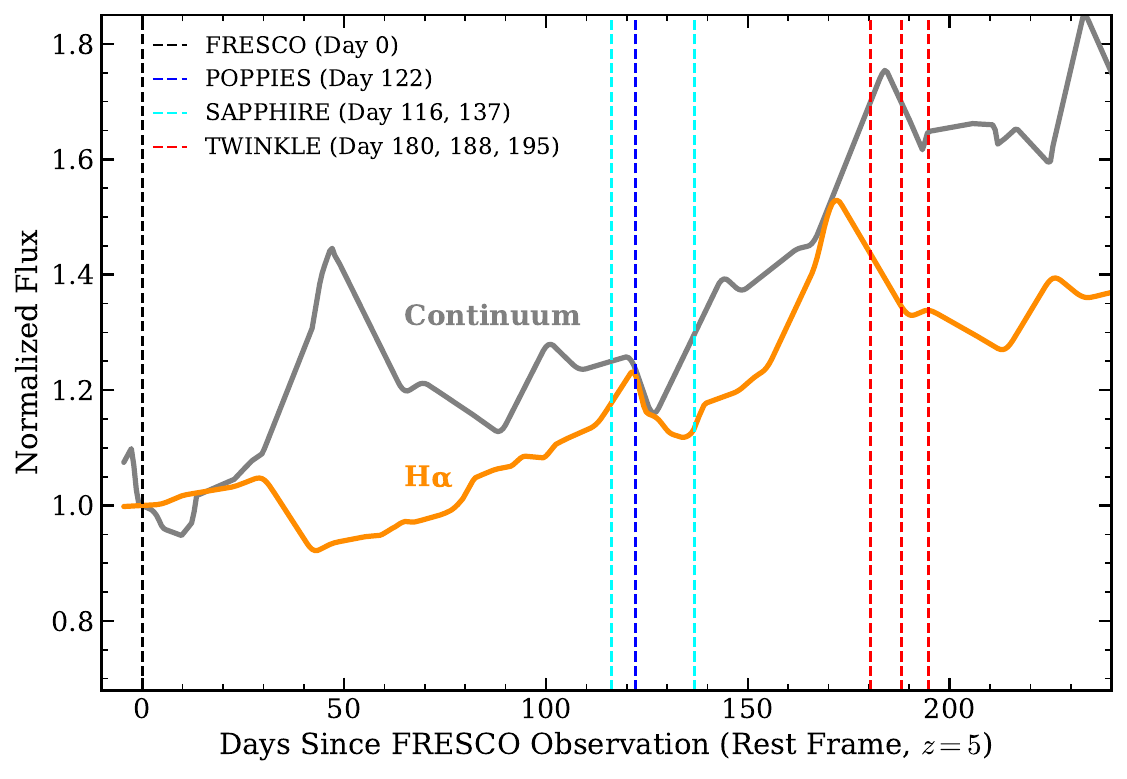}
\caption{An example SDSS-RM light curve of an AGN whose \ha\ luminosity is matched to that of a broad \ha\ emitter in our sample (RM17; \citealt{Shen24}), shown over a rest-frame window of $\sim$200 days. Vertical dashed lines mark the rest-frame time separations probed by our multi-epoch observations at $z = 5$. The flux variations seen in this luminosity-matched low-redshift AGN over the same rest-frame timescale illustrate the variability amplitude we would expect to detect if the high-$z$ sources were accreting similarly to typical local broad-line AGN.}
\label{fig:demo}
\end{figure}

\begin{figure*}
\centering

\begin{subfigure}{0.65\textwidth}
    \includegraphics[width=\linewidth]{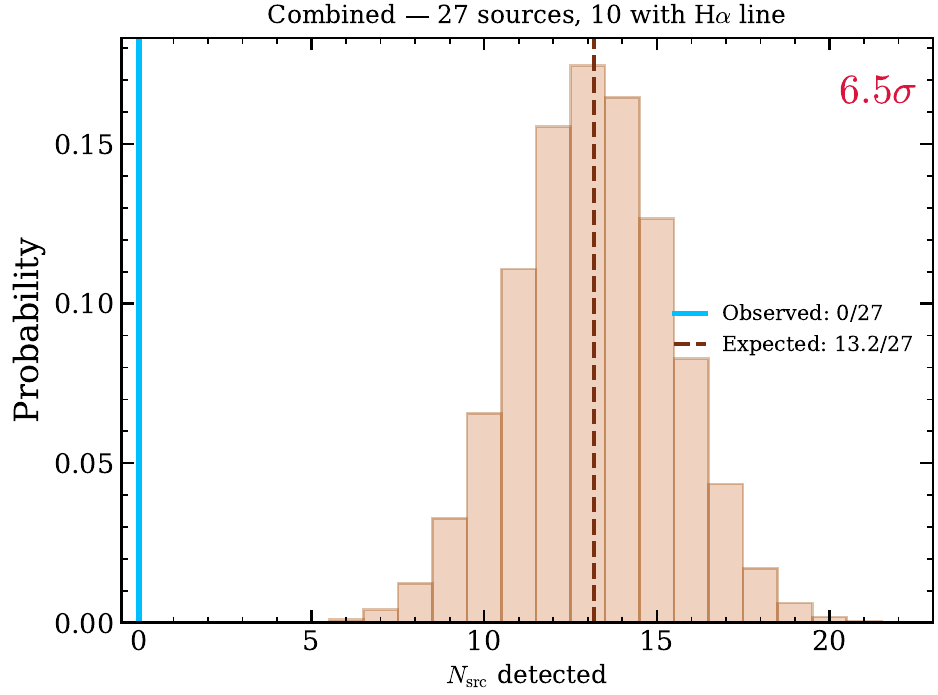}
\end{subfigure}
\begin{subfigure}{0.95\textwidth}
    \includegraphics[width=\linewidth]{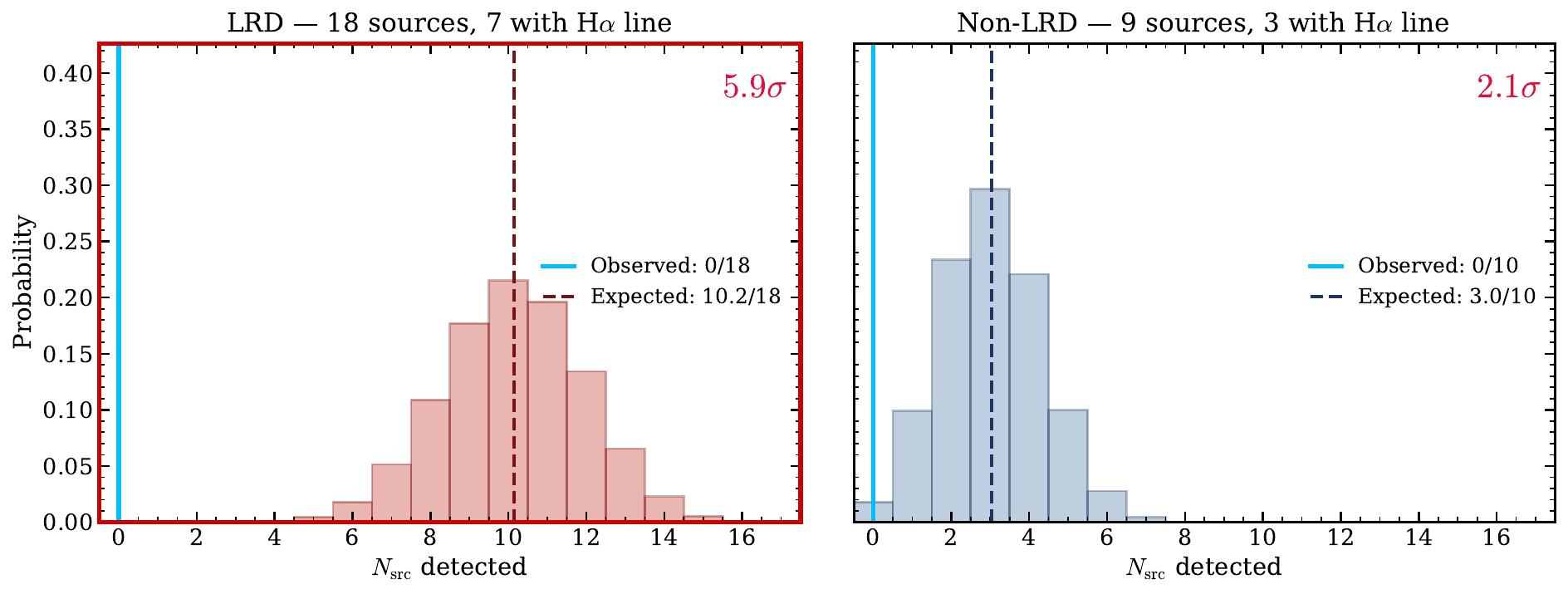}
\end{subfigure}
\caption{\textbf{SDSS-RM expected vs.\ observed detections.} Histograms show the probability distribution of the number of variable sources expected from $10^4$ Monte Carlo trials, drawing variability amplitudes from luminosity-matched SDSS-RM JAVELIN light curves at the rest-frame equivalent of our observational baseline. A source is counted as detected if the simulated $|\Delta F/F|$ exceeds its per-source $3\sigma$ threshold in the continuum or, where available, the \ha\ line. The blue line indicates the observed number of detections, the dashed line indicates the mean of the expected distribution from SDSS sample. Variability detection statistics are shown separately for the LRD-classified (left, red) and non-LRD (right, blue) subsamples, defined based on the sample selection described in Section~\ref{sec:lrd_selection}.}
\label{fig:sdss_rm_combined}
\end{figure*}

\begin{figure*}[!htb]
\centering
\includegraphics[width=0.95\textwidth]{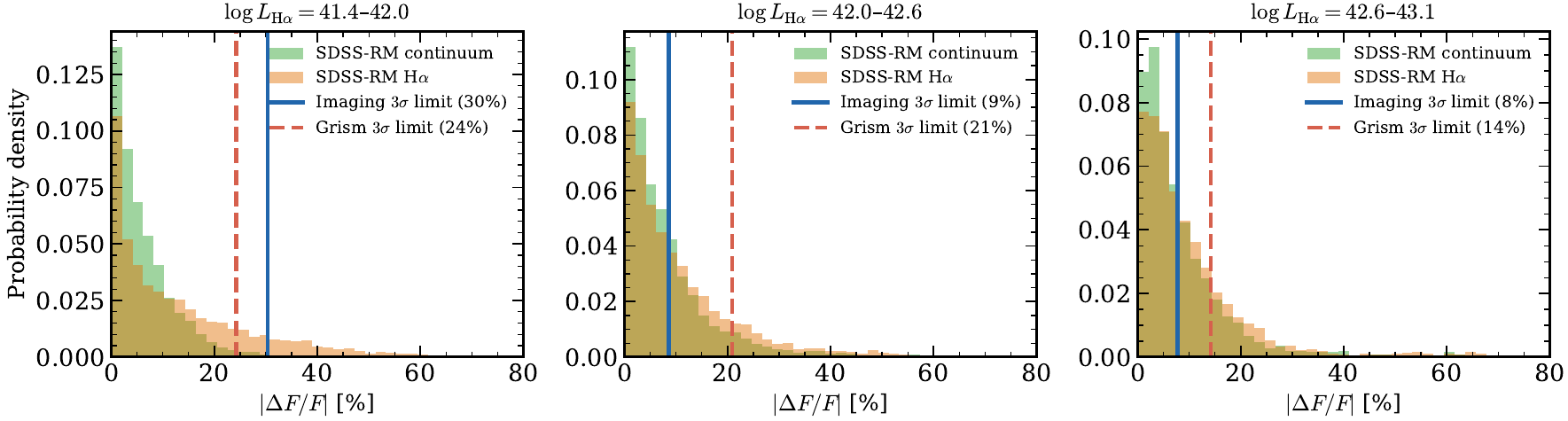}
\caption{Distributions of expected $|\Delta F/F|$ from $10^4$ Monte Carlo realizations of the SDSS-RM sample, in three bins of broad \ha\ luminosity. Green and orange histograms show the expected continuum and \ha\ line variability distributions for SDSS-RM sample, respectively. Vertical lines mark the median TWINKLE $3\sigma$ sensitivity threshold (imaging: blue solid; grism \ha: red dashed). }
\label{fig:sdss_sensitivity}
\end{figure*}

We match each of the 27 \ha\ emitters to SDSS-RM AGN by broad \ha\ luminosities. This choice provides a fully empirical comparison. The LRD sample spans $\log(L_{\mathrm{H}\alpha}/\mathrm{erg\,s^{-1}}) = 41.5$–$43.5$, and we require agreement within a factor of 2.0 with the SDSS-RM sample, for which broad \ha\ luminosities are derived from the \texttt{PrepSpec} mean line fluxes \citep{Shen16, Shen19}. This results in 49 unique matched SDSS-RM AGN spanning $\log(L_{\mathrm{H}\alpha}/\mathrm{erg\,s^{-1}}) = 41.3$–$42.9$. One luminous LRD (ID 1087388, $\log L_{\mathrm{H}\alpha} = 43.5$) exceeds the \ha\ luminosity range of SDSS-RM and is instead matched to the three nearest SDSS-RM sources in \ha\ luminosity.

With the matched sample, we use the JAVELIN-predicted light curves \citep{Zu11} to simulate two-epoch observations separated by the same rest-frame baselines as our data. JAVELIN models the sparsely sampled observed light curves as a damped random walk and generates densely sampled, gap-free predicted light curves over the full monitoring baseline. The observation baseline between FRESCO and TWINKLE Epoch 2 is 1126 observed-frame days, yielding $\sim$ 144–230 days for z = 3.9–6.8. For each LRD, we convert the matched SDSS-RM light curves to the rest frame and draw two random epochs separated by its corresponding rest-frame baseline, recording the resulting flux ratio between the two epochs. By repeating this simulation 10,000 times, we estimate the expected variability distribution for a standard AGN population at matched \ha\ luminosity. Since the SDSS-RM \ha\ sample lies at $z<0.6$ and spans a rest-frame baseline of $\sim$7 years, their light curves cover a wide dynamic range of variability states, allowing the simulation to capture a wide range of AGN variability states and flux conditions within the window of our LRD observations.

We apply this measurement to both photometric and spectroscopic variability analysis. Because the observed SDSS-RM light curves are sparsely sampled, we use JAVELIN-predicted light curves \citep{Zu11}, which provide densely sampled, gap-free realizations over the full monitoring baseline. For the full sample of 27 sources, we perform a simultaneous modeling that incorporates the available datasets, including 17 sources with photometric monitoring only, 10 sources with both photometric and spectroscopic coverage, and the corresponding line-profile modeling constraints. In both cases, we define the per-source detection probability $P_i$ as the fraction of Monte Carlo realizations in which the simulated fractional flux change exceeds the source-specific $3\sigma$ detection threshold. For photometric measurements, this threshold is the per-source $3\sigma$ imaging sensitivity listed in Table~\ref{tab:sources} (Section~\ref{sec:photometry}), while for spectroscopic sources it is set by the upper limits derived from the 2D spectral residuals (Section~\ref{sec:wfss}, Table~\ref{tab:sources}). Because each broad \ha\ emitter is treated as an independent source, the joint probability of observing zero detections across all sources under the null hypothesis that LRDs vary like the SDSS-RM sample is $P_{\rm null} = \prod_{i=1}^{N}(1 - P_i)$, which we convert to a Gaussian-equivalent significance. We repeat the same analysis for the LRD and non-LRD subsets defined in Section~\ref{sec:lrd_selection}. 

Based on our simulations, if our sample varied like the \ha-luminosity–matched SDSS-RM AGN, we would expect to detect $\sim13$ variable sources over the same rest-frame baseline. Instead, we detect none, corresponding to a $6.5\sigma$ deficit. For the LRD and non-LRD subsets, the expected number of detections are 10.2 and 3.0, with zero detections in both cases. These correspond to suppressed variability at $5.9\sigma$ and $2.1\sigma$ significance, respectively. The statistical difference is primarily driven by the sample sizes and the differing per-source sensitivities of the two subsets.


The non-detections are not sensitivity-limited. The per-source 3$\sigma$ detection thresholds range from $0.7\%$ to $54.5\%$ for photometric variability, with a median sensitivity of $9.0\%$, and from $2.7\%$ to $24.9\%$ for line variability (see Table \ref{tab:sources}). To quantify what we should have detected, we draw $10^4$ Monte Carlo realizations from SDSS-RM JAVELIN-predicted continuum and \ha\ light curves, matching each realization to sources in the same $L_{\mathrm{H\alpha}}$ bin and sampling epoch pairs separated by the rest-frame equivalent of our baseline ($\sim$166~days). Figure \ref{fig:sdss_sensitivity} shows the resulting $|\Delta F/F|$ distributions in three luminosity bins alongside our per-bin sensitivity thresholds. For $\log,L_{\mathrm{H\alpha}} > 42.0$ --- where the bulk of our sample lies, the median sensitivity falls well within the expected SDSS-RM distribution. The non-detections reflect a genuine suppression of variability, not a sensitivity floor.
\begin{figure*}[!htb]
\centering
\includegraphics[width=0.99\textwidth]{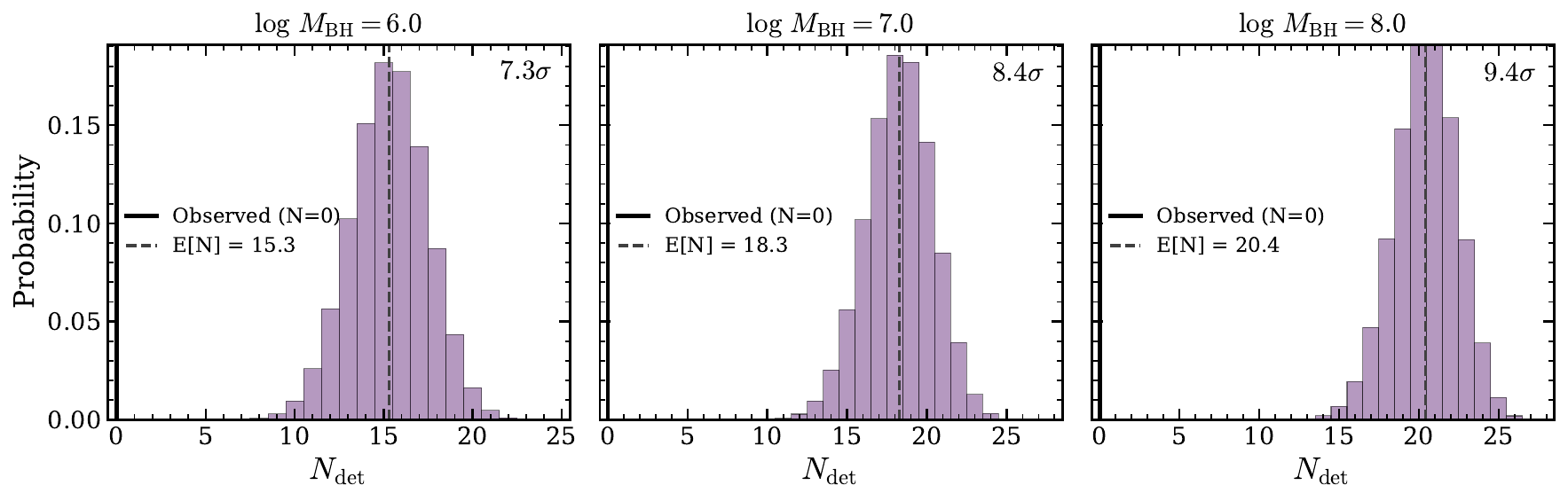}
\caption{Expected number of variable sources detectable by TWINKLE under the sub-Eddington accretion model with bolometric luminosity matched to our sample \citep{Secunda26}. This model is ruled out by $>7\sigma$ for the shown black hole masses.}
\label{fig:subedd}
\end{figure*}
We note that, in the SDSS-RM comparison, the variability amplitude inferred from the \ha\ light curves appears larger than that of the continuum, which is somewhat unexpected compared to the trends reported in Figure~3 of \citet{Shen24}. One possible explanation is that we are probing variability on a much shorter effective timescale, whereas \citet{Shen24} consider variability over the full SDSS-RM baseline. In this regime, the optical continuum may exhibit a longer effective damping timescale than the line emission, as high-frequency variability is partially smoothed out in the continuum due to disk reprocessing. We also note that interpolated light curves may introduce an additional bias toward underestimating variability amplitudes in realizations drawn from sparsely sampled data, although this effect would act in a conservative sense for our test and is unlikely to affect our conclusions.


\subsection{Comparison to Simulated AGN Light Curve Predictions}
\label{sec:theory}

To further test our conclusions, we compare our non-detections against mock sub-Eddington AGN light curves modeled by \citet{Secunda26}, using empirical relations calibrated on lower-redshift AGN light curves \citep[e.g.][]{Burke23}. The mock light curves span a range of black hole masses from $\log(M_\mathrm{BH}/M_\odot) = 6$--$8$ and are normalized to the median bolometric luminosity of our sample ($\log L_\mathrm{bol} = 43.6$ erg s$^{-1}$), derived from the \ha\ luminosities following the calibration of \citet{deGraaff25b} validated by \citet{Greene26}. We perform the simulation using the continuum channel only rather than combining the continuum and line channels. Unlike the SDSS-RM light curves in which continuum and line variability are physically correlated, the mock continuum and line light curves are generated as independent realizations, so combining them would overcount detections.

Across all mass bins, sub-Eddington models predict $15$--$20$ detectable variable sources, which are ruled out at $7.3$--$9.4\sigma$ given our observed null result (Figure \ref{fig:subedd}). This predicted count exceeds our empirical SDSS-RM benchmark of 13 expected detections (Section~\ref{sec:sdss}), which can be explained by the anticorrelation between AGN luminosity and variability amplitude \citep{Cristiani97, VandenBerk04, Wilhite08, MacLeod10}. \ha-matched SDSS quasars are intrinsically $\sim$10$\times$ more bolometrically luminous than our sources: standard AGN yield $L_{\rm bol}/L_{{\rm H}\alpha} \approx 170$ \citep{GreeneHo05}, whereas LRDs show $L_{\rm bol}/L_{{\rm H}\alpha} \approx 19$ \citep{deGraaff25b, Greene26}. Because more luminous AGN vary less, our \ha-matched comparison \textit{underestimates} the expected detection rate, and $N_{\rm exp} = 13$ is a lower limit on the expected detection count. Zero detections is anomalous regardless. 

The non-detections thus robustly disfavor standard sub-Eddington accretion as the dominant power source for high-redshift broad \ha\ emitters. The data are instead consistent with e.g., the presence of a dense reprocessing envelope that smooths intrinsic variability on rest-frame timescales of months. A super-Eddington accretion regime provides a natural framework for this picture, as radiation-pressure-driven outflows can self-consistently produce such a reprocessing envelope \citep[e.g.,][]{Liu25}. To compare with this scenario, we scale simulated light curves from a super-Eddington model derived by \citet{Secunda26} from radiation magnetohydrodynamic simulations of a $10^8~M_\odot$, $3~L_{\rm edd}$ AGN in \citet{Jiang19} and \citet{Jiang25}. One of these scaling prescriptions, in which the variability is scaled with $M_\mathrm{BH}$, is consistent with our observations, yielding zero detections of variable sources and placing the sources in the intermediate mass black hole regime ($\approx10^{3-5} M_{\rm{\odot}}$). The details of these scalings are discussed in the Appendix. Alternative scenarios, including SMSs, may also be consistent with the observed lack of short-term variability and are discussed in Section~\ref{discuss1}.


\section{Discussion}

None of the 27 sources show statistically significant variability over our observational baseline, either in the continuum or in the broad \ha\ emission line at significant tension ($\gtrsim7\sigma$) with empirical expectations for typical AGN (\S\ref{sec:results}). Our $\sim$200-day rest-frame baseline is comparable to the empirical DRW damping timescale expected for accretion disks around black holes of $M_\mathrm{BH} \sim 10^7$–$10^8\,M_\odot$ ($\tau_d \sim 50$–100 days from the empirical $M_\mathrm{BH}$–$\tau$ relation; \citealt{Burke21}). Therefore, the monitoring window is long enough to probe the characteristic variability timescale of such systems, indicating that these high-$z$ broad \ha\ emitters are simply not varying the way typical local broad-line AGN do. What might explain this result?

\subsection{Lack of Variability -- No Black Holes at All?}\label{discuss1}

\begin{figure*}
\centering
\includegraphics[width=0.99\textwidth]{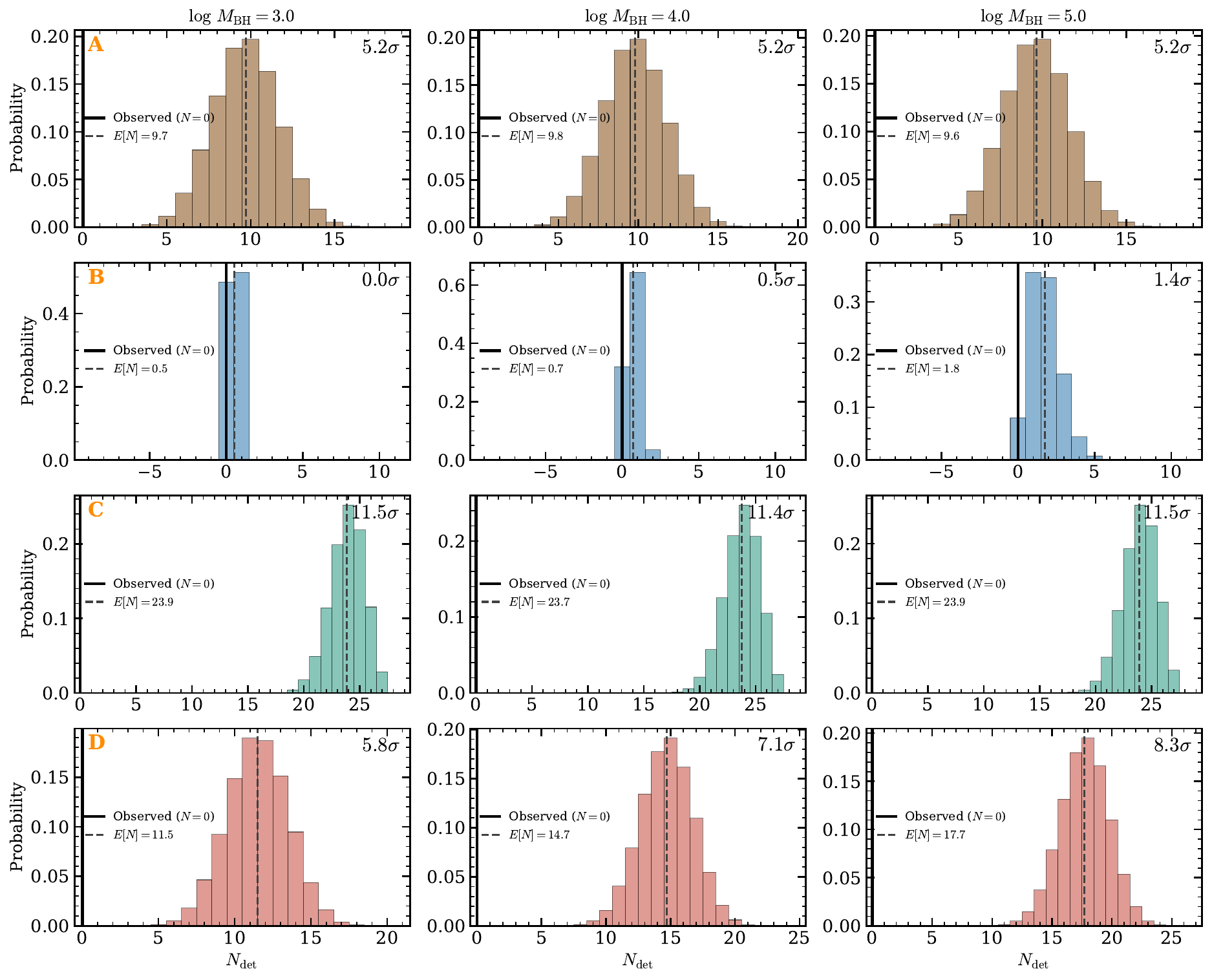}

\caption{Expected number of variable sources ($N_\mathrm{det}$) under super-Eddington accretion models using mock light curves from \citet{Secunda26}. Each row shows the $N_\mathrm{det}$ distribution from 10,000 Monte Carlo realizations. \textbf{(A)} Unscaled --- raw super-Eddington output ($\lambda=3\times$), no mass or luminosity correction. \textbf{(B)} Mass-scaled --- amplitude rescaled by $M_\mathrm{BH}$; detectability decreases toward lower masses.  \textbf{(C)} Luminosity-scaled --- amplitude rescaled by $L_\mathrm{obs}$. \textbf{(D)} Fully scaled --- amplitude rescaled by both $M_\mathrm{BH}$ and $L_\mathrm{obs}$. Of all these flavors of models, the data are most consistent with those shown in the second row -- super-Eddington ($L_{\rm{bol}}/L_{\rm{edd}}\approx3$) black holes in the IMBH regime ($\approx10^{3-5} M_{\rm{\odot}}$).}
\label{fig:super_edd_model}
\end{figure*}

Perhaps LRDs simply do not host AGN? An interpretation has been proposed in which LRDs are powered by short-lived SMSs \citep{Furtak23, Nandal26, Chisholm26}. This scenario can explain the broad Balmer lines (via electron scattering) and steep rest-optical slopes (as the blackbody emission of a $\approx5000-7000$ K star)  without any need to invoke black hole accretion.

Variability in SMS is expected, but on timescales far exceeding our baseline. Massive stars are known to exhibit both photospheric variability \citep[e.g.,][]{Bowman19, Lecoanet19} and line profile variability \citep[e.g.,][]{Prinja88, Fullerton96, Chene10} driven by stellar winds (see \citealt{Kudritzki2000} for a review), with the characteristic timescales mostly set by the stellar rotation rate. Extending this analogy to the SMS regime: SMSs can rotate rapidly \citep{Haemmerle18, Nandal25}, and at 10\% of the break-up rotation speed, the rotation period of a $10^4\,M_\odot$, $R \sim 10^3$\,au SMS is $t_\mathrm{rot} \sim 2\pi\sqrt{R^3/GM}/0.1 \sim 3000$ rest-frame years --- far beyond our $\sim$200-day baseline. If instead pulsations driven by accretion instabilities dominate, the relevant timescale is the dynamical time, $t_\mathrm{dyn} = \sqrt{R^3/GM} \sim 50$ years. Variability would then emerge on a few times $\,t_\mathrm{dyn}$, or $\sim$50--100 rest-frame years --- consistent with both our short-baseline non-detection \emph{and} the long-term flux variations revealed by the gravitational lensing time delay in \citet{zhang25century} probing over a century in rest-frame time.

\subsection{Lack of Variability -- A Super-Eddington Phase in Black Holes with Extended Envelopes?}\label{discuss3}

If LRDs indeed host AGN, their physics must diverge in significant ways relative to their local peers. Scattering-dominated emission \citep[e.g.,][]{Rusakov26} and super-Eddington accretion \citep[e.g.,][]{Liu25} in e.g., dense gas envelopes \citep[e.g.,][]{Kido25} as expected around BH*s remain viable. A large scattering optical depth from dense, ionized gas surrounding the black hole may either dilute the observed variability through scattered light or trap photons in the optically thick inflow under super-Eddington conditions. Furthermore, there may be no broad-line region whatsoever with the structure around the black hole instead paralleling the interiors and atmospheres of massive stars \citep[e.g.,][]{Cantiello25, Begelman26, Santarelli26,Liu26}. Indeed, \citet{Matthee26} recently demonstrated that LRD spectra may arise from dense flows around a central engine with P-Cygni, scattered profiles as seen in massive star winds and Type IIn supernovae. This finding is indeed reminiscent of super-Eddington flows \citep[e.g.,][]{Liu25}.

While the absence of variability in our sample is inconsistent with sub-Eddington accretion, a direct test against super-Eddington models is less straightforward, because the amplitude and timescales of super-Eddington variability --- and in particular how they scale with black hole mass and accretion luminosity --- are not fully understood.

Keeping this caveat in mind, to provide qualitative predictions against theoretical mock light curves (Section \ref{sec:theory}), we compare our observations against four scenarios: the raw simulated light curves from the \citet{Secunda26} simulation with no re-scaling applied, and three variants in which the variability amplitude is scaled by $M_\mathrm{BH}$, $L_\mathrm{bol}$, or both $M_\mathrm{BH}$ and $L_\mathrm{bol}$, following sub-Eddington empirical scaling relations (Figure~\ref{fig:super_edd_model}). Because these scaling relations have been calibrated only on sub-Eddington AGN and have not been tested in the $L/L_\mathrm{Edd} > 1$ regime, the scaled scenarios carry systematic uncertainties that are difficult to quantify. Some of these scenarios -- involving intermediate mass black holes ($\approx10^{3-5} M_{\rm{\odot}}$) and $L_{\rm{bol}}/L_{\rm{edd}}=3-300$-- are broadly consistent with our observations and predict $\sim$0 detections over our time baseline. However, this exercise is not intended to provide a definitive constraint on the black hole parameters, but rather to demonstrate that the variability non-detections are compatible with super-Eddington accretion. Precise limits on $\dot{m}$ from variability alone require independent knowledge of $M_\mathrm{BH}$ and $L/L_\mathrm{Edd}$ for this population, as well as a better understanding of how variability scales with mass and luminosity in the super-Eddington regime.

Encouragingly, this physical picture has been assembled from multiple directions. In super-Eddington accretion, the high mass-accretion rate increases the optical depth across the disk, pushing the photosphere to large radii of $\sim10^4$--$10^5\,r_g$ \citep{Secunda26, Liu25, Kido25}, consistent with the effective radii of $\approx1000$ au derived from blackbody fits to the optical continuum SEDs of LRDs \citep[e.g.,][]{deGraaff25b, Umeda25, Sun26}. \citet{Liu25} showed that this regime naturally produces a cool (T $\sim$ 5000\,K; e.g., \citealt{Umeda25, deGraaff25b, Sun26}), low-density ($\rho_\mathrm{ph} < 10^{-9}$ g cm$^{-3}$) photosphere as a direct consequence of the high accretion rate (see also \citealt{Kido25, Santarelli26, Roman-Garza26, Liu26}). At such low densities, a strong opacity discontinuity appears at the Balmer limit ($\lambda = 3645\,\text{\AA}$). This naturally generates both the prominent Balmer break and the red optical continuum observed in LRDs, without invoking stellar populations or external dust. The same extended photosphere that produces the prominent Balmer break also suppresses short-timescale variability on the timescale probed here. 

To quantify the dynamical timescale expected in this picture we adopt the blackbody parameters reported in \citet{Sun26} to describe the central engine of the median LRD ($R_{\rm{eff}}\approx1300$ au and $L_{\rm{bol}}\approx10^{44}$ erg s$^{-1}$). Assuming $L_{\rm{bol}}\approx L_{\rm{edd}}$ (for the reasons outlined in this section) yields $M_{\rm{BH}}\approx5\times10^{5} M_{\rm{\odot}}$. This implies a dynamical timescale of $t_{\rm{dyn}}=\sqrt{R^{3}/GM_{\rm{BH}}}$ of $\approx10$ years, far exceeding TWINKLE's baseline. Intriguingly, an LRD with exceptional gravitational lensing time delays spanning decades \citep[e.g.,][]{zhang25century} shows promising hints of variability consistent with these results.

Therefore, in the picture where extended envelopes reprocess the radiation from the black hole, the spectral and temporal properties of LRDs --- including Balmer breaks stronger than any stellar population model, red rest-optical continua, cool effective temperatures, X-ray weakness and the absence of short-timescale variability --- all emerge self-consistently as signatures of the same physical setting \citep[e.g.,][]{deGraaff25, deGraaff25b, Naidu25, Barro26,Wang26, Sun26}.

\subsection{Variability in LRDs and Non- LRDs}\label{discuss2}

The ``V-shaped" SED with a blue rest-UV slope transitioning to a red rest-optical slope has been identified in many LRDs and is widely adopted as the defining photometric criterion for LRD selection. However, it was originally defined as an empirical cut in a continuous distribution of SED shapes rather than a physically motivated boundary, and its physical origin and completeness have recently been questioned. Might ``non-LRD" (i.e., not V-shaped) broad-line sources also host central engines similar to LRDs? Perhaps these non-LRDs simply reside in luminous hosts that dilute the V-shape \citep[e.g.,][]{Sun26}. Or perhaps the non-LRDs are at a somewhat different evolutionary state (e.g., when the gas envelope has been thinned) or are being viewed along a low-density viewing angle \citep[e.g.,][]{Brazzini26, Madau26, Matthee26}?

Variability provides independent constraints on these hypotheses. If the LRD and non-LRD broad \ha\ emitters share the same underlying physics, one would expect them to exhibit similar variability. The TWINKLE observations may be consistent with this picture. We find no variability in either the 18 LRDs or the 9 non-LRDs. This is indeed what one might expect if a similar central engine were powering a large fraction of the broad-line AGN population observed by JWST. 

We caution that our LRD selection is incomplete by construction \citep{Hviding25}, so some fraction of the 9 non-LRDs may be genuine LRDs misclassified by photometry. The LRD fraction of 18/27 in our sample is thus a lower limit. While the blind grism selection is highly effective at producing a complete census of broad-line emitters, the lack of corresponding photometric or NIRSpec/PRISM coverage is the limiting factor to further probe the physics of V-shaped as well as other sub-populations.

\section{Summary}

TWINKLE delivers the first systematic time-domain spectrophotometric study of high-redshift broad \ha\ emitters. Combining the Cycle~1 FRESCO and Cycle~4 TWINKLE slitless spectroscopy programs with pure-parallel archival programs (SAPPHIRES, POPPIES), we monitor 27 broad \ha\ emitters at $z\sim4$--$7$ across four JWST epochs spanning $\sim$139--220 rest-frame days --- the longest rest-frame baseline currently available with JWST for a complete, \ha-flux-limited sample. Our key results are:
\begin{itemize}

\item \textbf{A complete slitless variability census.} We measure F444W continuum variability for all 27 sources. For the 10 sources with NIRCam WFSS coverage in F444W across multiple epochs, we construct 2D spectral residuals and apply a matched-filter upper limit to obtain per-source \ha\ line variability sensitivities of $2.7\%$--$28.4\%$. The slitless design ensures a complete and unbiased census to study their variability, while also sidestepping the systematics (e.g., flux losses, path loss corrections) inherent to repeated slit-spectroscopy. [ \S\ref{sec:wfss}, \S\ref{sec:photometry}]

\item \textbf{None vary -- these sources are different from local AGN.} We detect no significant variability in either continuum flux, \ha\ line emission, or \ha\ line shape. If these sources varied like typical reverberation mapping samples, of which the SDSS sample is representative, we would expect $13$ detectable fluctuations. Observing none yields a $6.5\sigma$ deficit, directly demonstrating that standard AGN calibrations (e.g., for black hole masses) may not apply to this population. [\S\ref{sec:results1}, \S\ref{sec:sdss}, Figures ~\ref{fig:line_profile}, \ref{fig:lc}]

\item \textbf{Sub-Eddington accretion robustly ruled out, hints of Super-Eddington intermediate mass black holes?} Mock light curves spanning $\log(M_\mathrm{BH}/M_\odot) = 6$--$8$ predict $15$--$20$ detections under sub-Eddington models, which are ruled out at $7.3$--$9.4\sigma$. Super-Eddington models involving an IMBH ($\approx10^{3-5} M_{\rm{\odot}}$) --- where the inflated $\approx1000$ au gas envelope dilutes variability timescales well beyond our baseline --- offers a plausible, albeit non-unique, explanation. [\S\ref{sec:theory}, Figure ~\ref{fig:subedd}]

\item \textbf{No AGN at all?} The lack of variability across our observed $\approx 5 - 7$ months baseline, but hints of variability on decade to century scales \citep[e.g.,][]{zhang25century} are fully consistent with there being no AGN whatsoever, in line with e.g., expectations for Supermassive Stars (SMS) that are candidates for the central engines of LRDs.

\item \textbf{V-shape LRDs and non-LRDs both show suppressed short-term variability.} Zero detections are observed in both the 18 LRDs ($5.9\sigma$ deficit, $10.2$ expected) and 9 non-LRDs ($2.1\sigma$, $3.0$ expected), suggesting suppressed variability is a shared property of broad-line AGN at $z\sim4$--$7$ rather than a feature unique to LRD classification. This may be consistent with a shared central engine across this population disguised by e.g., a relatively luminous host galaxy, viewing angle effects, or differing evolutionary states.
[\S\ref{sec:lrd_selection}, Figure ~\ref{fig:sdss_rm_combined}]

\end{itemize}

Taken together, these results firmly establish that high-redshift broad \ha\ emitters harbor central engines unlike any AGN population known at lower redshift. Their suppressed variability, uniform across V-shape and non-V-shape sources alike, demonstrates that standard AGN calibrations --- built on sub-Eddington accretion physics --- do not apply to these systems. TWINKLE has opened the first systematic time-domain spectroscopic window onto them. Longer baselines (e.g., via long-term monitoring, gravitational lensing, or local LRDs) and larger samples that capture the full diversity of distant broad emitters (such as rare phases of LRDs; \citealt{Fu25, Hviding26}) will be required to place them in an evolutionary context.

\section*{Acknowledgements}
This work is based on observations made with the NASA/ESA/CSA James Webb Space Telescope. The data were obtained from the Mikulski Archive for Space Telescopes at the Space Telescope Science Institute, which is operated by the Association of Universities for Research in Astronomy, Inc., under NASA contract NAS 5-03127 for JWST. RPN and ZL acknowledge funding from JWST program GO-7404. Support for this work was provided by NASA through the NASA Hubble Fellowship grant HST-HF2-51515.001-A awarded by the Space Telescope Science Institute, which is operated by the Association of Universities for Research in Astronomy, Incorporated, under NASA contract NAS5-26555. RPN thanks Neil Pappalardo and Jane Pappalardo for their generous support of the MIT Pappalardo Fellowships in Physics, and for their enthusiasm and encouragement for pursuing the earliest galaxies and black holes. JM and AT acknowledge funding by the European Union (ERC, AGENTS,  101076224). AdG acknowledges support from a Clay Fellowship awarded by the Smithsonian Astrophysical Observatory. DM acknowledges generous support from the Leonard and Jane Holmes Bernstein Professorship in Evolutionary Science. Support for program JWST-GO-01895, provided through a grant from the STScI under NASA contract NAS5-03127 is acknowledged. The Center for Computational Astrophysics at the Flatiron Institute is supported by the Simons Foundation. 

The authors acknowledge the program \#1895 (PI: P. Oesch), 5398 (PI: J. Kartaltepe) and 6434 (PI: E. Egami) for developing their observing program with a zero-exclusive-access period. This work also makes use of the JADES DR5 data release, which includes NIRCam imaging from JWST programs 1176, 1180, 1181, 1210, 1264, 1283, 1286, 1287, 1895, 1963, 2079, 2198, 2514, 2516, 2674, 3215, 3577, 3990, 4540, 4762, 5398, 5997, 6434, 6511, and 6541, as well as MIRI data from programs 1180, 1181, and 1207.

\software{\texttt{Astropy} \citep{Astropy13,Astropy18}, \texttt{grizli} \citep{Brammer19}}

\appendix
\label{appendix}

\section{Predictions from an $L_{5100}$-Matched SDSS Sample}\label{appendix1}

In the main text, we match each broad \ha\ emitter to the SDSS-RM sample based on their \ha\ luminosity, as this quantity provides the most direct observational constraint. Here we explore alternative matching scheme based on $L_{5100}$ to test whether our results remain unchanged.

\begin{figure*}
\centering
\begin{subfigure}{0.65\textwidth}
    \includegraphics[width=\linewidth]{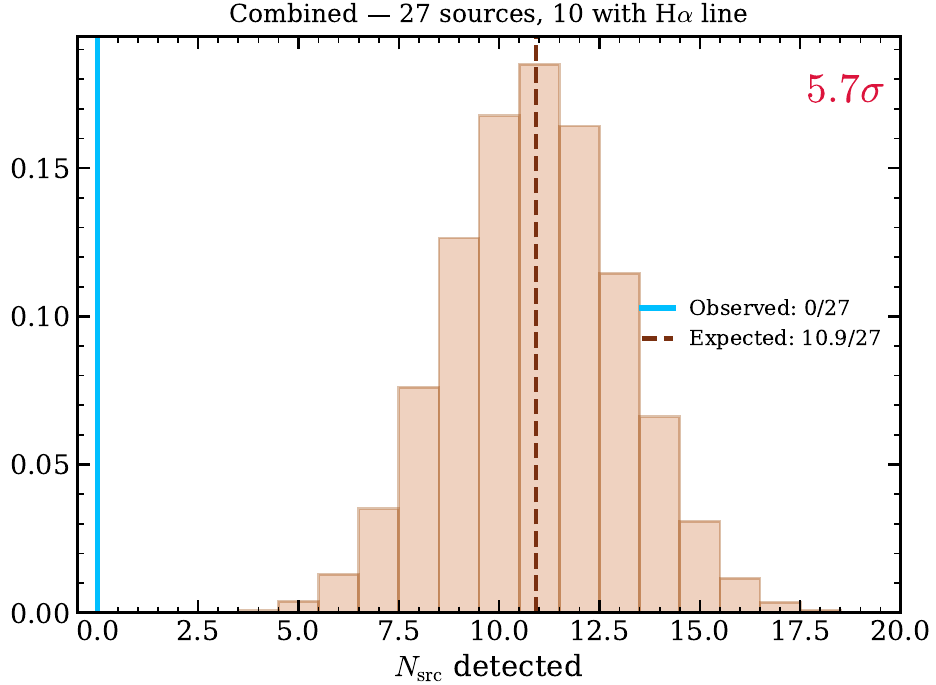}
\end{subfigure}
\begin{subfigure}{0.95\textwidth}
    \includegraphics[width=\linewidth]{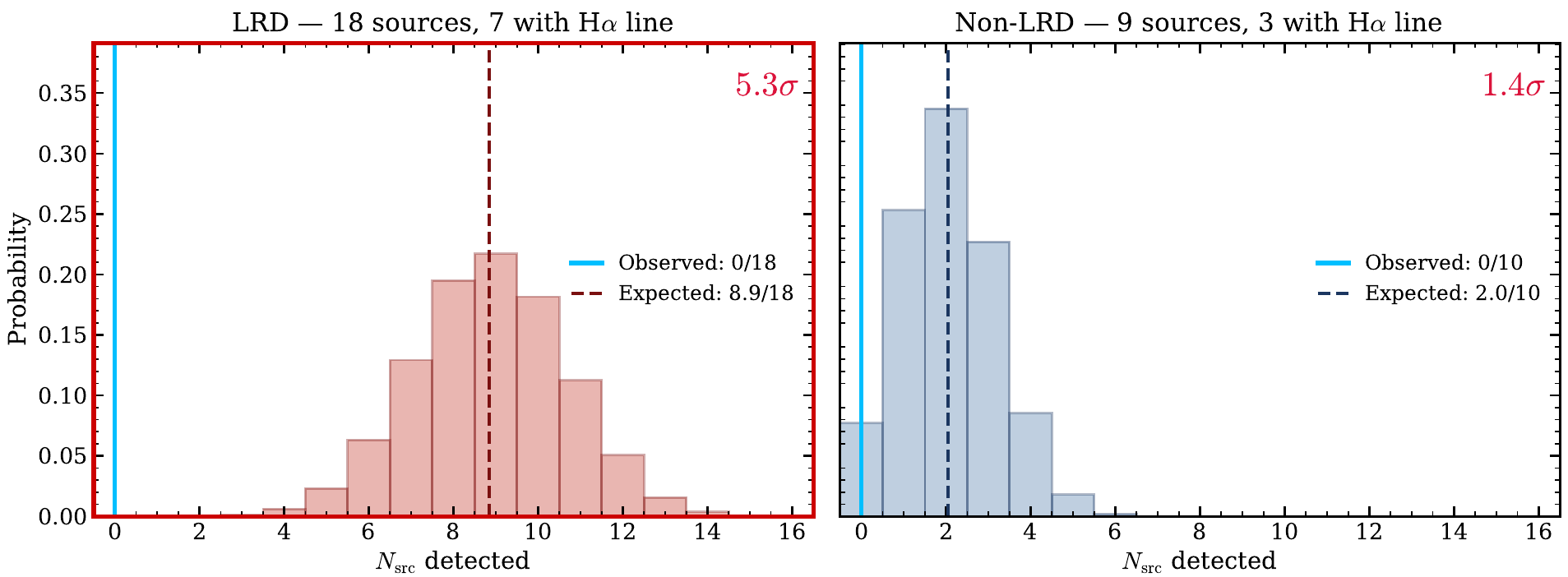}
\end{subfigure}
\caption{Same as Figure~\ref{fig:sdss_rm_combined}, but using an SDSS-RM control sample matched to our sources in $L_{5100}$.}
\label{fig:sdss_rm_combined_5100}
\end{figure*}

We follow Equation~(2) in \citet{deGraaff25b} to convert the \ha\ luminosities to $L_{5100}$, and use the same matching threshold as in Section~\ref{sec:sdss}, resulting in 160 SDSS-RM AGN within a factor of $2.0$ in luminosity. Among these, only sources at $z<0.56$ have \ha\ line detections. For higher-redshift AGN, where \ha\ is not covered, we use \hb\ as a proxy, since in typical AGN where both lines are coming from BLR, we would expect them to vary with comparable amplitudes. As shown in Figure~\ref{fig:sdss_rm_combined_5100}, regardless of the different matching method, the $L_{5100}$-matched control sample predicts $>10$ detectable variable sources. Our conclusion is therefore insensitive to the choice of luminosity proxy: the observed absence of variability in our high redshift broad \ha\ emitters is difficult to reconcile with the behavior of the local AGN population.

\section{Super-Eddington Model}\label{appendix2}
 In the standard sub-Eddington AGN picture, intrinsic variability amplitude and timescale scale with both luminosity and black hole mass \citep[e.g.,][]{Cristiani97, VandenBerk04, Wilhite08, MacLeod10}. These empirical relations are calibrated solely on sub-Eddington systems, however, and it is unclear whether they extend into the super-Eddington regime. To assess this uncertainty, we model the expected variability of a super-Eddington accretor using the mock light curves of \citet{Secunda26}, generated at a fixed black hole mass of $\log(M_\mathrm{BH}/M_\odot) = 8$. To apply these to our sample, we fix the bolometric luminosity to the median of our sources, $\log L_{\rm bol} = 43.6\ \rm erg\ s^{-1}$, and consider four treatments of the scaling: (i) \textit{unscaled}, in which the mock light curves are used directly without modification;  (ii)\textit{mass-scaled}, in which the variability is rescaled to account for the black hole mass of each source; (iii) \textit{luminosity-scaled}, in which the fixed bolometric luminosity ($\log L_{\rm bol} = 43.6\ \rm erg\ s^{-1}$) is used to derive the variability amplitude directly --- since luminosity is held constant across all mass bins, this yields a mass-independent prediction; and (iv) \textit{fully scaled}, applying both corrections simultaneously. We note that this framework is valid only up to $\log(M_\mathrm{BH}/M_\odot) \sim 5.5$: at fixed luminosity, higher black hole masses correspond to lower Eddington ratios, and above this threshold the system would enter the sub-Eddington regime, where the super-Eddington model no longer applies. One of these scenarios (e.g., mass-scaled) is broadly consistent with our observations and predicts $\sim 0 \,- 2$ detections over our time baseline. The implied Eddington ratio depends on the assumed $M_{\rm BH}$: at our median $\log L_{\rm bol} \approx 43.6$, assigning $M_{\rm BH} = 10^5\,M_\odot$ yields $\lambda_{\rm Edd} \approx 3$, while $M_{\rm BH} = 10^3\,M_\odot$ would require $\lambda_{\rm Edd} \approx 300$.

The \citet{Secunda26} models were not specifically calibrated for the black hole mass range of our sources, and the variability scaling relations in the super-Eddington regime remain poorly constrained. Our treatment of the four scaling schemes is therefore exploratory, intended only to demonstrate that the observed non-detections are physically compatible with a super-Eddington scenario rather than to uniquely identify it. Placing precise constraints on the accretion rate and black hole mass will require both a better understanding of variability in the super-Eddington regime and observations with longer rest-frame baselines and larger statistical samples.

\section{Calibration}\label{appendix3}

\begin{figure}[!htb]
\centering
\includegraphics[width=0.9\textwidth]{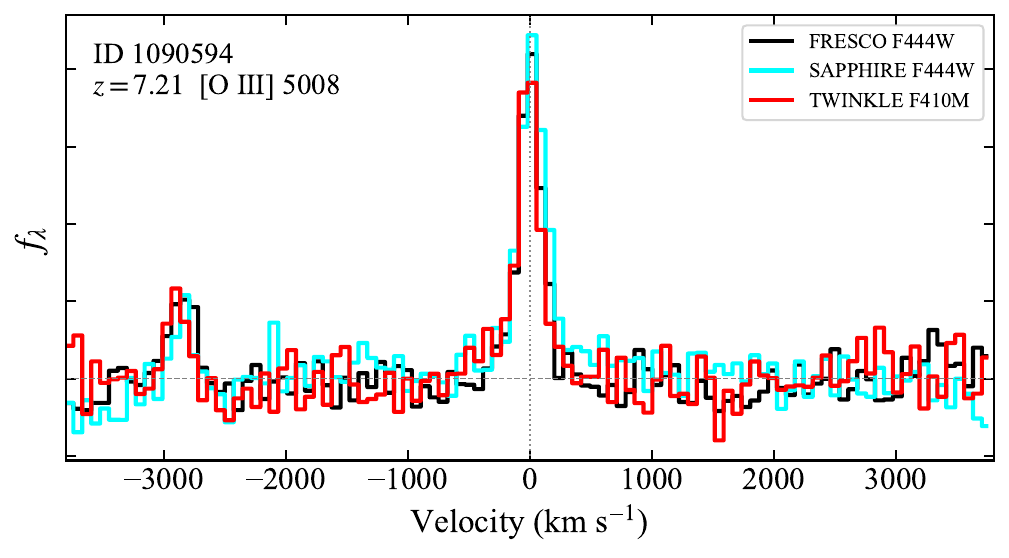}
\caption{Emission-line profile of the nearest \oiii\ emitter to ID 1090549 ($18.3\arcsec$ away) across FRESCO, SAPPHIRES, and TWINKLE Epoch 1. The similar offset in the SAPPHIRES epoch confirms a calibration artifact rather than intrinsic variability.}
\label{fig:oiii_offset}
\end{figure}

One object (ID 1090549) shows a systematic offset between the SAPPHIRES epoch and the other observations. This offset is likely a calibration artifact rather than intrinsic variability. We verify this by extracting the spectrum of the nearest \oiii\ emitter, located $18.3\arcsec$ away, and find a comparable velocity offset in the same epoch (Figure~\ref{fig:oiii_offset}). Physical considerations further support this conclusion: in standard AGN, continuum and broad-line variability arise from the compact broad-line region. The narrow-line region, by contrast, extends over $\gtrsim 100$ pc and responds to ionizing continuum changes on timescales of hundreds to thousands of years --- effectively constant over the rest-frame baseline probed here ($\lesssim 1$ yr). We attribute the offset to a ${\sim}0.5$--$1$ spectral-element shift in the position-dependent wavelength solution.

\begin{deluxetable*}{lcccrrrrc}
\tablecaption{Broad-line \ha\ emitters with per-source variability sensitivity \label{tab:sources}}
\tablewidth{0pt}
\tablehead{
  \colhead{JADES ID} & \colhead{R.A.} & \colhead{Decl.} & \colhead{$z$} &
  \colhead{$3\sigma_\mathrm{img,1}$} & \colhead{$3\sigma_\mathrm{img,2}$} &
  \colhead{$3\sigma_\mathrm{grism,1}$} & \colhead{$3\sigma_\mathrm{grism,2}$} &
  \colhead{$m_\mathrm{F444W}$} \\
  \colhead{} & \colhead{(deg)} & \colhead{(deg)} & \colhead{} &
  \colhead{(\%)} & \colhead{(\%)} &
  \colhead{(\%)} & \colhead{(\%)} &
  \colhead{(AB mag)}
}
\startdata
  1087315$^{a}$ & $189.334$ & $62.246$ & $3.91$ & $18.7$ & \nodata & $\cdots$ & $\cdots$ & $26.15$ \\
  1082263$^{a}$ & $189.213$ & $62.227$ & $3.98$ & $4.4$ & $4.8$ & $\cdots$ & $\cdots$ & $24.74$ \\
  1089568$^{a}$ & $189.152$ & $62.272$ & $4.05$ & $5.9$ & $5.6$ & $\cdots$ & $\cdots$ & $24.93$ \\
  1029154$^{a}$ & $189.159$ & $62.260$ & $4.17$ & $4.7$ & $4.7$ & $\cdots$ & $\cdots$ & $24.78$ \\
  1086784$^{a}$ & $189.306$ & $62.237$ & $4.40$ & $11.0$ & $10.9$ & $\cdots$ & $\cdots$ & $25.63$ \\
  1011836$^{b}$ & $189.221$ & $62.264$ & $4.41$ & $17.5$ & $16.0$ & $\cdots$ & $\cdots$ & $26.37$ \\
  1008411$^{a}$ & $189.211$ & $62.250$ & $4.41$ & $5.4$ & $5.6$ & $\cdots$ & $\cdots$ & $25.26$ \\
  1008671$^{a}$ & $189.162$ & $62.251$ & $4.41$ & $5.0$ & $5.7$ & $\cdots$ & $\cdots$ & $24.95$ \\
  1086855$^{a}$ & $189.287$ & $62.238$ & $4.41$ & $4.8$ & $4.9$ & $\cdots$ & $\cdots$ & $24.78$ \\
  1053757$^{b}$ & $189.270$ & $62.194$ & $4.45$ & $9.4$ & $5.7$ & $\cdots$ & $\cdots$ & $25.25$ \\
  1033320$^{a}$ & $189.126$ & $62.287$ & $4.48$ & $12.1$ & $12.3$ & $\cdots$ & $\cdots$ & $25.74$ \\
  1020621$^{b}$ & $189.123$ & $62.293$ & $4.68$ & $24.7$ & $24.0$ & $\cdots$ & $\cdots$ & $26.77$ \\
  1085355$^{a}$ & $189.094$ & $62.199$ & $4.88$ & $8.5$ & $8.6$ & $\cdots$ & $\cdots$ & $25.44$ \\
  1090253$^{a}$ & $189.286$ & $62.281$ & $5.09$ & $4.1$ & $4.1$ & 21.3 & 26.9 & $24.67$ \\
  1014406$^{a}$ & $189.072$ & $62.273$ & $5.14$ & $11.4$ & $14.4$ & 16.0 & 19.6 & $25.45$ \\
  1062309$^{b}$ & $189.249$ & $62.218$ & $5.17$ & $43.8$ & $40.0$ & 28.4 & 25.1 & $27.12$ \\
  1034620$^{a,\dagger}$ & $189.160$ & $62.296$ & $5.19$ & $3.8$ & $3.7$ & 11.9 & 13.5 & $24.30$ \\
  1090549$^{a}$ & $189.236$ & $62.286$ & $5.20$ & $15.3$ & $13.6$ & 24.9 & 23.6 & $25.93$ \\
  1188037$^{a}$ & $189.300$ & $62.212$ & $5.23$ & $8.2$ & $10.5$ & 16.7 & 14.8 & $24.87$ \\
  1077652$^{b}$ & $189.293$ & $62.199$ & $5.23$ & $37.9$ & $31.6$ & $\cdots$ & $\cdots$ & $26.59$ \\
  1013188$^{a}$ & $189.057$ & $62.269$ & $5.24$ & $8.8$ & \nodata & 20.9 & $\cdots$ & $25.42$ \\
  1003608$^{b}$ & $189.118$ & $62.236$ & $5.27$ & $45.7$ & $54.5$ & 19.3 & 23.4 & $27.28$ \\
  1020514$^{a}$ & $189.179$ & $62.293$ & $5.36$ & $5.1$ & $5.2$ & 8.2 & 9.0 & $24.80$ \\
  1087388$^{a}$ & $189.281$ & $62.247$ & $5.54$ & $0.7$ & $0.8$ & $2.7$ & $3.0$ & $22.99$ \\
  1001093$^{b}$ & $189.180$ & $62.225$ & $5.60$ & $40.2$ & $36.0$ & $\cdots$ & $\cdots$ & $26.98$ \\
  1061888$^{b}$ & $189.168$ & $62.217$ & $5.87$ & $21.2$ & $18.8$ & $\cdots$ & $\cdots$ & $26.42$ \\
  1010816$^{b}$ & $189.152$ & $62.260$ & $6.76$ & $7.9$ & $7.5$ & $\cdots$ & $\cdots$ & $25.30$ \\
\enddata
\tablecomments{Source IDs and coordinates are taken from the JADES v5.0 photometric catalog \citep{Robertson26}; $^\dagger$~source matched to the JADES v1.0 catalog (no v5.0 counterpart within $0\farcs5$; \citealt{Rieke23}); $^a$~NIRCam WFSS selection from CONGRESS+FRESCO \citep{matthee23, Zhang26}; $^b$~NIRSpec selection from \citet{Maiolino24}. (1) $3\sigma_{\mathrm{img},i}$ ($i=1,2$) denotes the $3\sigma$ F444W photometric sensitivity for TWINKLE Epoch~$i$ relative to FRESCO; (2) $3\sigma_{\mathrm{grism},i}$ ($i=1,2$) denotes the $3\sigma$ \ha\ line-flux sensitivity from NIRCam WFSS, with $\cdots$ indicating no coverage; (3) $m_{\mathrm{F444W}}$ is the JADES CIRC2 aperture AB magnitude ($r=0\farcs15$, aperture corrected).}
\end{deluxetable*}

\bibliography{sample631}{}

@ARTICLE{Jiang19,
       author = {{Jiang}, Yan-Fei and {Stone}, James M. and {Davis}, Shane W.},
        title = "{Super-Eddington Accretion Disks around Supermassive Black Holes}",
      journal = {\apj},
         year = 2019,
        month = aug,
       volume = {880},
       number = {2},
          eid = {67},
        pages = {67},
          doi = {10.3847/1538-4357/ab29ff},
archivePrefix = {arXiv},
       eprint = {1709.02845},
 primaryClass = {astro-ph.HE},
       adsurl = {https://ui.adsabs.harvard.edu/abs/2019ApJ...880...67J}
}

@ARTICLE{Chen26,
       author = {{Chen}, Yi-Xian and {Liu}, Hanpu and {Li}, Ruancun and {Wang}, Bingjie and {Ma}, Yilun and {Jiang}, Yan-Fei and {Greene}, Jenny E. and {Quataert}, Eliot and {Goodman}, Jeremy},
        title = "{Spectral Appearance of Self-gravitating AGN Disks Powered by Stellar Objects: Universal Effective Temperature in the Optical Continuum and Application to Little Red Dots}",
      journal = {arXiv e-prints},
         year = 2026,
        month = feb,
          eid = {arXiv:2602.06954},
        pages = {arXiv:2602.06954},
          doi = {10.48550/arXiv.2602.06954},
archivePrefix = {arXiv},
       eprint = {2602.06954},
 primaryClass = {astro-ph.HE},
       adsurl = {https://ui.adsabs.harvard.edu/abs/2026arXiv260206954C}
}

@ARTICLE{Zwick25,
       author = {{Zwick}, Lorenz and {Tiede}, Christopher and {Mayer}, Lucio},
        title = "{Little Red Dots as self-gravitating discs accreting on supermassive stars: Spectral appearance and formation pathway of the progenitors to direct collapse black holes}",
      journal = {arXiv e-prints},
         year = 2025,
        month = jul,
          eid = {arXiv:2507.22014},
        pages = {arXiv:2507.22014},
          doi = {10.48550/arXiv.2507.22014},
archivePrefix = {arXiv},
       eprint = {2507.22014},
 primaryClass = {astro-ph.GA},
       adsurl = {https://ui.adsabs.harvard.edu/abs/2025arXiv250722014Z}
}

@ARTICLE{degraaff25rubies,
       author = {{de Graaff}, Anna and {Brammer}, Gabriel and {Weibel}, Andrea and {Lewis}, Zach and {Maseda}, Michael V. and {Oesch}, Pascal A. and {Bezanson}, Rachel and {Boogaard}, Leindert A. and {Cleri}, Nikko J. and {Cooper}, Olivia R. and {Gottumukkala}, Rashmi and {Greene}, Jenny E. and {Hirschmann}, Michaela and {Hviding}, Raphael E. and {Katz}, Harley and {Labb{\'e}}, Ivo and {Leja}, Joel and {Matthee}, Jorryt and {McConachie}, Ian and {Miller}, Tim B. and {Naidu}, Rohan P. and {Price}, Sedona H. and {Rix}, Hans-Walter and {Setton}, David J. and {Suess}, Katherine A. and {Wang}, Bingjie and {Whitaker}, Katherine E. and {Williams}, Christina C.},
        title = "{RUBIES: A complete census of the bright and red distant Universe with JWST/NIRSpec}",
      journal = {\aap},
         year = 2025,
        month = may,
       volume = {697},
          eid = {A189},
        pages = {A189},
          doi = {10.1051/0004-6361/202452186},
archivePrefix = {arXiv},
       eprint = {2409.05948},
 primaryClass = {astro-ph.GA},
       adsurl = {https://ui.adsabs.harvard.edu/abs/2025A&A...697A.189D}
}

@ARTICLE{Chen25hostifany,
       author = {{Chen}, Chang-Hao and {Ho}, Luis C. and {Li}, Ruancun and {Zhuang}, Ming-Yang},
        title = "{The Host Galaxy (If Any) of the Little Red Dots}",
      journal = {\apj},
         year = 2025,
        month = apr,
       volume = {983},
       number = {1},
          eid = {60},
        pages = {60},
          doi = {10.3847/1538-4357/ada93a},
archivePrefix = {arXiv},
       eprint = {2411.04446},
 primaryClass = {astro-ph.GA},
       adsurl = {https://ui.adsabs.harvard.edu/abs/2025ApJ...983...60C}
}

@ARTICLE{Maiolino25qso1,
       author = {{Maiolino}, Roberto and {Uebler}, Hannah and {D'Eugenio}, Francesco and {Scholtz}, Jan and {Juodzbalis}, Ignas and {Ji}, Xihan and {Perna}, Michele and {Bromm}, Volker and {Dayal}, Pratika and {Koudmani}, Sophie and {Liu}, Boyuan and {Schneider}, Raffaella and {Sijacki}, Debora and {Valiante}, Rosa and {Trinca}, Alessandro and {Zhang}, Saiyang and {Volonteri}, Marta and {Inayoshi}, Kohei and {Carniani}, Stefano and {Nakajima}, Kimihiko and {Isobe}, Yuki and {Witstok}, Joris and {Jones}, Gareth C. and {Tacchella}, Sandro and {Arribas}, Santiago and {Bunker}, Andrew and {Cataldi}, Elisa and {Charlot}, Stephane and {Cresci}, Giovanni and {Curti}, Mirko and {Fabian}, Andrew C. and {Katz}, Harley and {Kumari}, Nimisha and {Laporte}, Nicolas and {Mazzolari}, Giovanni and {Robertson}, Brant and {Sun}, Fengwu and {Rodriguez Del Pino}, Bruno and {Venturi}, Giacomo},
        title = "{A black hole in a near-pristine galaxy 700 million years after the Big Bang}",
      journal = {arXiv e-prints},
         year = 2025,
        month = may,
          eid = {arXiv:2505.22567},
        pages = {arXiv:2505.22567},
          doi = {10.48550/arXiv.2505.22567},
archivePrefix = {arXiv},
       eprint = {2505.22567},
 primaryClass = {astro-ph.GA},
       adsurl = {https://ui.adsabs.harvard.edu/abs/2025arXiv250522567M}
}

@ARTICLE{Jones25,
       author = {{Jones}, Brenda L. and {Kocevski}, Dale D. and {Pacucci}, Fabio and {Taylor}, Anthony J. and {Finkelstein}, Steven L. and {Buchner}, Johannes and {Trump}, Jonathan R. and {Somerville}, Rachel S. and {Hirschmann}, Michaela and {Yung}, L.~Y. Aaron and {Barro}, Guillermo and {Bell}, Eric F. and {Bisigello}, Laura and {Calabro}, Antonello and {Cleri}, Nikko J. and {Dekel}, Avishai and {Dickinson}, Mark and {Gandolfi}, Giovanni and {Giavalisco}, Mauro and {Grogin}, Norman A. and {Inayoshi}, Kohei and {Kartaltepe}, Jeyhan S. and {Koekemoer}, Anton M. and {Napolitano}, Lorenzo and {Onoue}, Masafusa and {Ravindranath}, Swara and {Rodighiero}, Giulia and {Wilkins}, Stephen M.},
        title = "{The $M_{\rm BH}-M_{*}$ Relationship at $3<z<7$: Big Black Holes in Little Red Dots}",
      journal = {arXiv e-prints},
         year = 2025,
        month = oct,
          eid = {arXiv:2510.07376},
        pages = {arXiv:2510.07376},
          doi = {10.48550/arXiv.2510.07376},
archivePrefix = {arXiv},
       eprint = {2510.07376},
 primaryClass = {astro-ph.GA},
       adsurl = {https://ui.adsabs.harvard.edu/abs/2025arXiv251007376J}
}

@ARTICLE{Lambrides24,
       author = {{Lambrides}, Erini and {Garofali}, Kristen and {Larson}, Rebecca and {Ptak}, Andrew and {Chiaberge}, Marco and {Long}, Arianna S. and {Hutchison}, Taylor A. and {Norman}, Colin and {McKinney}, Jed and {Akins}, Hollis B. and {Berg}, Danielle A. and {Chisholm}, John and {Civano}, Francesca and {Cloonan}, Aidan P. and {Endsley}, Ryan and {Faisst}, Andreas L. and {Gilli}, Roberto and {Gillman}, Steven and {Hirschmann}, Michaela and {Kartaltepe}, Jeyhan S. and {Kocevski}, Dale D. and {Kokorev}, Vasily and {Pacucci}, Fabio and {Richardson}, Chris T. and {Stiavelli}, Massimo and {Whalen}, Kelly E.},
        title = "{The Case for Super-Eddington Accretion: Connecting Weak X-ray and UV Line Emission in JWST Broad-Line AGN During the First Gyr of Cosmic Time}",
      journal = {arXiv e-prints},
         year = 2024,
        month = sep,
          eid = {arXiv:2409.13047},
        pages = {arXiv:2409.13047},
          doi = {10.48550/arXiv.2409.13047},
archivePrefix = {arXiv},
       eprint = {2409.13047},
 primaryClass = {astro-ph.HE},
       adsurl = {https://ui.adsabs.harvard.edu/abs/2024arXiv240913047L}
}

@ARTICLE{Matthee26,
       author = {{Matthee}, Jorryt and {Torralba}, Alberto and {Pezzulli}, Gabriele and {Naidu}, Rohan P. and {Chisholm}, John and {Mascia}, Sara and {Greene}, Jenny E. and {Ishikawa}, Yuzo and {Gronke}, Max and {Wuyts}, Stijn and {Bordoloi}, Rongmon and {Brammer}, Gabriel and {Chang}, Seok-Jun and {Eilers}, Anna-Christina and {de Graaff}, Anna and {Hviding}, Raphael E. and {Iani}, Edoardo and {Illingworth}, Garth and {Kashino}, Daichi and {Labbe}, Ivo and {Ma}, Yilun and {Maseda}, Michael V. and {Meyer}, Romain and {Nelson}, Erica and {Oesch}, Pascal and {Xiao}, Mengyuan},
        title = "{The Engine and its Flows: Little Red Dot spectra are shaped by the column densities of their gas envelopes}",
      journal = {arXiv e-prints},
         year = 2026,
        month = mar,
          eid = {arXiv:2603.17667},
        pages = {arXiv:2603.17667},
          doi = {10.48550/arXiv.2603.17667},
archivePrefix = {arXiv},
       eprint = {2603.17667},
 primaryClass = {astro-ph.GA},
       adsurl = {https://ui.adsabs.harvard.edu/abs/2026arXiv260317667M}
}

@ARTICLE{Yan25,
       author = {{Yan}, Zu and {Inayoshi}, Kohei and {Chen}, Kejian and {Guo}, Jingsong},
        title = "{Balmer Transition Signatures from Gas-Enshrouded, Dust-Poor Active Galactic Nuclei}",
      journal = {arXiv e-prints},
         year = 2025,
        month = dec,
          eid = {arXiv:2512.11050},
        pages = {arXiv:2512.11050},
          doi = {10.48550/arXiv.2512.11050},
archivePrefix = {arXiv},
       eprint = {2512.11050},
 primaryClass = {astro-ph.GA},
       adsurl = {https://ui.adsabs.harvard.edu/abs/2025arXiv251211050Y}
}

@ARTICLE{Delvecchio25,
       author = {{Delvecchio}, I. and {Daddi}, E. and {Magnelli}, B. and {Elbaz}, D. and {Giavalisco}, M. and {Traina}, A. and {Lanzuisi}, G. and {Akins}, H.~B. and {Belli}, S. and {Casey}, C.~M. and {Gentile}, F. and {Gruppioni}, C. and {Pozzi}, F. and {Zamorani}, G.},
        title = "{Active galactic nuclei-heated dust revealed in ``little red dots''}",
      journal = {\aap},
         year = 2025,
        month = dec,
       volume = {704},
          eid = {A313},
        pages = {A313},
          doi = {10.1051/0004-6361/202557164},
archivePrefix = {arXiv},
       eprint = {2509.07100},
 primaryClass = {astro-ph.GA},
       adsurl = {https://ui.adsabs.harvard.edu/abs/2025A&A...704A.313D}
}

@ARTICLE{Lyubarskii97,
       author = {{Lyubarskii}, Yu. E.},
        title = "{Flicker noise in accretion discs}",
      journal = {\mnras},
         year = 1997,
        month = dec,
       volume = {292},
       number = {3},
        pages = {679-685},
          doi = {10.1093/mnras/292.3.679},
       adsurl = {https://ui.adsabs.harvard.edu/abs/1997MNRAS.292..679L}
}

@ARTICLE{Secunda24,
       author = {{Secunda}, Amy and {Jiang}, Yan-Fei and {Greene}, Jenny E.},
        title = "{Simulating X-Ray Reverberation in the Ultraviolet-emitting Regions of Active Galactic Nuclei Accretion Disks with Three-dimensional Multifrequency Radiation Magnetohydrodynamic Simulations}",
      journal = {\apjl},
         year = 2024,
        month = apr,
       volume = {965},
       number = {2},
          eid = {L29},
        pages = {L29},
          doi = {10.3847/2041-8213/ad34b0},
archivePrefix = {arXiv},
       eprint = {2311.10820},
 primaryClass = {astro-ph.HE},
       adsurl = {https://ui.adsabs.harvard.edu/abs/2024ApJ...965L..29S}
}

@ARTICLE{Beard25,
       author = {{Beard}, Max and {McHardy}, Ian and {Horne}, Keith and {Cackett}, Edward and {Vincentelli}, Federico and {Venancio Hernandez Santisteban}, Juan and {Miller}, Jake and {Dhillon}, Vikram and {Knapen}, Johan and {Littlefair}, Stuart and {Kynoch}, Daniel and {Breedt}, Elm{\'e} and {Shen}, Yue and {Gelbord}, Jonathan},
        title = "{Testing disc reprocessing models for AGN optical variability by comparison of X-ray and optical power spectra of NGC 4395}",
      journal = {arXiv e-prints},
         year = 2025,
        month = jan,
          eid = {arXiv:2501.02664},
        pages = {arXiv:2501.02664},
          doi = {10.48550/arXiv.2501.02664},
archivePrefix = {arXiv},
       eprint = {2501.02664},
 primaryClass = {astro-ph.HE},
       adsurl = {https://ui.adsabs.harvard.edu/abs/2025arXiv250102664B}
}

@ARTICLE{Burke23,
       author = {{Burke}, Colin J. and {Shen}, Yue and {Liu}, Xin and {Natarajan}, Priyamvada and {Caplar}, Neven and {Bellovary}, Jillian M. and {Wang}, Z. Franklin},
        title = "{Dwarf AGNs from variability for the origins of seeds (DAVOS): Intermediate-mass black hole demographics from optical synoptic surveys}",
      journal = {\mnras},
         year = 2023,
        month = jan,
       volume = {518},
       number = {2},
        pages = {1880-1904},
          doi = {10.1093/mnras/stac2478},
archivePrefix = {arXiv},
       eprint = {2207.04092},
 primaryClass = {astro-ph.GA},
       adsurl = {https://ui.adsabs.harvard.edu/abs/2023MNRAS.518.1880B}
}

@ARTICLE{Jiang25,
       author = {{Jiang}, Yan-Fei and {Blaes}, Omer and {Kaul}, Ish and {Zhang}, Lizhong},
        title = "{Radiation and Magnetic Pressure Support in Accretion Disks Around Supermassive Black Holes and the Physical Origin of the Extreme-ultraviolet to Soft X-Ray Spectrum}",
      journal = {\apj},
         year = 2025,
        month = jul,
       volume = {988},
       number = {1},
          eid = {43},
        pages = {43},
          doi = {10.3847/1538-4357/addecb},
archivePrefix = {arXiv},
       eprint = {2505.09671},
 primaryClass = {astro-ph.HE},
       adsurl = {https://ui.adsabs.harvard.edu/abs/2025ApJ...988...43J}
}

@ARTICLE{Umeda25,
       author = {{Umeda}, Hiroya and {Inayoshi}, Kohei and {Harikane}, Yuichi and {Murase}, Kohta},
        title = "{A Black-Hole Envelope Interpretation for Cosmological Demographics of Little Red Dots}",
      journal = {arXiv e-prints},
         year = 2025,
        month = dec,
          eid = {arXiv:2512.04208},
        pages = {arXiv:2512.04208},
          doi = {10.48550/arXiv.2512.04208},
archivePrefix = {arXiv},
       eprint = {2512.04208},
 primaryClass = {astro-ph.GA},
       adsurl = {https://ui.adsabs.harvard.edu/abs/2025arXiv251204208U}
}

@ARTICLE{Zu11,
       author = {{Zu}, Ying and {Kochanek}, C.~S. and {Peterson}, Bradley M.},
        title = "{An Alternative Approach to Measuring Reverberation Lags in Active Galactic Nuclei}",
      journal = {\apj},
         year = 2011,
        month = jul,
       volume = {735},
       number = {2},
          eid = {80},
        pages = {80},
          doi = {10.1088/0004-637X/735/2/80},
archivePrefix = {arXiv},
       eprint = {1008.0641},
 primaryClass = {astro-ph.CO},
       adsurl = {https://ui.adsabs.harvard.edu/abs/2011ApJ...735...80Z}
}

@ARTICLE{Horne86,
       author = {{Horne}, K.},
        title = "{An optimal extraction algorithm for CCD spectroscopy.}",
      journal = {\pasp},
         year = 1986,
        month = jun,
       volume = {98},
        pages = {609-617},
          doi = {10.1086/131801},
       adsurl = {https://ui.adsabs.harvard.edu/abs/1986PASP...98..609H}
}

@ARTICLE{Roman-Garza26,
       author = {{Roman-Garza}, J. and {Schaerer}, D. and {Charbonnel}, C. and {Fragos}, T. and {Cenci}, E. and {Marques-Chaves}, R. and {Oesch}, P. and {Xiao}, M.},
        title = "{A quasi-star is born: formation and evolution of accreting quasi-stars as a metallicity-independent pathway to Little Red Dots}",
      journal = {arXiv e-prints},
         year = 2026,
        month = mar,
          eid = {arXiv:2603.21714},
        pages = {arXiv:2603.21714},
          doi = {10.48550/arXiv.2603.21714},
archivePrefix = {arXiv},
       eprint = {2603.21714},
 primaryClass = {astro-ph.SR},
       adsurl = {https://ui.adsabs.harvard.edu/abs/2026arXiv260321714R}
}

@ARTICLE{Stone25,
       author = {{Stone}, Zachary and {Shen}, Yue and {Zhuang}, Ming-Yang and {Hu}, Lei and {Pierel}, Justin and {Li}, Junyao and {Burgasser}, Adam J. and {Greene}, Jenny E. and {Pan}, Zhiwei and {Shapley}, Alice E. and {Sun}, Fengwu and {Venkatraman}, Padmavathi and {Wang}, Feige},
        title = "{NEXUS: A Search for Nuclear Variability with the First Two JWST NIRCam Epochs}",
      journal = {arXiv e-prints},
         year = 2025,
        month = sep,
          eid = {arXiv:2509.19585},
        pages = {arXiv:2509.19585},
          doi = {10.48550/arXiv.2509.19585},
archivePrefix = {arXiv},
       eprint = {2509.19585},
 primaryClass = {astro-ph.GA},
       adsurl = {https://ui.adsabs.harvard.edu/abs/2025arXiv250919585S}
}

@ARTICLE{Shen24,
       author = {{Shen}, Yue and {Grier}, Catherine J. and {Horne}, Keith and {Stone}, Zachary and {Li}, Jennifer I. and {Yang}, Qian and {Homayouni}, Yasaman and {Trump}, Jonathan R. and {Anderson}, Scott F. and {Brandt}, W.~N. and {Hall}, Patrick B. and {Ho}, Luis C. and {Jiang}, Linhua and {Petitjean}, Patrick and {Schneider}, Donald P. and {Tao}, Charling and {Donnan}, Fergus. R. and {AlSayyad}, Yusra and {Bershady}, Matthew A. and {Blanton}, Michael R. and {Bizyaev}, Dmitry and {Bundy}, Kevin and {Chen}, Yuguang and {Davis}, Megan C. and {Dawson}, Kyle and {Fan}, Xiaohui and {Greene}, Jenny E. and {Gr{\"o}ller}, Hannes and {Guo}, Yucheng and {Ibarra-Medel}, H{\'e}ctor and {Jiang}, Yuanzhe and {Keenan}, Ryan P. and {Kollmeier}, Juna A. and {Lejoly}, Cassandra and {Li}, Zefeng and {de la Macorra}, Axel and {Moe}, Maxwell and {Nie}, Jundan and {Rossi}, Graziano and {Smith}, Paul S. and {Tee}, Wei Leong and {Weijmans}, Anne-Marie and {Xu}, Jiachuan and {Yue}, Minghao and {Zhou}, Xu and {Zhou}, Zhimin and {Zou}, Hu},
        title = "{The Sloan Digital Sky Survey Reverberation Mapping Project: Key Results}",
      journal = {\apjs},
         year = 2024,
        month = jun,
       volume = {272},
       number = {2},
          eid = {26},
        pages = {26},
          doi = {10.3847/1538-4365/ad3936},
archivePrefix = {arXiv},
       eprint = {2305.01014},
 primaryClass = {astro-ph.GA},
       adsurl = {https://ui.adsabs.harvard.edu/abs/2024ApJS..272...26S}
}

@ARTICLE{Zhang26,
       author = {{Zhang}, Junyu and {Egami}, Eiichi and {Sun}, Fengwu and {Lin}, Xiaojing and {Lyu}, Jianwei and {Zhu}, Yongda and {Rinaldi}, Pierluigi and {Sun}, Yang and {Bunker}, Andrew J. and {Bhatawdekar}, Rachana and {Helton}, Jakob M. and {Maiolino}, Roberto and {Ma}, Zheng and {Robertson}, Brant and {Tacchella}, Sandro and {Venturi}, Giacomo and {Williams}, Christina C. and {Willott}, Chris},
        title = "{Abundant Population of Broad H{\ensuremath{\alpha}} Emitters in the GOODS-N Field Revealed by CONGRESS, FRESCO, and JADES}",
      journal = {\apj},
         year = 2026,
        month = feb,
       volume = {997},
       number = {2},
          eid = {250},
        pages = {250},
          doi = {10.3847/1538-4357/ae2681},
archivePrefix = {arXiv},
       eprint = {2505.02895},
 primaryClass = {astro-ph.GA},
       adsurl = {https://ui.adsabs.harvard.edu/abs/2026ApJ...997..250Z}
      
}

@ARTICLE{Brazzini26,
       author = {{Brazzini}, M. and {D'Eugenio}, F. and {Maiolino}, R. and {Lyu}, J. and {DeCoursey}, C. and {{\"U}bler}, H. and {Ji}, X. and {Juod{\v{z}}balis}, I. and {Scholtz}, J. and {Jones}, G.~C. and {Hainline}, K. and {Dalla Bont{\`a}}, E. and {{\'e}rez-Gonz{\'a}lez}, P.~G. P and {Geris}, S. and {Harshan}, A. and {Feruglio}, C. and {Bischetti}, M. and {Mazzolari}, G. and {Rieke}, G. and {Alberts}, S. and {Trefoloni}, B. and {Carniani}, S. and {Parlanti}, E. and {Marconi}, A. and {Risaliti}, G. and {Ramos Almeida}, C. and {Rinaldi}, P. and {Perna}, M. and {Zamora}, S. and {Lamperti}, I. and {Venturi}, G. and {Cresci}, G. and {Bunker}, Andrew J. and {Ivey}, L.~R.},
        title = "{The Little Blue and Red Dots Rosetta Stones: Non-Gaussian broad lines, hot dust, and X-ray weakness}",
      journal = {arXiv e-prints},
         year = 2026,
        month = jan,
          eid = {arXiv:2601.22214},
        pages = {arXiv:2601.22214},
          doi = {10.48550/arXiv.2601.22214},
archivePrefix = {arXiv},
       eprint = {2601.22214},
 primaryClass = {astro-ph.GA},
       adsurl = {https://ui.adsabs.harvard.edu/abs/2026arXiv260122214B}
}

@ARTICLE{Matthee25,
       author = {{Matthee}, Jorryt and {Naidu}, Rohan P. and {Kotiwale}, Gauri and {Furtak}, Lukas J. and {Kramarenko}, Ivan and {Mackenzie}, Ruari and {Greene}, Jenny and {Adamo}, Angela and {Bouwens}, Rychard J. and {Di Cesare}, Claudia and {Eilers}, Anna-Christina and {de Graaff}, Anna and {Heintz}, Kasper E. and {Kashino}, Daichi and {Maseda}, Michael V. and {Tacchella}, Sandro and {Torralba}, Alberto},
        title = "{Environmental Evidence for Overly Massive Black Holes in Low-mass Galaxies and a Black Hole{\textendash}Halo Mass Relation at z {\ensuremath{\sim}} 5}",
      journal = {\apj},
         year = 2025,
        month = aug,
       volume = {988},
       number = {2},
          eid = {246},
        pages = {246},
          doi = {10.3847/1538-4357/ade886},
archivePrefix = {arXiv},
       eprint = {2412.02846},
 primaryClass = {astro-ph.GA},
       adsurl = {https://ui.adsabs.harvard.edu/abs/2025ApJ...988..246M}
}

@ARTICLE{Pizzati25,
       author = {{Pizzati}, Elia and {Hennawi}, Joseph F. and {Schaye}, Joop and {Eilers}, Anna-Christina and {Huang}, Jiamu and {Schindler}, Jan-Torge and {Wang}, Feige},
        title = "{'Little red dots' cannot reside in the same dark matter haloes as comparably luminous unobscured quasars}",
      journal = {\mnras},
         year = 2025,
        month = jun,
       volume = {539},
       number = {4},
        pages = {2910-2925},
          doi = {10.1093/mnras/staf660},
archivePrefix = {arXiv},
       eprint = {2409.18208},
 primaryClass = {astro-ph.GA},
       adsurl = {https://ui.adsabs.harvard.edu/abs/2025MNRAS.539.2910P}
}

@ARTICLE{Lin26,
       author = {{Lin}, Xiaojing and {Fan}, Xiaohui and {Sun}, Fengwu and {Zhang}, Junyu and {Egami}, Eiichi and {Helton}, Jakob M. and {Wang}, Feige and {Zhang}, Haowen and {Bunker}, Andrew J. and {Cai}, Zheng and {Ji}, Zhiyuan and {Jin}, Xiangyu and {Maiolino}, Roberto and {Pudoka}, Maria Anne and {Rinaldi}, Pierluigi and {Robertson}, Brant and {Tacchella}, Sandro and {Tee}, Wei Leong and {Sun}, Yang and {Willmer}, Christopher N.~A. and {Willott}, Chris and {Zhu}, Yongda},
        title = "{The Large-scale Environments of Low-luminosity AGNs at 3.9 < z < 6 and Implications for Their Host Dark Matter Halos from a Complete NIRCam Grism Redshift Survey}",
      journal = {\apj},
         year = 2026,
        month = jan,
       volume = {997},
       number = {1},
          eid = {61},
        pages = {61},
          doi = {10.3847/1538-4357/ae1eef},
archivePrefix = {arXiv},
       eprint = {2505.02896},
 primaryClass = {astro-ph.GA},
       adsurl = {https://ui.adsabs.harvard.edu/abs/2026ApJ...997...61L}
}

@ARTICLE{Williams24,
       author = {{Williams}, Christina C. and {Alberts}, Stacey and {Ji}, Zhiyuan and {Hainline}, Kevin N. and {Lyu}, Jianwei and {Rieke}, George and {Endsley}, Ryan and {Suess}, Katherine A. and {Sun}, Fengwu and {Johnson}, Benjamin D. and {Florian}, Michael and {Shivaei}, Irene and {Rujopakarn}, Wiphu and {Baker}, William M. and {Bhatawdekar}, Rachana and {Boyett}, Kristan and {Bunker}, Andrew J. and {Cameron}, Alex J. and {Carniani}, Stefano and {Charlot}, Stephane and {Curtis-Lake}, Emma and {DeCoursey}, Christa and {de Graaff}, Anna and {Egami}, Eiichi and {Eisenstein}, Daniel J. and {Gibson}, Justus L. and {Hausen}, Ryan and {Helton}, Jakob M. and {Maiolino}, Roberto and {Maseda}, Michael V. and {Nelson}, Erica J. and {P{\'e}rez-Gonz{\'a}lez}, Pablo G. and {Rieke}, Marcia J. and {Robertson}, Brant E. and {Saxena}, Aayush and {Tacchella}, Sandro and {Willmer}, Christopher N.~A. and {Willott}, Chris J.},
        title = "{The Galaxies Missed by Hubble and ALMA: The Contribution of Extremely Red Galaxies to the Cosmic Census at 3 < z < 8}",
      journal = {\apj},
         year = 2024,
        month = jun,
       volume = {968},
       number = {1},
          eid = {34},
        pages = {34},
          doi = {10.3847/1538-4357/ad3f17},
archivePrefix = {arXiv},
       eprint = {2311.07483},
 primaryClass = {astro-ph.GA},
       adsurl = {https://ui.adsabs.harvard.edu/abs/2024ApJ...968...34W}
}

@ARTICLE{Liu26,
       author = {{Liu}, Hanpu and {Jiang}, Yan-Fei and {Quataert}, Eliot and {Greene}, Jenny E. and {Ma}, Yilun and {Lin}, Xiaojing},
        title = "{Synthetic Spectral Library of Optically Thick Atmospheres for Little Red Dots}",
      journal = {arXiv e-prints},
         year = 2026,
        month = mar,
          eid = {arXiv:2603.02317},
        pages = {arXiv:2603.02317},
archivePrefix = {arXiv},
       eprint = {2603.02317},
 primaryClass = {astro-ph.GA},
       adsurl = {https://ui.adsabs.harvard.edu/abs/2026arXiv260302317L}
}

@ARTICLE{Setton25b,
       author = {{Setton}, David J. and {Greene}, Jenny E. and {de Graaff}, Anna and {Ma}, Yilun and {Leja}, Joel and {Matthee}, Jorryt and {Bezanson}, Rachel and {Boogaard}, Leindert A. and {Cleri}, Nikko J. and {Katz}, Harley and {Labbe}, Ivo and {Maseda}, Michael V. and {McConachie}, Ian and {Miller}, Tim B. and {Price}, Sedona H. and {Suess}, Katherine A. and {van Dokkum}, Pieter and {Wang}, Bingjie and {Weibel}, Andrea and {Whitaker}, Katherine E. and {Williams}, Christina C.},
        title = "{Little Red Dots at an Inflection Point: Ubiquitous V-shaped Turnover Consistently Occurs at the Balmer Limit}",
      journal = {\apj},
         year = 2025,
        month = dec,
       volume = {995},
       number = {1},
          eid = {118},
        pages = {118},
          doi = {10.3847/1538-4357/ae1500},
archivePrefix = {arXiv},
       eprint = {2411.03424},
 primaryClass = {astro-ph.GA},
       adsurl = {https://ui.adsabs.harvard.edu/abs/2025ApJ...995..118S}
}

@ARTICLE{Setton25,
       author = {{Setton}, David J. and {Greene}, Jenny E. and {Spilker}, Justin S. and {Williams}, Christina C. and {Labb{\'e}}, Ivo and {Ma}, Yilun and {Wang}, Bingjie and {Whitaker}, Katherine E. and {Leja}, Joel and {de Graaff}, Anna and {Alberts}, Stacey and {Bezanson}, Rachel and {Boogaard}, Leindert A. and {Brammer}, Gabriel and {Cutler}, Sam E. and {Cleri}, Nikko J. and {Cooper}, Olivia R. and {Dayal}, Pratika and {Fujimoto}, Seiji and {Furtak}, Lukas J. and {Goulding}, Andy D. and {Hirschmann}, Michaela and {Kokorev}, Vasily and {Maseda}, Michael V. and {McConachie}, Ian and {Matthee}, Jorryt and {Miller}, Tim B. and {Naidu}, Rohan P. and {Oesch}, Pascal A. and {Pan}, Richard and {Price}, Sedona H. and {Suess}, Katherine A. and {Weaver}, John R. and {Xiao}, Mengyuan and {Zhang}, Yunchong and {Zitrin}, Adi},
        title = "{A Confirmed Deficit of Hot and Cold Dust Emission in the Most Luminous Little Red Dots}",
      journal = {\apjl},
         year = 2025,
        month = sep,
       volume = {991},
       number = {1},
          eid = {L10},
        pages = {L10},
          doi = {10.3847/2041-8213/ade78b},
archivePrefix = {arXiv},
       eprint = {2503.02059},
 primaryClass = {astro-ph.GA},
       adsurl = {https://ui.adsabs.harvard.edu/abs/2025ApJ...991L..10S}
}

@ARTICLE{Naidu25,
       author = {{Naidu}, Rohan P. and {Matthee}, Jorryt and {Katz}, Harley and {de Graaff}, Anna and {Oesch}, Pascal and {Smith}, Aaron and {Greene}, Jenny E. and {Brammer}, Gabriel and {Weibel}, Andrea and {Hviding}, Raphael and {Chisholm}, John and {Labb\textbackslash'e}, Ivo and {Simcoe}, Robert A. and {Witten}, Callum and {Atek}, Hakim and {Baggen}, Josephine F.~W. and {Belli}, Sirio and {Bezanson}, Rachel and {Boogaard}, Leindert A. and {Bose}, Sownak and {Covelo-Paz}, Alba and {Dayal}, Pratika and {Fudamoto}, Yoshinobu and {Furtak}, Lukas J. and {Giovinazzo}, Emma and {Goulding}, Andy and {Gronke}, Max and {Heintz}, Kasper E. and {Hirschmann}, Michaela and {Illingworth}, Garth and {Inoue}, Akio K. and {Johnson}, Benjamin D. and {Leja}, Joel and {Leonova}, Ecaterina and {McConachie}, Ian and {Maseda}, Michael V. and {Natarajan}, Priyamvada and {Nelson}, Erica and {Setton}, David J. and {Shivaei}, Irene and {Sobral}, David and {Stefanon}, Mauro and {Tacchella}, Sandro and {Toft}, Sune and {Torralba}, Alberto and {van Dokkum}, Pieter and {van der Wel}, Arjen and {Volonteri}, Marta and {Walter}, Fabian and {Wang}, Bingjie and {Watson}, Darach},
        title = "{A ``Black Hole Star'' Reveals the Remarkable Gas-Enshrouded Hearts of the Little Red Dots}",
      journal = {arXiv e-prints},
         year = 2025,
        month = mar,
          eid = {arXiv:2503.16596},
        pages = {arXiv:2503.16596},
          doi = {10.48550/arXiv.2503.16596},
archivePrefix = {arXiv},
       eprint = {2503.16596},
 primaryClass = {astro-ph.GA},
       adsurl = {https://ui.adsabs.harvard.edu/abs/2025arXiv250316596N}
}

@ARTICLE{Xiao25,
       author = {{Xiao}, Mengyuan and {Oesch}, Pascal A. and {Bing}, Longji and {Elbaz}, David and {Matthee}, Jorryt and {Fudamoto}, Yoshinobu and {Fujimoto}, Seiji and {Marques-Chaves}, Rui and {Williams}, Christina C. and {Dessauges-Zavadsky}, Miroslava and {Valentino}, Francesco and {Brammer}, Gabriel and {Covelo-Paz}, Alba and {Daddi}, Emanuele and {Fynbo}, Johan P.~U. and {Gillman}, Steven and {Ginolfi}, Michele and {Giovinazzo}, Emma and {Greene}, Jenny E. and {Gu}, Qiusheng and {Illingworth}, Garth and {Inayoshi}, Kohei and {Kokorev}, Vasily and {Meyer}, Romain A. and {Naidu}, Rohan P. and {Reddy}, Naveen A. and {Schaerer}, Daniel and {Shapley}, Alice and {Stefanon}, Mauro and {Steinhardt}, Charles L. and {Setton}, David J. and {Vestergaard}, Marianne and {Wang}, Tao},
        title = "{No [C II] or dust detection in two Little Red Dots at z$_{spec}$> 7}",
      journal = {\aap},
         year = 2025,
        month = aug,
       volume = {700},
          eid = {A231},
        pages = {A231},
          doi = {10.1051/0004-6361/202554361},
archivePrefix = {arXiv},
       eprint = {2503.01945},
 primaryClass = {astro-ph.GA},
       adsurl = {https://ui.adsabs.harvard.edu/abs/2025A&A...700A.231X}
}

@ARTICLE{Yue24,
       author = {{Yue}, Minghao and {Eilers}, Anna-Christina and {Ananna}, Tonima Tasnim and {Panagiotou}, Christos and {Kara}, Erin and {Miyaji}, Takamitsu},
        title = "{Stacking X-Ray Observations of ``Little Red Dots'': Implications for Their Active Galactic Nucleus Properties}",
      journal = {\apjl},
         year = 2024,
        month = oct,
       volume = {974},
       number = {2},
          eid = {L26},
        pages = {L26},
          doi = {10.3847/2041-8213/ad7eba},
archivePrefix = {arXiv},
       eprint = {2404.13290},
 primaryClass = {astro-ph.GA},
       adsurl = {https://ui.adsabs.harvard.edu/abs/2024ApJ...974L..26Y}
}

@ARTICLE{Hviding25,
       author = {{Hviding}, Raphael E. and {de Graaff}, Anna and {Miller}, Tim B. and {Setton}, David J. and {Greene}, Jenny E. and {Labb{\'e}}, Ivo and {Brammer}, Gabriel and {Bezanson}, Rachel and {Boogaard}, Leindert A. and {Cleri}, Nikko J. and {Leja}, Joel and {Maseda}, Michael V. and {McConachie}, Ian and {Matthee}, Jorryt and {Naidu}, Rohan P. and {Oesch}, Pascal A. and {Wang}, Bingjie and {Whitaker}, Katherine E. and {Williams}, Christina C.},
        title = "{RUBIES: A spectroscopic census of little red dots: All point sources with v-shaped continua have broad lines}",
      journal = {\aap},
         year = 2025,
        month = oct,
       volume = {702},
          eid = {A57},
        pages = {A57},
          doi = {10.1051/0004-6361/202555816},
archivePrefix = {arXiv},
       eprint = {2506.05459},
 primaryClass = {astro-ph.GA},
       adsurl = {https://ui.adsabs.harvard.edu/abs/2025A&A...702A..57H}
}

@ARTICLE{Greene26,
       author = {{Greene}, Jenny E. and {Setton}, David J. and {Furtak}, Lukas J. and {Naidu}, Rohan P. and {Volonteri}, Marta and {Dayal}, Pratika and {Labbe}, Ivo and {van Dokkum}, Pieter and {Bezanson}, Rachel and {Brammer}, Gabriel and {Cutler}, Sam E. and {Glazebrook}, Karl and {de Graaff}, Anna and {Hirschmann}, Michaela and {Hviding}, Raphael E. and {Kokorev}, Vasily and {Leja}, Joel and {Liu}, Hanpu and {Ma}, Yilun and {Matthee}, Jorryt and {Nanayakkara}, Themiya and {Oesch}, Pascal A. and {Pan}, Richard and {Price}, Sedona H. and {Spilker}, Justin S. and {Wang}, Bingjie and {Weaver}, John R. and {Whitaker}, Katherine E. and {Williams}, Christina C. and {Zitrin}, Adi},
        title = "{What You See Is What You Get: Empirically Measured Bolometric Luminosities of Little Red Dots}",
      journal = {\apj},
         year = 2026,
        month = jan,
       volume = {996},
       number = {2},
          eid = {129},
        pages = {129},
          doi = {10.3847/1538-4357/ae1836},
archivePrefix = {arXiv},
       eprint = {2509.05434},
 primaryClass = {astro-ph.GA},
       adsurl = {https://ui.adsabs.harvard.edu/abs/2026ApJ...996..129G}
}

@ARTICLE{Inayoshi24,
       author = {{Inayoshi}, Kohei and {Ichikawa}, Kohei},
        title = "{Birth of Rapidly Spinning, Overmassive Black Holes in the Early Universe}",
      journal = {\apjl},
         year = 2024,
        month = oct,
       volume = {973},
       number = {2},
          eid = {L49},
        pages = {L49},
          doi = {10.3847/2041-8213/ad74e2},
archivePrefix = {arXiv},
       eprint = {2402.14706},
 primaryClass = {astro-ph.GA},
       adsurl = {https://ui.adsabs.harvard.edu/abs/2024ApJ...973L..49I}
}

@ARTICLE{Kokorev24,
       author = {{Kokorev}, Vasily and {Caputi}, Karina I. and {Greene}, Jenny E. and {Dayal}, Pratika and {Trebitsch}, Maxime and {Cutler}, Sam E. and {Fujimoto}, Seiji and {Labb{\'e}}, Ivo and {Miller}, Tim B. and {Iani}, Edoardo and {Navarro-Carrera}, Rafael and {Rinaldi}, Pierluigi},
        title = "{A Census of Photometrically Selected Little Red Dots at 4 < z < 9 in JWST Blank Fields}",
      journal = {\apj},
         year = 2024,
        month = jun,
       volume = {968},
       number = {1},
          eid = {38},
        pages = {38},
          doi = {10.3847/1538-4357/ad4265},
archivePrefix = {arXiv},
       eprint = {2401.09981},
 primaryClass = {astro-ph.GA},
       adsurl = {https://ui.adsabs.harvard.edu/abs/2024ApJ...968...38K}
}

@ARTICLE{Eisenstein26,
       author = {{Eisenstein}, Daniel J. and {Willott}, Chris and {Alberts}, Stacey and {Arribas}, Santiago and {Bonaventura}, Nina and {Bunker}, Andrew J. and {Cameron}, Alex J. and {Carniani}, Stefano and {Charlot}, Stephane and {Curtis-Lake}, Emma and {D'Eugenio}, Francesco and {Ferruit}, Pierre and {Giardino}, Giovanna and {Hainline}, Kevin and {Hausen}, Ryan and {Jakobsen}, Peter and {Johnson}, Benjamin D. and {Maiolino}, Roberto and {Rauscher}, Bernard J. and {Rieke}, Marcia and {Rieke}, George and {Rix}, Hans-Walter and {Robertson}, Brant and {Stark}, Daniel P. and {Tacchella}, Sandro and {Williams}, Christina C. and {Willmer}, Christopher N.~A. and {Baker}, William M. and {Baum}, Stefi and {Bhatawdekar}, Rachana and {Boyett}, Kristan and {Chen}, Zuyi and {Chevallard}, Jacopo and {Circosta}, Chiara and {Curti}, Mirko and {Danhaive}, A. Lola and {DeCoursey}, Christa and {Endsley}, Ryan and {de Graaff}, Anna and {Dressler}, Alan and {Egami}, Eiichi and {Helton}, Jakob M. and {Hviding}, Raphael E. and {Ji}, Zhiyuan and {Jones}, Gareth C. and {Kumari}, Nimisha and {L{\"u}tzgendorf}, Nora and {Laseter}, Isaac and {Looser}, Tobias J. and {Lyu}, Jianwei and {Maseda}, Michael V. and {Nelson}, Erica and {Parlanti}, Eleonora and {Perna}, Michele and {Pusk{\'a}s}, D{\'a}vid and {Rawle}, Tim and {Rodr{\'\i}guez Del Pino}, Bruno and {Rujopakarn}, Wiphu and {Sandles}, Lester and {Saxena}, Aayush and {Scholtz}, Jan and {Sharpe}, Katherine and {Shivaei}, Irene and {Silcock}, Maddie S. and {Simmonds}, Charlotte and {Skarbinski}, Maya and {Smit}, Renske and {Stone}, Meredith and {Suess}, Katherine A. and {Sun}, Fengwu and {Tang}, Mengtao and {Topping}, Michael W. and {{\"U}bler}, Hannah and {Villanueva}, Natalia C. and {Wallace}, Imaan E.~B. and {Whitler}, Lily and {Witstok}, Joris and {Woodrum}, Charity},
        title = "{Overview of the JWST Advanced Deep Extragalactic Survey (JADES)}",
      journal = {\apjs},
         year = 2026,
        month = mar,
       volume = {283},
       number = {1},
          eid = {6},
        pages = {6},
          doi = {10.3847/1538-4365/ae3163},
archivePrefix = {arXiv},
       eprint = {2306.02465},
 primaryClass = {astro-ph.GA},
       adsurl = {https://ui.adsabs.harvard.edu/abs/2026ApJS..283....6E}
}

@ARTICLE{McHardy06,
       author = {{McHardy}, I.~M. and {Koerding}, E. and {Knigge}, C. and {Uttley}, P. and {Fender}, R.~P.},
        title = "{Active galactic nuclei as scaled-up Galactic black holes}",
      journal = {\nat},
         year = 2006,
        month = dec,
       volume = {444},
       number = {7120},
        pages = {730-732},
          doi = {10.1038/nature05389},
archivePrefix = {arXiv},
       eprint = {astro-ph/0612273},
 primaryClass = {astro-ph},
       adsurl = {https://ui.adsabs.harvard.edu/abs/2006Natur.444..730M}
}

@ARTICLE{Kara25,
       author = {{Kara}, Erin and {Garc{\'\i}a}, Javier},
        title = "{Supermassive Black Holes in X-Rays: From Standard Accretion to Extreme Transients}",
      journal = {\araa},
         year = 2025,
        month = aug,
       volume = {63},
       number = {1},
        pages = {379-430},
          doi = {10.1146/annurev-astro-071221-052844},
archivePrefix = {arXiv},
       eprint = {2503.22791},
 primaryClass = {astro-ph.HE},
       adsurl = {https://ui.adsabs.harvard.edu/abs/2025ARA&A..63..379K}
}

@ARTICLE{deGraaff25b,
       author = {{de Graaff}, Anna and {Hviding}, Raphael E. and {Naidu}, Rohan P. and {Greene}, Jenny E. and {Miller}, Tim B. and {Leja}, Joel and {Matthee}, Jorryt and {Brammer}, Gabriel and {Katz}, Harley and {Bezanson}, Rachel and {Boogaard}, Leindert A. and {Bose}, Sownak and {Chisholm}, John and {Cleri}, Nikko J. and {Dayal}, Pratika and {Feldmann}, Robert and {Fudamoto}, Yoshinobu and {Fujimoto}, Seiji and {Furtak}, Lukas J. and {Glazebrook}, Karl and {Gottumukkala}, Rashmi and {Heintz}, Kasper E. and {Kokorev}, Vasily and {Labbe}, Ivo and {Maseda}, Michael V. and {McConachie}, Ian and {Nanayakkara}, Themiya and {Nelson}, Erica and {Nowaczyk}, Przemys{\l}aw and {Oesch}, Pascal A. and {Rix}, Hans-Walter and {Setton}, David J. and {Torralba}, Alberto and {Walter}, Fabian and {Wang}, Bingjie and {Weibel}, Andrea and {van der Wel}, Arjen},
        title = "{Little Red Dots host Black Hole Stars: A unified family of gas-reddened AGN revealed by JWST/NIRSpec spectroscopy}",
      journal = {arXiv e-prints},
         year = 2025,
        month = nov,
          eid = {arXiv:2511.21820},
        pages = {arXiv:2511.21820},
          doi = {10.48550/arXiv.2511.21820},
archivePrefix = {arXiv},
       eprint = {2511.21820},
 primaryClass = {astro-ph.GA},
       adsurl = {https://ui.adsabs.harvard.edu/abs/2025arXiv251121820D}
}

@ARTICLE{Labbe25,
       author = {{Labbe}, Ivo and {Greene}, Jenny E. and {Bezanson}, Rachel and {Fujimoto}, Seiji and {Furtak}, Lukas J. and {Goulding}, Andy D. and {Matthee}, Jorryt and {Naidu}, Rohan P. and {Oesch}, Pascal A. and {Atek}, Hakim and {Brammer}, Gabriel and {Chemerynska}, Iryna and {Coe}, Dan and {Cutler}, Sam E. and {Dayal}, Pratika and {Feldmann}, Robert and {Franx}, Marijn and {Glazebrook}, Karl and {Leja}, Joel and {Maseda}, Michael and {Marchesini}, Danilo and {Nanayakkara}, Themiya and {Nelson}, Erica J. and {Pan}, Richard and {Papovich}, Casey and {Price}, Sedona H. and {Suess}, Katherine A. and {Wang}, Bingjie and {Weaver}, John R. and {Whitaker}, Katherine E. and {Williams}, Christina C. and {Zitrin}, Adi},
        title = "{UNCOVER: Candidate Red Active Galactic Nuclei at 3 < z < 7 with JWST and ALMA}",
      journal = {\apj},
         year = 2025,
        month = jan,
       volume = {978},
       number = {1},
          eid = {92},
        pages = {92},
          doi = {10.3847/1538-4357/ad3551},
archivePrefix = {arXiv},
       eprint = {2306.07320},
 primaryClass = {astro-ph.GA},
       adsurl = {https://ui.adsabs.harvard.edu/abs/2025ApJ...978...92L}
}

@ARTICLE{CoveloPaz25,
       author = {{Covelo-Paz}, Alba and {Giovinazzo}, Emma and {Oesch}, Pascal A. and {Meyer}, Romain A. and {Weibel}, Andrea and {Brammer}, Gabriel and {Fudamoto}, Yoshinobu and {Kerutt}, Josephine and {Lin}, Jamie and {Matharu}, Jasleen and {Naidu}, Rohan P. and {Velichko}, Anna and {Bollo}, Victoria and {Bouwens}, Rychard and {Chisholm}, John and {Illingworth}, Garth D. and {Kramarenko}, Ivan and {Magee}, Daniel and {Maseda}, Michael and {Matthee}, Jorryt and {Nelson}, Erica and {Reddy}, Naveen and {Schaerer}, Daniel and {Stefanon}, Mauro and {Xiao}, Mengyuan},
        title = "{An H{\ensuremath{\alpha}} view of galaxy buildup in the first 2 Gyr: Luminosity functions at z {\ensuremath{\sim}} 4{\ensuremath{-}}6.5 from NIRCam/grism spectroscopy}",
      journal = {\aap},
         year = 2025,
        month = feb,
       volume = {694},
          eid = {A178},
        pages = {A178},
          doi = {10.1051/0004-6361/202452363},
archivePrefix = {arXiv},
       eprint = {2409.17241},
 primaryClass = {astro-ph.GA},
       adsurl = {https://ui.adsabs.harvard.edu/abs/2025A&A...694A.178C}
}

@ARTICLE{Kokubo25,
       author = {{Kokubo}, Mitsuru and {Harikane}, Yuichi},
        title = "{Challenging the Active Galactic Nucleus Scenario for JWST/NIRSpec Little Red Dot and Non─Little Red Dot Broad H{\ensuremath{\alpha}} Emitters in Light of Nondetection of NIRCam Photometric Variability and X-Ray}",
      journal = {\apj},
         year = 2025,
        month = dec,
       volume = {995},
       number = {1},
          eid = {24},
        pages = {24},
          doi = {10.3847/1538-4357/ae119e},
archivePrefix = {arXiv},
       eprint = {2407.04777},
 primaryClass = {astro-ph.GA},
       adsurl = {https://ui.adsabs.harvard.edu/abs/2025ApJ...995...24K}
}

@ARTICLE{Witten25,
       author = {{Witten}, Callum and {McClymont}, William and {Laporte}, Nicolas and {Roberts-Borsani}, Guido and {Sijacki}, Debora and {Tacchella}, Sandro and {Simmonds}, Charlotte and {Katz}, Harley and {Ellis}, Richard S. and {Witstok}, Joris and {Maiolino}, Roberto and {Ji}, Xihan and {Hayes}, Billy R. and {Looser}, Tobias J. and {D'Eugenio}, Francesco},
        title = "{Rising from the ashes: evidence of old stellar populations and rejuvenation events in the very early Universe}",
      journal = {\mnras},
         year = 2025,
        month = feb,
       volume = {537},
       number = {1},
        pages = {112-126},
          doi = {10.1093/mnras/staf001},
archivePrefix = {arXiv},
       eprint = {2407.07937},
 primaryClass = {astro-ph.GA},
       adsurl = {https://ui.adsabs.harvard.edu/abs/2025MNRAS.537..112W}
}

@ARTICLE{Morishita25,
       author = {{Morishita}, Takahiro and {Liu}, Zhaoran and {Stiavelli}, Massimo and {Treu}, Tommaso and {Trenti}, Michele and {Chartab}, Nima and {Roberts-Borsani}, Guido and {Vulcani}, Benedetta and {Bergamini}, Pietro and {Castellano}, Marco and {Grillo}, Claudio},
        title = "{Accelerated Emergence of Evolved Galaxies in Early Overdensities at z {\ensuremath{\sim}} 5.7}",
      journal = {\apj},
         year = 2025,
        month = apr,
       volume = {982},
       number = {2},
          eid = {153},
        pages = {153},
          doi = {10.3847/1538-4357/adb30f},
archivePrefix = {arXiv},
       eprint = {2408.10980},
 primaryClass = {astro-ph.GA},
       adsurl = {https://ui.adsabs.harvard.edu/abs/2025ApJ...982..153M}
}

@ARTICLE{Wang25,
       author = {{Wang}, Bingjie and {de Graaff}, Anna and {Davies}, Rebecca L. and {Greene}, Jenny E. and {Leja}, Joel and {Brammer}, Gabriel B. and {Goulding}, Andy D. and {Miller}, Tim B. and {Suess}, Katherine A. and {Weibel}, Andrea and {Williams}, Christina C. and {Bezanson}, Rachel and {Boogaard}, Leindert A. and {Cleri}, Nikko J. and {Hirschmann}, Michaela and {Katz}, Harley and {Labb{\'e}}, Ivo and {Maseda}, Michael V. and {Matthee}, Jorryt and {McConachie}, Ian and {Naidu}, Rohan P. and {Oesch}, Pascal A. and {Rix}, Hans-Walter and {Setton}, David J. and {Whitaker}, Katherine E.},
        title = "{RUBIES: JWST/NIRSpec Confirmation of an Infrared-luminous, Broad-line Little Red Dot with an Ionized Outflow}",
      journal = {\apj},
         year = 2025,
        month = may,
       volume = {984},
       number = {2},
          eid = {121},
        pages = {121},
          doi = {10.3847/1538-4357/adc1ca},
archivePrefix = {arXiv},
       eprint = {2403.02304},
 primaryClass = {astro-ph.GA},
       adsurl = {https://ui.adsabs.harvard.edu/abs/2025ApJ...984..121W}
}

@ARTICLE{Furtak24,
       author = {{Furtak}, Lukas J. and {Labb{\'e}}, Ivo and {Zitrin}, Adi and {Greene}, Jenny E. and {Dayal}, Pratika and {Chemerynska}, Iryna and {Kokorev}, Vasily and {Miller}, Tim B. and {Goulding}, Andy D. and {de Graaff}, Anna and {Bezanson}, Rachel and {Brammer}, Gabriel B. and {Cutler}, Sam E. and {Leja}, Joel and {Pan}, Richard and {Price}, Sedona H. and {Wang}, Bingjie and {Weaver}, John R. and {Whitaker}, Katherine E. and {Atek}, Hakim and {Bogd{\'a}n}, {\'A}kos and {Charlot}, St{\'e}phane and {Curtis-Lake}, Emma and {van Dokkum}, Pieter and {Endsley}, Ryan and {Feldmann}, Robert and {Fudamoto}, Yoshinobu and {Fujimoto}, Seiji and {Glazebrook}, Karl and {Juneau}, St{\'e}phanie and {Marchesini}, Danilo and {Maseda}, Micheal V. and {Nelson}, Erica and {Oesch}, Pascal A. and {Plat}, Ad{\`e}le and {Setton}, David J. and {Stark}, Daniel P. and {Williams}, Christina C.},
        title = "{A high black-hole-to-host mass ratio in a lensed AGN in the early Universe}",
      journal = {\nat},
         year = 2024,
        month = apr,
       volume = {628},
       number = {8006},
        pages = {57-61},
          doi = {10.1038/s41586-024-07184-8},
archivePrefix = {arXiv},
       eprint = {2308.05735},
 primaryClass = {astro-ph.GA},
       adsurl = {https://ui.adsabs.harvard.edu/abs/2024Natur.628...57F}
}

@ARTICLE{Wang24a,
       author = {{Wang}, Bingjie and {Leja}, Joel and {de Graaff}, Anna and {Brammer}, Gabriel B. and {Weibel}, Andrea and {van Dokkum}, Pieter and {Baggen}, Josephine F.~W. and {Suess}, Katherine A. and {Greene}, Jenny E. and {Bezanson}, Rachel and {Cleri}, Nikko J. and {Hirschmann}, Michaela and {Labb{\'e}}, Ivo and {Matthee}, Jorryt and {McConachie}, Ian and {Naidu}, Rohan P. and {Nelson}, Erica and {Oesch}, Pascal A. and {Setton}, David J. and {Williams}, Christina C.},
        title = "{RUBIES: Evolved Stellar Populations with Extended Formation Histories at z {\ensuremath{\sim}} 7{\textendash}8 in Candidate Massive Galaxies Identified with JWST/NIRSpec}",
      journal = {\apjl},
         year = 2024,
        month = jul,
       volume = {969},
       number = {1},
          eid = {L13},
        pages = {L13},
          doi = {10.3847/2041-8213/ad55f7},
archivePrefix = {arXiv},
       eprint = {2405.01473},
 primaryClass = {astro-ph.GA},
       adsurl = {https://ui.adsabs.harvard.edu/abs/2024ApJ...969L..13W}
}

@ARTICLE{deGraaff25a,
       author = {{de Graaff}, Anna and {Setton}, David J. and {Brammer}, Gabriel and {Cutler}, Sam and {Suess}, Katherine A. and {Labb{\'e}}, Ivo and {Leja}, Joel and {Weibel}, Andrea and {Maseda}, Michael V. and {Whitaker}, Katherine E. and {Bezanson}, Rachel and {Boogaard}, Leindert A. and {Cleri}, Nikko J. and {De Lucia}, Gabriella and {Franx}, Marijn and {Greene}, Jenny E. and {Hirschmann}, Michaela and {Matthee}, Jorryt and {McConachie}, Ian and {Naidu}, Rohan P. and {Oesch}, Pascal A. and {Price}, Sedona H. and {Rix}, Hans-Walter and {Valentino}, Francesco and {Wang}, Bingjie and {Williams}, Christina C.},
        title = "{Efficient formation of a massive quiescent galaxy at redshift 4.9}",
      journal = {Nature Astronomy},
         year = 2025,
        month = feb,
       volume = {9},
        pages = {280-292},
          doi = {10.1038/s41550-024-02424-3},
archivePrefix = {arXiv},
       eprint = {2404.05683},
 primaryClass = {astro-ph.GA},
       adsurl = {https://ui.adsabs.harvard.edu/abs/2025NatAs...9..280D}
}

@ARTICLE{Burke25,
       author = {{Burke}, Colin J. and {Stone}, Zachary and {Shen}, Yue and {Jiang}, Yan-Fei},
        title = "{Too Quiet for Comfort: Local Little Red Dots Lack Variability over Decades}",
      journal = {arXiv e-prints},
         year = 2025,
        month = nov,
          eid = {arXiv:2511.16082},
        pages = {arXiv:2511.16082},
          doi = {10.48550/arXiv.2511.16082},
archivePrefix = {arXiv},
       eprint = {2511.16082},
 primaryClass = {astro-ph.GA},
       adsurl = {https://ui.adsabs.harvard.edu/abs/2025arXiv251116082B}
}

@ARTICLE{Burke21,
       author = {{Burke}, Colin J. and {Shen}, Yue and {Blaes}, Omer and {Gammie}, Charles F. and {Horne}, Keith and {Jiang}, Yan-Fei and {Liu}, Xin and {McHardy}, Ian M. and {Morgan}, Christopher W. and {Scaringi}, Simone and {Yang}, Qian},
        title = "{A characteristic optical variability time scale in astrophysical accretion disks}",
      journal = {Science},
         year = 2021,
        month = aug,
       volume = {373},
       number = {6556},
        pages = {789-792},
          doi = {10.1126/science.abg9933},
archivePrefix = {arXiv},
       eprint = {2108.05389},
 primaryClass = {astro-ph.GA},
       adsurl = {https://ui.adsabs.harvard.edu/abs/2021Sci...373..789B}
}

@ARTICLE{King24,
       author = {{King}, Andrew},
        title = "{The black hole masses of high-redshift QSOs}",
      journal = {\mnras},
         year = 2024,
        month = jun,
       volume = {531},
       number = {1},
        pages = {550-553},
          doi = {10.1093/mnras/stae1171},
archivePrefix = {arXiv},
       eprint = {2404.16832},
 primaryClass = {astro-ph.GA},
       adsurl = {https://ui.adsabs.harvard.edu/abs/2024MNRAS.531..550K}
}

@ARTICLE{Lupi24,
       author = {{Lupi}, Alessandro and {Trinca}, Alessandro and {Volonteri}, Marta and {Dotti}, Massimo and {Mazzucchelli}, Chiara},
        title = "{Size matters: are we witnessing super-Eddington accretion in high-redshift black holes from JWST?}",
      journal = {\aap},
         year = 2024,
        month = sep,
       volume = {689},
          eid = {A128},
        pages = {A128},
          doi = {10.1051/0004-6361/202451249},
archivePrefix = {arXiv},
       eprint = {2406.17847},
 primaryClass = {astro-ph.HE},
       adsurl = {https://ui.adsabs.harvard.edu/abs/2024A&A...689A.128L}
}

@ARTICLE{Shen19,
       author = {{Shen}, Yue and {Hall}, Patrick B. and {Horne}, Keith and {Zhu}, Guangtun and {McGreer}, Ian and {Simm}, Torben and {Trump}, Jonathan R. and {Kinemuchi}, Karen and {Brandt}, W.~N. and {Green}, Paul J. and {Grier}, C.~J. and {Guo}, Hengxiao and {Ho}, Luis C. and {Homayouni}, Yasaman and {Jiang}, Linhua and {I-Hsiu Li}, Jennifer and {Morganson}, Eric and {Petitjean}, Patrick and {Richards}, Gordon T. and {Schneider}, Donald P. and {Starkey}, D.~A. and {Wang}, Shu and {Chambers}, Ken and {Kaiser}, Nick and {Kudritzki}, Rolf-Peter and {Magnier}, Eugene and {Waters}, Christopher},
        title = "{The Sloan Digital Sky Survey Reverberation Mapping Project: Sample Characterization}",
      journal = {\apjs},
         year = 2019,
        month = apr,
       volume = {241},
       number = {2},
          eid = {34},
        pages = {34},
          doi = {10.3847/1538-4365/ab074f},
archivePrefix = {arXiv},
       eprint = {1810.01447},
 primaryClass = {astro-ph.GA},
       adsurl = {https://ui.adsabs.harvard.edu/abs/2019ApJS..241...34S}
}

@ARTICLE{Shen16,
       author = {{Shen}, Yue and {Horne}, Keith and {Grier}, C.~J. and {Peterson}, Bradley M. and {Denney}, Kelly D. and {Trump}, Jonathan R. and {Sun}, Mouyuan and {Brandt}, W.~N. and {Kochanek}, Christopher S. and {Dawson}, Kyle S. and {Green}, Paul J. and {Greene}, Jenny E. and {Hall}, Patrick B. and {Ho}, Luis C. and {Jiang}, Linhua and {Kinemuchi}, Karen and {McGreer}, Ian D. and {Petitjean}, Patrick and {Richards}, Gordon T. and {Schneider}, Donald P. and {Strauss}, Michael A. and {Tao}, Charling and {Wood-Vasey}, W.~M. and {Zu}, Ying and {Pan}, Kaike and {Bizyaev}, Dmitry and {Ge}, Jian and {Oravetz}, Daniel and {Simmons}, Audrey},
        title = "{The Sloan Digital Sky Survey Reverberation Mapping Project: First Broad-line H{\ensuremath{\beta}} and Mg II Lags at z {\ensuremath{\gtrsim}}  0.3 from Six-month Spectroscopy}",
      journal = {\apj},
         year = 2016,
        month = feb,
       volume = {818},
       number = {1},
          eid = {30},
        pages = {30},
          doi = {10.3847/0004-637X/818/1/30},
archivePrefix = {arXiv},
       eprint = {1510.02802},
 primaryClass = {astro-ph.GA},
       adsurl = {https://ui.adsabs.harvard.edu/abs/2016ApJ...818...30S}
}

@ARTICLE{deGraaff25,
       author = {{de Graaff}, Anna and {Rix}, Hans-Walter and {Naidu}, Rohan P. and {Labb{\'e}}, Ivo and {Wang}, Bingjie and {Leja}, Joel and {Matthee}, Jorryt and {Katz}, Harley and {Greene}, Jenny E. and {Hviding}, Raphael E. and {Baggen}, Josephine and {Bezanson}, Rachel and {Boogaard}, Leindert A. and {Brammer}, Gabriel and {Dayal}, Pratika and {van Dokkum}, Pieter and {Goulding}, Andy D. and {Hirschmann}, Michaela and {Maseda}, Michael V. and {McConachie}, Ian and {Miller}, Tim B. and {Nelson}, Erica and {Oesch}, Pascal A. and {Setton}, David J. and {Shivaei}, Irene and {Weibel}, Andrea and {Whitaker}, Katherine E. and {Williams}, Christina C.},
        title = "{A remarkable ruby: Absorption in dense gas, rather than evolved stars, drives the extreme Balmer break of a little red dot at z = 3.5}",
      journal = {\aap},
         year = 2025,
        month = sep,
       volume = {701},
          eid = {A168},
        pages = {A168},
          doi = {10.1051/0004-6361/202554681},
archivePrefix = {arXiv},
       eprint = {2503.16600},
 primaryClass = {astro-ph.GA},
       adsurl = {https://ui.adsabs.harvard.edu/abs/2025A&A...701A.168D}
}

@ARTICLE{Torralba2026,
  author = {Torralba, Alberto and Matthee, Jorryt and Weibel, Andrea and Naidu, Rohan P. and Ma, Yilun and Cloonan, Aidan P. and Desai, Aayush and de Graaff, Anna and Greene, Jenny E. and Jespersen, Christian Kragh and Kramarenko, Ivan G. and Mascia, Sara and Oesch, Pascal A. and Sun, Wendy Q. and Williams, Christina C.},
  title = {A Black Hole Star at Cosmic Noon: Extreme Balmer break, photospheric continuum, and broad absorption by thick winds in a Little Red Dot at z=1.7},
  journal = {arXiv e-prints},
  year = {2026},
  month = mar,
  eid = {arXiv:2603.28335},
  pages = {arXiv:2603.28335},
  archivePrefix = {arXiv},
  eprint = {2603.28335},
  primaryClass = {astro-ph.GA},
  doi = {10.48550/arXiv.2603.28335}
}

@ARTICLE{Torralba25,
       author = {{Torralba}, Alberto and {Matthee}, Jorryt and {Pezzulli}, Gabriele and {Naidu}, Rohan P. and {Ishikawa}, Yuzo and {Brammer}, Gabriel B. and {Chang}, Seok-Jun and {Chisholm}, John and {de Graaff}, Anna and {D'Eugenio}, Francesco and {Di Cesare}, Claudia and {Eilers}, Anna-Christina and {Greene}, Jenny E. and {Gronke}, Max and {Iani}, Edoardo and {Kokorev}, Vasily and {Kotiwale}, Gauri and {Kramarenko}, Ivan and {Ma}, Yilun and {Mascia}, Sara and {Navarrete}, Benjam{\'\i}n and {Nelson}, Erica and {Oesch}, Pascal and {Simcoe}, Robert A. and {Wuyts}, Stijn},
        title = "{The warm outer layer of a Little Red Dot as the source of [Fe II] and collisional Balmer lines with scattering wings}",
      journal = {arXiv e-prints},
         year = 2025,
        month = sep,
          eid = {arXiv:2510.00103},
        pages = {arXiv:2510.00103},
          doi = {10.48550/arXiv.2510.00103},
archivePrefix = {arXiv},
       eprint = {2510.00103},
 primaryClass = {astro-ph.GA},
       adsurl = {https://ui.adsabs.harvard.edu/abs/2025arXiv251000103T}
}

@ARTICLE{Inayoshi25,
       author = {{Inayoshi}, Kohei and {Maiolino}, Roberto},
        title = "{Extremely Dense Gas around Little Red Dots and High-redshift Active Galactic Nuclei: A Nonstellar Origin of the Balmer Break and Absorption Features}",
      journal = {\apjl},
         year = 2025,
        month = feb,
       volume = {980},
       number = {2},
          eid = {L27},
        pages = {L27},
          doi = {10.3847/2041-8213/adaebd},
archivePrefix = {arXiv},
       eprint = {2409.07805},
 primaryClass = {astro-ph.GA},
       adsurl = {https://ui.adsabs.harvard.edu/abs/2025ApJ...980L..27I}
}

@ARTICLE{bertin1996,
       author = {{Bertin}, E. and {Arnouts}, S.},
        title = "{SExtractor: Software for source extraction.}",
      journal = {\aaps},
         year = 1996,
        month = jun,
       volume = {117},
        pages = {393-404},
          doi = {10.1051/aas:1996164},
       adsurl = {https://ui.adsabs.harvard.edu/abs/1996A&AS..117..393B}
}

@ARTICLE{Kashino23_grism,
       author = {{Kashino}, Daichi and {Lilly}, Simon J. and {Matthee}, Jorryt and {Eilers}, Anna-Christina and {Mackenzie}, Ruari and {Bordoloi}, Rongmon and {Simcoe}, Robert A.},
        title = "{EIGER. I. A Large Sample of [O III]-emitting Galaxies at 5.3 < z < 6.9 and Direct Evidence for Local Reionization by Galaxies}",
      journal = {\apj},
         year = 2023,
        month = jun,
       volume = {950},
       number = {1},
          eid = {66},
        pages = {66},
          doi = {10.3847/1538-4357/acc588},
archivePrefix = {arXiv},
       eprint = {2211.08254},
 primaryClass = {astro-ph.GA},
       adsurl = {https://ui.adsabs.harvard.edu/abs/2023ApJ...950...66K}
}

@ARTICLE{Sun23_grism,
       author = {{Sun}, Fengwu and {Egami}, Eiichi and {Pirzkal}, Nor and {Rieke}, Marcia and {Baum}, Stefi and {Boyer}, Martha and {Boyett}, Kristan and {Bunker}, Andrew J. and {Cameron}, Alex J. and {Curti}, Mirko and {Eisenstein}, Daniel J. and {Gennaro}, Mario and {Greene}, Thomas P. and {Jaffe}, Daniel and {Kelly}, Doug and {Koekemoer}, Anton M. and {Kumari}, Nimisha and {Maiolino}, Roberto and {Maseda}, Michael and {Perna}, Michele and {Rest}, Armin and {Robertson}, Brant E. and {Schlawin}, Everett and {Smit}, Renske and {Stansberry}, John and {Sunnquist}, Ben and {Tacchella}, Sandro and {Williams}, Christina C. and {Willmer}, Christopher N.~A.},
        title = "{First Sample of H{\ensuremath{\alpha}}+[O III]{\ensuremath{\lambda}}5007 Line Emitters at z > 6 Through JWST/NIRCam Slitless Spectroscopy: Physical Properties and Line-luminosity Functions}",
      journal = {\apj},
         year = 2023,
        month = aug,
       volume = {953},
       number = {1},
          eid = {53},
        pages = {53},
          doi = {10.3847/1538-4357/acd53c},
archivePrefix = {arXiv},
       eprint = {2209.03374},
 primaryClass = {astro-ph.GA},
       adsurl = {https://ui.adsabs.harvard.edu/abs/2023ApJ...953...53S}
}

@ARTICLE{Liu24,
       author = {{Liu}, Zhaoran and {Morishita}, Takahiro and {Kodama}, Tadayuki},
        title = "{Characterizing Dust Extinction and Spatially Resolved Paschen-{\ensuremath{\alpha}} Emission within 97 Galaxies at 1 < z < 1.6 with JWST NIRCam Slitless Spectroscopy}",
      journal = {\apj},
         year = 2026,
        month = feb,
       volume = {998},
       number = {2},
          eid = {203},
        pages = {203},
          doi = {10.3847/1538-4357/ae3184},
archivePrefix = {arXiv},
       eprint = {2406.11188},
 primaryClass = {astro-ph.GA},
       adsurl = {https://ui.adsabs.harvard.edu/abs/2026ApJ...998..203L}
}

@software{Bushouse23,
       author = {{Bushouse}, Howard and {Eisenhamer}, Jonathan and {Dencheva}, Nadia and {Davies}, James and {Greenfield}, Perry and {Morrison}, Jane and {Hodge}, Phil and {Simon}, Bernie and {Grumm}, David and {Droettboom}, Michael and {Slavich}, Edward and {Sosey}, Megan and {Pauly}, Tyler and {Miller}, Todd and {Jedrzejewski}, Robert and {Hack}, Warren and {Davis}, David and {Crawford}, Steven and {Law}, David and {Gordon}, Karl and {Regan}, Michael and {Cara}, Mihai and {MacDonald}, Ken and {Bradley}, Larry and {Shanahan}, Clare and {Jamieson}, William and {Teodoro}, Mairan and {Williams}, Thomas and {Pena-Guerrero}, Maria and {Graham}, Brett and {Molter}, Edward and {Brandt}, Timothy and {Hayes}, Christian and {Cooper}, Rachel and {Clarke}, Melanie and {Filippazzo}, Joseph},
        title = "{JWST Calibration Pipeline}",
         year = 2025,
        month = jul,
          eid = {10.5281/zenodo.6984365},
          doi = {10.5281/zenodo.6984365},
      version = {1.19.1},
    publisher = {Zenodo},
       adsurl = {https://ui.adsabs.harvard.edu/abs/2023zndo...6984365B}
}

@ARTICLE{Sun25,
       author = {{Sun}, Fengwu and {Fudamoto}, Yoshinobu and {Lin}, Xiaojing and {Helton}, Jakob M. and {Hsiao}, Tiger Yu-Yang and {Egami}, Eiichi and {Akhtarkavan}, Arshia and {Bunker}, Andrew J. and {Cai}, Zheng and {DeCoursey}, Christa and {Eisenstein}, Daniel J. and {Fan}, Xiaohui and {Harikane}, Yuichi and {Ji}, Zhiyuan and {Jin}, Xiangyu and {Liu}, Weizhe and {Liu}, Yichen and {Ma}, Zheng and {Maiolino}, Roberto and {Ouchi}, Masami and {Tee}, Wei Leong and {Wang}, Feige and {Willmer}, Christopher N.~A. and {Wu}, Yunjing and {Xu}, Yi and {Yang}, Jinyi and {Zhang}, Junyu and {Zhu}, Yongda},
        title = "{Slitless Areal Pure-Parallel HIgh-Redshift Emission Survey (SAPPHIRES): Early Data Release of Deep JWST/NIRCam Images and Spectra in MACS J0416 Parallel Field}",
      journal = {arXiv e-prints},
         year = 2025,
        month = mar,
          eid = {arXiv:2503.15587},
        pages = {arXiv:2503.15587},
          doi = {10.48550/arXiv.2503.15587},
archivePrefix = {arXiv},
       eprint = {2503.15587},
 primaryClass = {astro-ph.GA},
       adsurl = {https://ui.adsabs.harvard.edu/abs/2025arXiv250315587S}
}

@ARTICLE{Secunda26,
       author = {{Secunda}, Amy and {Somerville}, Rachel S. and {Jiang}, Yan-Fei and {Greene}, Jenny E. and {Furtak}, Lukas J. and {Zitrin}, Adi},
        title = "{Do Little Red Dots Vary?}",
      journal = {\apj},
         year = 2026,
        month = jan,
       volume = {996},
       number = {1},
          eid = {6},
        pages = {6},
          doi = {10.3847/1538-4357/ae1f08},
archivePrefix = {arXiv},
       eprint = {2509.03571},
 primaryClass = {astro-ph.GA},
       adsurl = {https://ui.adsabs.harvard.edu/abs/2026ApJ...996....6S}
}

@ARTICLE{Wang26,
       author = {{Wang}, Bingjie and {Leja}, Joel and {Labbe}, Ivo and {Greene}, Jenny E. and {Liu}, Hanpu and {de Graaff}, Anna and {Hviding}, Raphael E. and {Matthee}, Jorryt and {Quataert}, Eliot and {Bezanson}, Rachel and {Boogaard}, Leindert A. and {Brammer}, Gabriel and {Burgasser}, Adam J. and {Chen}, Yi-Xian and {Cleri}, Nikko J. and {Cutler}, Sam E. and {Dayal}, Pratika and {Furtak}, Lukas J. and {Fujimoto}, Seiji and {Glazebrook}, Karl and {Goulding}, Andy D. and {Helton}, Jakob M. and {Hirschmann}, Michaela and {Jiang}, Yan-Fei and {Kokorev}, Vasily and {Ma}, Yilun and {Miller}, Tim B. and {Naidu}, Rohan P. and {Oesch}, Pascal and {Pan}, Richard and {Papovich}, Casey and {Price}, Sedona H. and {Rix}, Hans-Walter and {Setton}, David J. and {Sun}, Wendy Q. and {Weaver}, John R. and {Whitaker}, Katherine E. and {Zitrin}, Adi},
        title = "{Water absorption confirms cool atmospheres in two little red dots}",
      journal = {arXiv e-prints},
         year = 2026,
        month = feb,
          eid = {arXiv:2602.06024},
        pages = {arXiv:2602.06024},
          doi = {10.48550/arXiv.2602.06024},
archivePrefix = {arXiv},
       eprint = {2602.06024},
 primaryClass = {astro-ph.GA},
       adsurl = {https://ui.adsabs.harvard.edu/abs/2026arXiv260206024W}
}

@ARTICLE{Furtak25,
       author = {{Furtak}, Lukas J. and {Secunda}, Amy R. and {Greene}, Jenny E. and {Zitrin}, Adi and {Labb{\'e}}, Ivo and {Golubchik}, Miriam and {Bezanson}, Rachel and {Kokorev}, Vasily and {Atek}, Hakim and {Brammer}, Gabriel B. and {Chemerynska}, Iryna and {Cutler}, Sam E. and {Dayal}, Pratika and {Feldmann}, Robert and {Fujimoto}, Seiji and {Glazebrook}, Karl and {Leja}, Joel and {Ma}, Yilun and {Matthee}, Jorryt and {Naidu}, Rohan P. and {Nelson}, Erica J. and {Oesch}, Pascal A. and {Pan}, Richard and {Price}, Sedona H. and {Suess}, Katherine A. and {Wang}, Bingjie and {Weaver}, John R. and {Whitaker}, Katherine E.},
        title = "{Investigating photometric and spectroscopic variability in the multiply imaged little red dot A2744-QSO1}",
      journal = {\aap},
         year = 2025,
        month = jun,
       volume = {698},
          eid = {A227},
        pages = {A227},
          doi = {10.1051/0004-6361/202554110},
archivePrefix = {arXiv},
       eprint = {2502.07875},
 primaryClass = {astro-ph.GA},
       adsurl = {https://ui.adsabs.harvard.edu/abs/2025A&A...698A.227F}
}

@ARTICLE{Ji25,
       author = {{Ji}, Xihan and {Maiolino}, Roberto and {{\"U}bler}, Hannah and {Scholtz}, Jan and {D'Eugenio}, Francesco and {Sun}, Fengwu and {Perna}, Michele and {Turner}, Hannah and {Carniani}, Stefano and {Arribas}, Santiago and {Bennett}, Jake S. and {Bunker}, Andrew and {Charlot}, St{\'e}phane and {Cresci}, Giovanni and {Curti}, Mirko and {Egami}, Eiichi and {Fabian}, Andy and {Inayoshi}, Kohei and {Isobe}, Yuki and {Jones}, Gareth and {Juod{\v{z}}balis}, Ignas and {Kumari}, Nimisha and {Lyu}, Jianwei and {Mazzolari}, Giovanni and {Parlanti}, Eleonora and {Robertson}, Brant and {Rodr{\'\i}guez Del Pino}, Bruno and {Schneider}, Raffaella and {Sijacki}, Debora and {Tacchella}, Sandro and {Trinca}, Alessandro and {Valiante}, Rosa and {Venturi}, Giacomo and {Volonteri}, Marta and {Willott}, Chris and {Witten}, Callum and {Witstok}, Joris},
        title = "{BlackTHUNDER ─ A non-stellar Balmer break in a black hole-dominated little red dot at z = 7.04}",
      journal = {\mnras},
         year = 2025,
        month = dec,
       volume = {544},
       number = {4},
        pages = {3900-3935},
          doi = {10.1093/mnras/staf1867},
archivePrefix = {arXiv},
       eprint = {2501.13082},
 primaryClass = {astro-ph.GA},
       adsurl = {https://ui.adsabs.harvard.edu/abs/2025MNRAS.544.3900J}
}

@ARTICLE{Rusakov26,
       author = {{Rusakov}, V. and {Watson}, D. and {Nikopoulos}, G.~P. and {Brammer}, G. and {Gottumukkala}, R. and {Harvey}, T. and {Heintz}, K.~E. and {Damgaard}, R. and {Sim}, S.~A. and {Sneppen}, A. and {Vijayan}, A.~P. and {Adams}, N. and {Austin}, D. and {Conselice}, C.~J. and {Goolsby}, C.~M. and {Toft}, S. and {Witstok}, J.},
        title = "{Little red dots as young supermassive black holes in dense ionized cocoons}",
      journal = {\nat},
         year = 2026,
        month = jan,
       volume = {649},
       number = {8097},
        pages = {574-579},
          doi = {10.1038/s41586-025-09900-4},
archivePrefix = {arXiv},
       eprint = {2503.16595},
 primaryClass = {astro-ph.GA},
       adsurl = {https://ui.adsabs.harvard.edu/abs/2026Natur.649..574R}
}

@ARTICLE{Zhang25a,
       author = {{Zhang}, Zijian and {Jiang}, Linhua and {Liu}, Weiyang and {Ho}, Luis C.},
        title = "{Analysis of Multi-epoch JWST Images of {\ensuremath{\sim}}300 Little Red Dots: Tentative Detection of Variability in a Minority of Sources}",
      journal = {\apj},
         year = 2025,
        month = may,
       volume = {985},
       number = {1},
          eid = {119},
        pages = {119},
          doi = {10.3847/1538-4357/adcb3e},
archivePrefix = {arXiv},
       eprint = {2411.02729},
 primaryClass = {astro-ph.GA},
       adsurl = {https://ui.adsabs.harvard.edu/abs/2025ApJ...985..119Z}
}

@ARTICLE{Begelman26,
       author = {{Begelman}, Mitchell C. and {Dexter}, Jason},
        title = "{Little Red Dots as Late-stage Quasi-stars}",
      journal = {\apj},
         year = 2026,
        month = jan,
       volume = {996},
       number = {1},
          eid = {48},
        pages = {48},
          doi = {10.3847/1538-4357/ae274a},
archivePrefix = {arXiv},
       eprint = {2507.09085},
 primaryClass = {astro-ph.GA},
       adsurl = {https://ui.adsabs.harvard.edu/abs/2026ApJ...996...48B}
}

@ARTICLE{Chang26,
       author = {{Chang}, Seok-Jun and {Gronke}, Max and {Matthee}, Jorryt and {Mason}, Charlotte},
        title = "{Impact of resonance, Raman, and Thomson scattering on hydrogen line formation in Little Red Dots}",
      journal = {\mnras},
         year = 2026,
        month = feb,
       volume = {545},
       number = {4},
          eid = {staf2131},
        pages = {staf2131},
          doi = {10.1093/mnras/staf2131},
archivePrefix = {arXiv},
       eprint = {2508.08768},
 primaryClass = {astro-ph.GA},
       adsurl = {https://ui.adsabs.harvard.edu/abs/2026MNRAS.545f2131C}
}

@ARTICLE{Cantiello25,
       author = {{Cantiello}, Matteo and {Hassan}, Jake B. and {Perna}, Rosalba and {Armitage}, Philip J. and {Begelman}, Mitchell C. and {Jiang}, Yan-Fei and {Ryu}, Taeho and {Townsend}, Richard H.~D.},
        title = "{Pulsational Instability of Quasi-Stars: Interpreting the Variability of Little Red Dots}",
      journal = {arXiv e-prints},
         year = 2025,
        month = dec,
          eid = {arXiv:2512.17997},
        pages = {arXiv:2512.17997},
          doi = {10.48550/arXiv.2512.17997},
archivePrefix = {arXiv},
       eprint = {2512.17997},
 primaryClass = {astro-ph.HE},
       adsurl = {https://ui.adsabs.harvard.edu/abs/2025arXiv251217997C}
}

@ARTICLE{Santarelli26,
       author = {{Santarelli}, Andrew D. and {Farag}, Ebraheem and {Bellinger}, Earl P. and {Natarajan}, Priyamvada and {Naidu}, Rohan P. and {Campbell}, Claire B. and {Caplan}, Matthew E.},
        title = "{Evolutionary Tracks and Spectral Properties of Quasi-stars and Their Correlation with Little Red Dots}",
      journal = {\apjl},
         year = 2026,
        month = feb,
       volume = {998},
       number = {1},
          eid = {L4},
        pages = {L4},
          doi = {10.3847/2041-8213/ae3713},
archivePrefix = {arXiv},
       eprint = {2510.17952},
 primaryClass = {astro-ph.GA},
       adsurl = {https://ui.adsabs.harvard.edu/abs/2026ApJ...998L...4S}
}

@ARTICLE{Sun26,
       author = {{Sun}, Wendy Q. and {Naidu}, Rohan P. and {Matthee}, Jorryt and {de Graaff}, Anna and {Chisholm}, John and {Greene}, Jenny E. and {Oesch}, Pascal A. and {Torralba}, Alberto and {Hviding}, Raphael E. and {Brammer}, Gabriel and {Simcoe}, Robert A. and {Bose}, Sownak and {Bouwens}, Rychard and {Dayal}, Pratika and {Eilers}, Anna-Christina and {Fei}, Qinyue and {Furtak}, Lukas J. and {Gottumukkala}, Rashmi and {Goulding}, Andy and {Heintz}, Kasper E. and {Hirschmann}, Michaela and {Kokorev}, Vasily and {Leja}, Joel and {Liu}, Zhaoran and {Natarajan}, Priyamvada and {Santarelli}, Andrew D. and {Setton}, David J. and {Smith}, Aaron and {Tacchella}, Sandro and {Volonteri}, Marta and {Walter}, Fabian and {Weibel}, Andrea and {Williams}, Christina C.},
        title = "{Little Red Dot $-$ Host Galaxy $=$ Black Hole Star: A Gas-Enshrouded Heart at the Center of Every Little Red Dot}",
      journal = {arXiv e-prints},
         year = 2026,
        month = jan,
          eid = {arXiv:2601.20929},
        pages = {arXiv:2601.20929},
          doi = {10.48550/arXiv.2601.20929},
archivePrefix = {arXiv},
       eprint = {2601.20929},
 primaryClass = {astro-ph.GA},
       adsurl = {https://ui.adsabs.harvard.edu/abs/2026arXiv260120929S}
}

@ARTICLE{Astropy18,
       author = {{Astropy Collaboration} and {Price-Whelan}, A.~M. and {Sip{\H{o}}cz}, B.~M. and {G{\"u}nther}, H.~M. and {Lim}, P.~L. and {Crawford}, S.~M. and {Conseil}, S. and {Shupe}, D.~L. and {Craig}, M.~W. and {Dencheva}, N. and {Ginsburg}, A. and {VanderPlas}, J.~T. and {Bradley}, L.~D. and {P{\'e}rez-Su{\'a}rez}, D. and {de Val-Borro}, M. and {Aldcroft}, T.~L. and {Cruz}, K.~L. and {Robitaille}, T.~P. and {Tollerud}, E.~J. and {Ardelean}, C. and {Babej}, T. and {Bach}, Y.~P. and {Bachetti}, M. and {Bakanov}, A.~V. and {Bamford}, S.~P. and {Barentsen}, G. and {Barmby}, P. and {Baumbach}, A. and {Berry}, K.~L. and {Biscani}, F. and {Boquien}, M. and {Bostroem}, K.~A. and {Bouma}, L.~G. and {Brammer}, G.~B. and {Bray}, E.~M. and {Breytenbach}, H. and {Buddelmeijer}, H. and {Burke}, D.~J. and {Calderone}, G. and {Cano Rodr{\'\i}guez}, J.~L. and {Cara}, M. and {Cardoso}, J.~V.~M. and {Cheedella}, S. and {Copin}, Y. and {Corrales}, L. and {Crichton}, D. and {D'Avella}, D. and {Deil}, C. and {Depagne}, {\'E}. and {Dietrich}, J.~P. and {Donath}, A. and {Droettboom}, M. and {Earl}, N. and {Erben}, T. and {Fabbro}, S. and {Ferreira}, L.~A. and {Finethy}, T. and {Fox}, R.~T. and {Garrison}, L.~H. and {Gibbons}, S.~L.~J. and {Goldstein}, D.~A. and {Gommers}, R. and {Greco}, J.~P. and {Greenfield}, P. and {Groener}, A.~M. and {Grollier}, F. and {Hagen}, A. and {Hirst}, P. and {Homeier}, D. and {Horton}, A.~J. and {Hosseinzadeh}, G. and {Hu}, L. and {Hunkeler}, J.~S. and {Ivezi{\'c}}, {\v{Z}}. and {Jain}, A. and {Jenness}, T. and {Kanarek}, G. and {Kendrew}, S. and {Kern}, N.~S. and {Kerzendorf}, W.~E. and {Khvalko}, A. and {King}, J. and {Kirkby}, D. and {Kulkarni}, A.~M. and {Kumar}, A. and {Lee}, A. and {Lenz}, D. and {Littlefair}, S.~P. and {Ma}, Z. and {Macleod}, D.~M. and {Mastropietro}, M. and {McCully}, C. and {Montagnac}, S. and {Morris}, B.~M. and {Mueller}, M. and {Mumford}, S.~J. and {Muna}, D. and {Murphy}, N.~A. and {Nelson}, S. and {Nguyen}, G.~H. and {Ninan}, J.~P. and {N{\"o}the}, M. and {Ogaz}, S. and {Oh}, S. and {Parejko}, J.~K. and {Parley}, N. and {Pascual}, S. and {Patil}, R. and {Patil}, A.~A. and {Plunkett}, A.~L. and {Prochaska}, J.~X. and {Rastogi}, T. and {Reddy Janga}, V. and {Sabater}, J. and {Sakurikar}, P. and {Seifert}, M. and {Sherbert}, L.~E. and {Sherwood-Taylor}, H. and {Shih}, A.~Y. and {Sick}, J. and {Silbiger}, M.~T. and {Singanamalla}, S. and {Singer}, L.~P. and {Sladen}, P.~H. and {Sooley}, K.~A. and {Sornarajah}, S. and {Streicher}, O. and {Teuben}, P. and {Thomas}, S.~W. and {Tremblay}, G.~R. and {Turner}, J.~E.~H. and {Terr{\'o}n}, V. and {van Kerkwijk}, M.~H. and {de la Vega}, A. and {Watkins}, L.~L. and {Weaver}, B.~A. and {Whitmore}, J.~B. and {Woillez}, J. and {Zabalza}, V. and {Astropy Contributors}},
        title = "{The Astropy Project: Building an Open-science Project and Status of the v2.0 Core Package}",
      journal = {\aj},
         year = 2018,
        month = sep,
       volume = {156},
       number = {3},
          eid = {123},
        pages = {123},
          doi = {10.3847/1538-3881/aabc4f},
archivePrefix = {arXiv},
       eprint = {1801.02634},
 primaryClass = {astro-ph.IM},
       adsurl = {https://ui.adsabs.harvard.edu/abs/2018AJ....156..123A}
}

@ARTICLE{Astropy13,
       author = {{Astropy Collaboration} and {Robitaille}, Thomas P. and {Tollerud}, Erik J. and {Greenfield}, Perry and {Droettboom}, Michael and {Bray}, Erik and {Aldcroft}, Tom and {Davis}, Matt and {Ginsburg}, Adam and {Price-Whelan}, Adrian M. and {Kerzendorf}, Wolfgang E. and {Conley}, Alexander and {Crighton}, Neil and {Barbary}, Kyle and {Muna}, Demitri and {Ferguson}, Henry and {Grollier}, Fr{\'e}d{\'e}ric and {Parikh}, Madhura M. and {Nair}, Prasanth H. and {Unther}, Hans M. and {Deil}, Christoph and {Woillez}, Julien and {Conseil}, Simon and {Kramer}, Roban and {Turner}, James E.~H. and {Singer}, Leo and {Fox}, Ryan and {Weaver}, Benjamin A. and {Zabalza}, Victor and {Edwards}, Zachary I. and {Azalee Bostroem}, K. and {Burke}, D.~J. and {Casey}, Andrew R. and {Crawford}, Steven M. and {Dencheva}, Nadia and {Ely}, Justin and {Jenness}, Tim and {Labrie}, Kathleen and {Lim}, Pey Lian and {Pierfederici}, Francesco and {Pontzen}, Andrew and {Ptak}, Andy and {Refsdal}, Brian and {Servillat}, Mathieu and {Streicher}, Ole},
        title = "{Astropy: A community Python package for astronomy}",
      journal = {\aap},
         year = 2013,
        month = oct,
       volume = {558},
          eid = {A33},
        pages = {A33},
          doi = {10.1051/0004-6361/201322068},
archivePrefix = {arXiv},
       eprint = {1307.6212},
 primaryClass = {astro-ph.IM},
       adsurl = {https://ui.adsabs.harvard.edu/abs/2013A&A...558A..33A}
}

@ARTICLE{Labbe23,
       author = {{Labb{\'e}}, Ivo and {van Dokkum}, Pieter and {Nelson}, Erica and {Bezanson}, Rachel and {Suess}, Katherine A. and {Leja}, Joel and {Brammer}, Gabriel and {Whitaker}, Katherine and {Mathews}, Elijah and {Stefanon}, Mauro and {Wang}, Bingjie},
        title = "{A population of red candidate massive galaxies  600 Myr after the Big Bang}",
      journal = {\nat},
         year = 2023,
        month = apr,
       volume = {616},
       number = {7956},
        pages = {266-269},
          doi = {10.1038/s41586-023-05786-2},
archivePrefix = {arXiv},
       eprint = {2207.12446},
 primaryClass = {astro-ph.GA},
       adsurl = {https://ui.adsabs.harvard.edu/abs/2023Natur.616..266L}
}

@ARTICLE{Fukugita96,
       author = {{Fukugita}, M. and {Ichikawa}, T. and {Gunn}, J.~E. and {Doi}, M. and {Shimasaku}, K. and {Schneider}, D.~P.},
        title = "{The Sloan Digital Sky Survey Photometric System}",
      journal = {\aj},
         year = 1996,
        month = apr,
       volume = {111},
        pages = {1748},
          doi = {10.1086/117915},
       adsurl = {https://ui.adsabs.harvard.edu/abs/1996AJ....111.1748F}
}

@ARTICLE{oke83,
       author = {{Oke}, J.~B. and {Gunn}, J.~E.},
        title = "{Secondary standard stars for absolute spectrophotometry.}",
      journal = {\apj},
         year = 1983,
        month = mar,
       volume = {266},
        pages = {713-717},
          doi = {10.1086/160817},
       adsurl = {https://ui.adsabs.harvard.edu/abs/1983ApJ...266..713O}
}

@ARTICLE{Matthee24,
       author = {{Matthee}, Jorryt and {Naidu}, Rohan P. and {Brammer}, Gabriel and {Chisholm}, John and {Eilers}, Anna-Christina and {Goulding}, Andy and {Greene}, Jenny and {Kashino}, Daichi and {Labbe}, Ivo and {Lilly}, Simon J. and {Mackenzie}, Ruari and {Oesch}, Pascal A. and {Weibel}, Andrea and {Wuyts}, Stijn and {Xiao}, Mengyuan and {Bordoloi}, Rongmon and {Bouwens}, Rychard and {van Dokkum}, Pieter and {Illingworth}, Garth and {Kramarenko}, Ivan and {Maseda}, Michael V. and {Mason}, Charlotte and {Meyer}, Romain A. and {Nelson}, Erica J. and {Reddy}, Naveen A. and {Shivaei}, Irene and {Simcoe}, Robert A. and {Yue}, Minghao},
        title = "{Little Red Dots: An Abundant Population of Faint Active Galactic Nuclei at z {\ensuremath{\sim}} 5 Revealed by the EIGER and FRESCO JWST Surveys}",
      journal = {\apj},
         year = 2024,
        month = mar,
       volume = {963},
       number = {2},
          eid = {129},
        pages = {129},
          doi = {10.3847/1538-4357/ad2345},
archivePrefix = {arXiv},
       eprint = {2306.05448},
 primaryClass = {astro-ph.GA},
       adsurl = {https://ui.adsabs.harvard.edu/abs/2024ApJ...963..129M}
}

@ARTICLE{Tee25,
       author = {{Tee}, Wei Leong and {Fan}, Xiaohui and {Wang}, Feige and {Yang}, Jinyi},
        title = "{Lack of Rest-frame Ultraviolet Variability in Little Red Dots Based on HST and JWST Observations}",
      journal = {\apjl},
         year = 2025,
        month = apr,
       volume = {983},
       number = {1},
          eid = {L26},
        pages = {L26},
          doi = {10.3847/2041-8213/adc5e3},
archivePrefix = {arXiv},
       eprint = {2412.05242},
 primaryClass = {astro-ph.GA},
       adsurl = {https://ui.adsabs.harvard.edu/abs/2025ApJ...983L..26T}
}

@ARTICLE{Liu25,
       author = {{Liu}, Hanpu and {Jiang}, Yan-Fei and {Quataert}, Eliot and {Greene}, Jenny E. and {Ma}, Yilun},
        title = "{The Balmer Break and Optical Continuum of Little Red Dots from Super-Eddington Accretion}",
      journal = {\apj},
         year = 2025,
        month = nov,
       volume = {994},
       number = {1},
          eid = {113},
        pages = {113},
          doi = {10.3847/1538-4357/ae0c19},
archivePrefix = {arXiv},
       eprint = {2507.07190},
 primaryClass = {astro-ph.GA},
       adsurl = {https://ui.adsabs.harvard.edu/abs/2025ApJ...994..113L}
}

@ARTICLE{Madau26,
       author = {{Madau}, Piero and {Maiolino}, Roberto},
        title = "{Little Red Dots as Obscured Little Blue Dots: A Super-Eddington Unification Model}",
      journal = {arXiv e-prints},
         year = 2026,
        month = feb,
          eid = {arXiv:2602.22386},
        pages = {arXiv:2602.22386},
          doi = {10.48550/arXiv.2602.22386},
archivePrefix = {arXiv},
       eprint = {2602.22386},
 primaryClass = {astro-ph.GA},
       adsurl = {https://ui.adsabs.harvard.edu/abs/2026arXiv260222386M}
}

@ARTICLE{Kido25,
       author = {{Kido}, Daisaburo and {Ioka}, Kunihito and {Hotokezaka}, Kenta and {Inayoshi}, Kohei and {Irwin}, Christopher M.},
        title = "{Black hole envelopes in Little Red Dots}",
      journal = {\mnras},
         year = 2025,
        month = dec,
       volume = {544},
       number = {4},
        pages = {3407-3416},
          doi = {10.1093/mnras/staf1898},
archivePrefix = {arXiv},
       eprint = {2505.06965},
 primaryClass = {astro-ph.HE},
       adsurl = {https://ui.adsabs.harvard.edu/abs/2025MNRAS.544.3407K}
}
\bibliographystyle{aasjournal}

\end{document}